\pdfoutput=1
\documentclass[fleqn,usenatbib,useAMS]{mnras}

\usepackage{newtxtext,newtxmath}

\usepackage[T1]{fontenc}
\usepackage{ae,aecompl}

\usepackage{graphicx}	
\usepackage{amsmath}	
\usepackage{amssymb}	
\usepackage{multicol}        
\usepackage{bm}		
\usepackage{pdflscape}	
\usepackage{epstopdf}
\usepackage[T1]{fontenc}
\usepackage{ae,aecompl}

\newcommand{\ms}[1]{#1}

\newcommand{\beq}{\begin{equation}}
\newcommand{\beqa}{\begin{eqnarray}}
\newcommand{\eeq}{\end{equation}}
\newcommand{\eeqa}{\end{eqnarray}}

\newcommand{\simgt}{\lower.5ex\hbox{$\; \buildrel > \over \sim \;$}}
\newcommand{\simlt}{\lower.5ex\hbox{$\; \buildrel < \over \sim \;$}}

\newcommand{\bd}[1]{\mbox{\boldmath $#1$}}

\title[Joint analysis with tSZ, X rays and WL]{
Probing Cosmology and Cluster Astrophysics with Multi-Wavelength Surveys~I.~ Correlation Statistics}

\author[M. Shirasaki, E. T. Lau \& D. Nagai]{
Masato Shirasaki$^{1}$\thanks{E-mail: masato.shirasaki@nao.ac.jp},
Erwin T. Lau$^{2}$
and Daisuke Nagai$^{3,4}$
\\
$^{1}$National Astronomical Observatory of Japan, 
Mitaka, Tokyo 181-8588, Japan\\
$^{2}$Department of Physics, University of Miami, Coral Gables, FL 33124, USA\\
$^{3}$Department of Physics, Yale University, New Haven, CT 06520, USA\\
$^{4}$Yale Center for Astronomy and Astrophysics, Yale University, New Haven, CT 06520,USA
}

\date{Accepted XXX. Received YYY; in original form ZZZ}

\pubyear{2019}

\hypersetup{draft}

\begin{document}
\label{firstpage}
\pagerange{\pageref{firstpage}--\pageref{lastpage}}
\maketitle

\begin{abstract}
Upcoming multi-wavelength astronomical surveys will soon discover all massive galaxy clusters and provide unprecedented constraints on cosmology and cluster astrophysics. 
In this paper, we investigate the constraining power of the multi-band cluster surveys, through a joint analysis of three observables associated with clusters of galaxies, including thermal Sunyaev-Zel'dovich (tSZ) effect in cosmic microwave background (CMB), X-ray emission of ionized gas, and gravitational weak lensing effect of background galaxies by the cluster's gravitational potential. We develop a theoretical framework to predict and interpret two-point correlation statistics among the three observables using a semi-analytic model of intracluster medium (ICM) and halo-based approach. 
In this work, we show that the auto- and cross-angular power spectra in tSZ, X-ray and lensing statistics from upcoming missions (eROSITA, CMB-S4, and LSST) can help break the degeneracy between cosmology and ICM physics. These correlation statistics are less sensitive to selection biases, and are able to probe ICM physics in distant, faint and small clusters that are otherwise difficult to be detected individually. 
We show that the correlation statistics are able to provide cosmological constraints 
\ms{comparable} 
to the conventional cluster abundance measurements, while constraining cluster astrophysics at the same time. Our results indicate that the correlation statistics can significantly enhance the scientific returns of upcoming multi-wavelength cluster surveys.
\end{abstract}



\section{INTRODUCTION}

Upcoming multi-wavelength cosmological surveys will provide unprecedented images of the distribution of dark matter, gas and stars in the universe. Clusters of galaxies are the most massive gravitationally bound objects in the universe. They form at the highest density peaks in the large-scale structure of the universe. Their abundance traces the growth of structure of the universe, providing unique constraints on cosmology \citep[e.g.][for review]{allen11}. 

One of the primary challenges facing this field lies in accurately calibrating the scaling relations between cluster masses and the X-ray, thermal Sunyaev-Zel'dovich (tSZ) effect, and lensing observables \citep[e.g.][for a review]{pratt19}. The major astrophysical uncertainties include (i) the effects of feedback from supernovae (SNe) and active galactic nuclei (AGN) \citep[e.g.][]{puchwein_etal08,planelles_etal14,lebrun_etal17}, (ii) non-thermal pressure due to merger-induced gas motions \citep[e.g.,][]{lau09,battaglia12,nelson14b,shi_komatsu14,shi15}, which introduce biases in hydrostatic mass estimates \citep[e.g.][]{nagai07a,lau13,nelson14a,shi16,biffi16}, and (iii) gas clumping in galaxy cluster outskirts, which biases the density and temperature profile measurements in the outskirts of clusters \citep[e.g.][for  a review]{nagai11,vazza13,rasia14,avestruz16,walker19}.

With upcoming large-scale galaxy cluster surveys, astrophysical systematics such as feedback from SNe and AGN, non-thermal pressure support, and gas clumping, need to be controlled for the large samples of clusters in upcoming X-ray and microwave. In the X-ray band, the upcoming eROSITA mission will potentially improve cluster cosmological constraints by detecting very large number ($\sim10^5$) of galaxy clusters out to redshifts $z>1$ \citep{merloni12}. In the microwave, ongoing ground-based experiments by the Atacama Cosmology Telescope (ACT) and the South Pole Telescope (SPT) will perform comprehensive studies of galaxy clusters through the tSZ effect \citep[e.g.][]{hilton18,chown18,bocquet19}, while the next-generation ground-based experiment of CMB, the CMB-S4, will obtain precise measurements of the properties of ICM via tSZ effect and the mass distribution via CMB lensing \citep{CMBS4-19}. In the optical regime, ongoing galaxy imaging surveys will measure the lensing effect of large-scale structures with a few percent level accuracy \citep[e.g.][]{hildebrandt17,troxel18,hikage19},
which will be further improved with future surveys with the Wide Field Infrared Survey Telescope (WFIRST\footnote{\url{https://wfirst.gsfc.nasa.gov/}}), the Large Synoptic Survey Telescope (LSST\footnote{\url{https://www.lsst.org/}}), and the Euclid satellite\footnote{\url{http://sci.esa.int/euclid/}}.

Multi-wavelength correlation statistics (i.e., correlation function or its Fourier-space equivalent, the power spectrum) are promising in providing robust statistical constraints on cluster astrophysics \citep[e.g.][]{shaw10,battaglia10} and cosmology \citep[e.g.][]{komatsu02,shirasaki16,bolliet18,makiya19}. 
Correlation statistics are complementary to the conventional cluster abundance measurements, as they are less sensitive to selection biases, and are able to probe distant, faint and small clusters which will not be resolved on an individual-by-individual basis. In particular, given the large amount datasets coming online, cross-correlation or clusters across different wavebands will allow us to probe the baryonic and dark matter content in galaxy clusters that will improve our understanding of cluster astrophysics and cosmology \citep[e.g.][]{vanwaerbeke14,hill14,battaglia15,ma15,hojjati17,osato18,lakey19}.

This paper is the first part of a series of papers where we investigate the use of multi-wavelength correlation statistics in constraining cosmology and cluster astrophysics. 
In this paper, we introduce a framework for jointly constraining cluster astrophysics and cosmology using multi-wavelength angular power spectra of galaxy clusters. 
Specifically, we use a semi-analytic model of ICM based from \citet{shaw10,flender17}, which provide physical prescriptions for ICM profiles in dark matter halos, including gas physics relevant to the tSZ and X-ray observables, such as cluster cool-cores,  non-thermal pressure. In this paper, we extend the previous model by incorporating the effects of gas clumping in cluster outskirts. 
We then provide Fisher forecasts on cluster astrophysics and cosmology from the multi-wavelength angular power spectrum analysis for upcoming surveys, in particular eROSITA in X-ray, CMB-S4 in microwave, and LSST in optical. Our forthcoming projects include validating our modeling of the correlation statistics with hydrodynamical simulations and applying our model to observational datasets.

This paper is organized as follows.
In \S\ref{sec:obs}, we provide an overview of galaxy cluster observables in X-ray, microwave, and lensing.
We present our semi-analytic model of ICM in \S\ref{sec:SAM} and then describe a theoretical framework for computing two-point auto- and cross- correlations of the galaxy cluster observables in \S\ref{sec:powerspec}.
In \S\ref{sec:method}, we summarize the statistical uncertainties of two-point correlations from our Fisher analysis to forecast the parameter constraints of cosmology and ICM physics in upcoming surveys. 
\S\ref{sec:res} presents the main results in this paper, which includes (i) the expected parameter constraints in a hypothetical set of future multi-wavelength surveys with a large sky coverage of 20,000 square degrees,
(ii) the dependence of survey configurations and analysis methods on our Fisher forecast, and
(iii) possible systematic uncertainties induced by the statistical fluctuation in 
reconstruction method of tSZ effects and mis-estimation of lensing systematics.
We discuss the current issues in our modeling of two-point correlation analysis in \S\ref{sec:caveat}, 
and conclude this paper in \S\ref{sec:conclusions}.

\section{OBSERVABLES}\label{sec:obs}

Throughout this paper, we assume General Relativity under a Friedmann Robertson Walker (FRW) metric 
with a spatially flat curvature. Within this framework, the governing equation of late-time expansion of the universe 
is given by
\footnotesize
\begin{align}
H(a) &= \frac{{\rm d}\ln a}{{\rm d}t}= H_{0}\Bigg\{
\Omega_{{\rm m}0}a^{-3}+\Omega_{\rm DE, 0}
\exp\left[-3\int_{1}^{a} \frac{{\rm d}a^{\prime}}{a^{\prime}} (1+w_{\rm DE}(a^{\prime}))\right]
\Bigg\}^{1/2}, \label{eq:Hubble}
\end{align}
\normalsize
where 
$a$ is the scale factor,
$H(a)$ is known as the Hubble parameter and 
$H_{0}=100h\, {\rm km}\, {\rm s}^{-1}\, {\rm Mpc}^{-1}$ is the present value of $H(a)$.
Note that we work with the usual normalization $a=1$ at present.
The $\Omega_{\rm m0}a^{-3}$ term in Eq.~(\ref{eq:Hubble}) represents the contribution from cosmic matter density,
while the second term in the right-hand side arises from dark energy causing the accelerated expansion of the universe at $a\simgt0.5$.
The equation of state of dark energy is specified as $w_{\rm DE}(a) = P_{\rm DE}(a)/\rho_{\rm DE}(a)$, 
where $P_{\rm DE}$ and $\rho_{\rm DE}$ denote pressure and density of dark energy, respectively.
Ongoing and future surveys of large-scale structure of the universe aim at measuring $w_{\rm DE}(a)$ at different epoch with high precision and exploring a possible deviation from the cosmological constant with $w_{\rm DE}(a) = -1$.
In this sense, we adopt the simplest model of dark energy with constant value of $w_{0}$, i.e. $w_{\rm DE}(a) = w_0$, 
and study the constraining power of $w_{0}$ using several observables accessible to upcoming multi-wavelength astronomical surveys. 

\subsection{Thermal Sunyaev-Zel'dovich effect}
The thermal Sunyaev-Zel'dovich (tSZ) effect probes the thermal pressure of hot electrons in galaxy clusters. 
At frequency $\nu$, the change in CMB temperature by the tSZ effect is expressed as
\beq
\frac{\Delta T}{T_0} = g(x) y, \label{eq:T-y}
\eeq
where $T_{0}=2.725\, {\rm K}$ is the CMB temperature \citep{2009ApJ...707..916F}, 
$g(x) = x{\rm coth}(x/2)-4$ with $x=h_{\rm P}\nu/k_{\rm B}T_{0}$,
$h_{\rm P}$ and $k_{\rm B}$ are the Planck constant and the Boltzmann constant, respectively\footnote{In this paper, we ignore the relativistic correction for $g(x)$ which is only important for the tSZ effects in most massive galaxy clusters \citep{1998ApJ...502....7I,2012MNRAS.426..510C}.}.
Compton parameter $y$ is obtained as the integral of the electron pressure $P_{\rm e}$ along a line of sight:
\beq
y(\bd{\theta}) = \int_0^{\chi_H}\, \frac{{\rm d}\chi}{1+z}\, \frac{k_{\rm B}\sigma_{\rm T}}{m_{\rm e}c^2} P_{\rm e}\left(f_{\rm K}(\chi)\bd{\theta}, z(\chi)\right), \label{eq:tSZ_y}
\eeq
where $\sigma_{\rm T}$ is the Thomson cross section, $\chi$ is the comoving radial distance to redshift $z$, 
$f_{\rm K}(\chi)$ is the comoving angular diameter distance, and $\chi_{H}$ is the comoving distance up to $z\rightarrow \infty$.

\subsection{X-ray surface brightness}

Another probe of ICM is X-ray surface brightness, defined as the number of X-ray photons per unit time per unit area per solid angle
in a given range of energy $E_{\rm min}$ to $E_{\rm max}$. 
The cumulative X-ray surface brightness from ionized gas along a line of sight is given by
\beq
x(\bd{\theta}) = \frac{1}{4\pi}\int_0^{\chi_H}\, \frac{{\rm d}\chi}{1+z}\, \frac{\epsilon_{\rm X}\left(f_{\rm K}\bd{\theta}, z\right)}{(1+z)^4}\, 
n_{\rm e}\left(f_{\rm K}\bd{\theta}, z\right)
n_{\rm H}\left(f_{\rm K}\bd{\theta}, z\right), \label{eq:XSB}
\eeq
where
$\epsilon_{\rm X}$ represents the X-ray emissivity for the energy band of interest,
$n_{\rm e}$ and $n_{\rm H}\approx n_{\rm e}/1.2$ are the number density of electrons and protons, respectively.
The X-ray emissivity depends on the temperature of plasma $T$ and gas metalicity $Z$. 
To be specific, $\epsilon_{\rm X}$ is computed as
\beq
\epsilon_{\rm X}({\bd r}, z) = \int_{E_{\rm min}(1+z)}^{E_{\rm max}(1+z)}\, {\rm d}E\, \Lambda\left(T({\bd r}, z), Z, E\right), \label{eq:emis_X}
\eeq
where $\Lambda(T, Z, E)$ is the volume emissivity (the number of photons emitted per unit volume, time, and energy range)
at energy $E$ for a plasma of temperature $T$.
The receiver's frame energy range is set as $[E_{\rm min},E_{\rm max}] = [0.5,2.0]$~keV. 
In this paper, we use the atomic database $\tt AtomDB$\footnote{\url{http://www.atomdb.org/download.php}} to compute \ms{$\Lambda(T, Z, E)$} with $Z=0.3Z_{\odot}$, where $Z_{\odot}$ is the solar metallicity.
We include Galactic absorption $\exp[-\sigma(E) N_{\rm H}]$ where $\sigma(E)$ is the \ms{ effective photoelectric absorption cross section of hydrogen atom,  taken from \cite{mm83}}.  The hydrogen column density is set to be $N_{\rm H} = 2.5 \times 10^{20} {\rm cm^{-2}}$, which is estimated as the averaged $N_{\rm H}$ for the full sky with the Galaxy masked\footnote{\url{https://lambda.gsfc.nasa.gov/product/foreground/fg_HI_get.cfm}}.
Note that we define the surface brightness in units of ${\rm photons}\, {\rm s^{-1}}{\rm cm^{-2}}{\rm arcsec}^{-2}$ 
throughout this paper.

\subsection{Weak gravitational lensing in galaxy's image}

Weak gravitational lensing offers a direct probe of the mass distribution of galaxy clusters.
Relevant quantity of weak lensing effect in galaxy imaging surveys is known as convergence field $\kappa$,
which is related to 
the second derivative of the gravitational potential $\Phi$ \citep{Bartelmann2001}.
Using the Poisson equation and the Born approximation, 
one can express the weak lensing convergence field as the weighted 
integral of matter overdensity field $\delta_{\rm m}(\bd{x})$:
\beq
\kappa(\bd{\theta})
= \int_{0}^{\chi_{H}} {\rm d}\chi \, q(\chi)\, \delta_{\rm m}(\chi,f_{\rm K}(\chi)\bd{\theta}), \label{eq:kappa_delta}
\eeq
where $q(\chi)$ is called lensing kernel.
For a given redshift distribution of source galaxies,
the lensing kernel is expressed as
\beq
q(\chi) = \frac{3}{2}\left( \frac{H_{0}}{c}\right)^2 \Omega_{\rm m0} \frac{f_{\rm K}(\chi)}{a(\chi)}\, \int_{\chi}^{\chi_{H}} {\rm d}\chi^{\prime} p(\chi^{\prime})\frac{f_{\rm K}(\chi^{\prime}-\chi)}{f_{\rm K}(\chi^{\prime})}, \label{eq:lens_kernel}
\eeq
where $p(\chi)$ represents the redshift distribution of source galaxies
normalized to $\int {\rm d}\chi \, p(\chi) =1$.

\section{A semi-analytic model of intracluster medium}\label{sec:SAM}

To predict the statistical property of three fields defined by Eqs.~(\ref{eq:tSZ_y}), (\ref{eq:XSB}) and (\ref{eq:kappa_delta}), 
we adopt a physically motivated semi-analytic ICM model as developed in \citet{shaw10, flender17}. This ICM model is the modified version of the one originally proposed in \citet{ostriker05}.
The updated features from the original model include a radially-dependent non-thermal pressure component 
(e.g., turbulence or bulk flows) \citep{nelson14b} and gas-cooling effect in cluster cores to provide a better description of recent X-ray data \citep{flender17}. 

\subsection{Dark matter distribution}

The ICM model assumes that the gas inside cluster-sized dark matter halos initially follows the underlying dark matter density distributions which is modeled as Navarro-Frenk-White (NFW) profile \citep{NFW}:
\beq
\rho_{\rm DM}(r) = \frac{\rho_s}{\left(r/r_s\right)\left(1+r/r_s\right)^2},
\eeq
where $r_s$ and $\rho_s$ are a scale radius and scale density, respectively.
These parameters $r_s$ and $\rho_s$ are related to
the concentration $c_{\Delta}=r_{\Delta}/r_s$, through the definitions of the halo mass 
\begin{align}
M_{\Delta} \equiv& \frac{4\pi}{3} \Delta \, \rho_{\rm crit}(z) r^3_{\Delta}, \label{eq:SOmass_def} \\
M_{\Delta} =& \int_{0}^{r_{\Delta}}\, {\rm d}x\, 4\pi x^{2}\rho_{\rm NFW}(x), \label{eq:SOmass_NFW}
\end{align}
where $\rho_{\rm crit}(z)$ is the critical density in the universe at redshift $z$ and $\Delta$ is the overdensity parameter.
Eqs~(\ref{eq:SOmass_def}) and (\ref{eq:SOmass_NFW}) reduce to
\beq
\frac{\rho_s}{\rho_{\rm crit}} = \frac{\Delta}{3}\frac{c^3_{\Delta}}{\left[\ln (1+c_{\Delta})-c_{\Delta}/(1+c_{\Delta})\right]}.
\eeq
Various overdensity parameters have been adopted in the literature. In this paper, we define the virial overdensity parameter 
as in \citet{1998ApJ...495...80B}:
\beq
\Delta_{\rm vir}(z) = 18\pi^2+82(\Omega_{\rm M}(z)-1)-39(\Omega_{\rm M}(z)-1)^2, \label{eq:delta_v}
\eeq
where $\Omega_{\rm M}(z) \equiv \Omega_{\rm m0}(1+z)^{3}/[\Omega_{\rm m0}(1+z)^{3}+(1-\Omega_{\rm m0})]$.
Throughout this paper, we set the fiducial mass to be the virial mass with Eq.~(\ref{eq:delta_v}).
For the concentration parameter, we adopt the model developed in \citet[][DK15 hereafter]{2015ApJ...799..108D}, which
has been calibrated with dark-matter-only $N$-body simulations. The halo concentration can be affected by baryonic processes such as the star formation and additional AGN feedback \citep[e.g.][]{2004ApJ...616...16G, duffy10, 2012MNRAS.424.1244F, shirasaki18}.
We include baryonic effects in halo concentration by varying the normalization of halo concentration:
\beq
c_{\rm vir}(z, M) = A_{C} \, c_{\rm vir, DK15}(z,M), \label{eq:c_z_M}
\eeq
where $c_{\rm vir} = r_{\rm vir}/r_s$ and $c_{\rm vir, DK15}$ is the predicted concentration by the DK15 model. We use the fitting formula from \citet{2003ApJ...584..702H} for the conversion between masses of different spherical overdensities. 

\begin{table*}
\caption{
	The list of free parameters involved in the ICM model in this paper.
	\label{tb:params_ICM}
	}
\scalebox{0.80}[0.80]{
\begin{tabular}{@{}llcc}
\hline
\hline
Name & Physical meaning & Reference & Fiducial value \\ 
\hline
Dark matter structure & & & \\
\hline
$A_C$ & Normalization in redshift-mass-concentration relation & Eq.~(\ref{eq:c_z_M}) & 1.0 \\
\hline
Adiabatic index of ICM & & & \\
\hline
$\tilde{\Gamma}$ & Index in inner region at present & Eq.~(\ref{eq:Gamma_ICM}) & 0.10 \\
$\gamma$ & Redshift evolution of gas cooling effects & Eq.~(\ref{eq:Gamma_ICM}) & 1.72 \\
$x_{\rm break}$ & Size of cooling core region (in units of $r_{500}$) & Eq.~(\ref{eq:Gamma_ICM}) & 0.195 \\
\hline
Feedback processes & & & \\
\hline
$\epsilon_{\rm DM}$ & Energy introduced during major halo mergers (in the unit of total energy of an NFW halo)& Eq.~(\ref{eq:E_balance}) & 0.01 \\
$\epsilon_{f}$ & Energy injected by feedback from SNe and AGN (in units of rest mass energy of stars $M_\star c^2$) & Eq.~(\ref{eq:E_balance}) & $4\times10^{-6}$ \\
\hline
Stellar mass fraction & & & \\
\hline
$f_{*}$ & Stellar mass fraction at $M=3\times10^{14}\, M_{\odot}$ & Eq.~(\ref{eq:fstar}) & 0.026 \\
$S_{*}$ & Dependence of halo mass & Eq.~(\ref{eq:fstar}) & 0.12 \\
\hline
Non-thermal pressure & & & \\
\hline
$\gamma_{\rm nt}$ & Radial slope index & Eq.~(\ref{eq:P_nonth}) & 1.628 \\
$A_{\rm nt}$ & Overall amplitude of the fraction of non-thermal pressure & Eq.~(\ref{eq:P_nonth}) & 0.452 \\
$B_{\rm nt}$ & Scale radius of non-thermal pressure (in the unit of $r_{\rm 200m}$) & Eq.~(\ref{eq:P_nonth}) & 0.841 \\
\hline
Gas clumping & & & \\
\hline
$C_0$ & Amplitude of gas clumping effects & Eq.~(\ref{eq:cfac}) & 0.90 \\
$\alpha_C$ & Shape parameter of clumping profile & Eq.~(\ref{eq:cfac}) & 1.00 \\
$\beta_C$ & Shape parameter of clumping profile & Eq.~(\ref{eq:cfac}) & 6.00 \\
$\gamma_{C}$ & Shape parameter of clumping profile & Eq.~(\ref{eq:cfac}) & 3.00 \\
\hline
\end{tabular}
}
\end{table*}

\subsection{Gas distribution}\label{app_sub:ICM}

Given the gravitational potential determined by NFW profile, we then assume the gas inside every halo
re-distributes so that its state can be described by hydrostatic equilibrium. 
The governing equation of hydrostatic equilibrium is given by
\beq
\frac{{\rm d}P_{\rm tot}}{{\rm d}r} = -\rho_{g}\frac{{\rm d}\Phi}{{\rm d}r}, \label{eq:HSE}
\eeq
where $P_{\rm tot}$ is the total (thermal+non-thermal) pressure of gas, $\rho_g$ is the gas density,
and $\Phi$ represents the NFW gravitational potential.
We further assume a polytrope solution of Eq.~(\ref{eq:HSE}), which is written as
\begin{align}
P_{\rm tot}(r) =& P_{0} \, \theta^{n+1}(r), \\
\rho_g(r) =& \rho_0 \, \theta^{n}(r),
\end{align}
where $n$ is the polytropic index and $\theta(r)$ is the polytropic variable
\beq
\theta(r) = 1+\frac{\Gamma-1}{\Gamma}\frac{\rho_0}{P_0}\left(\Phi_0 - \Phi(r)\right). \label{eq:theta}
\eeq
In Eq.~(\ref{eq:theta}), $\Phi_0$ is the NFW potential at the halo centre, and $\Gamma=1+1/n$ is the polytropic exponent.

Cosmological hydrodynamical simulations suggest that the ICM follows $\Gamma\simeq 1.2$ except in high-density core \citep{ostriker05,shaw10,2012ApJ...758...75B}.
Gas cooling due to some strong feedback at the cluster core can change the polytropic exponent. 
The gas density in the inner region, relative to the critical density of the universe, is shown to be evolving with time  \citep{2013ApJ...774...23M}.
Hence we adopt a two-value polytropic model, one for the outskirts, and one for the core:
\beq
\Gamma(r,z) = \left\{
\begin{array}{cc}
1.2 & (r/r_{500} > x_{\rm break})  \\
\tilde{\Gamma}(1+z)^{\gamma} &  (\rm otherwise)\\
\end{array}
\right. \label{eq:Gamma_ICM}
\eeq
where $r_{500}$ is the spherical over-density radius with $\Delta=500$ in Eq.~(\ref{eq:SOmass_def})
and $\tilde{\Gamma}$, $\gamma$ are free parameters in our ICM model, and $x_{\rm break}$ is fixed to be $0.195$, which defines the cluster core region that is dominated by cooling and feedback physics.  

The two unknowns, i.e. the normalizations in gas pressure profile $P_{0}$ and gas density profile $\rho_0$, are solved by two conditions.
The first condition describes the energy balance between initial and final states of gas, given as
\beq
E_{g,f} = E_{g,i}+\epsilon_{\rm DM}|E_{\rm DM}|+\epsilon_{f}M_{*}c^2+\Delta E_{p}, \label{eq:E_balance}
\eeq
where the left hand side of the equation is the final energy of the ICM,
and the right hand side includes the initial energy of the ICM (denoted as $E_{g,i}$)
and the work done by gas as it expands relative to its initial state ($\Delta E_{p}$).
The two terms of $\epsilon_{\rm DM}|E_{\rm DM}|$ and $\epsilon_{f}M_{*}c^2$ in Eq.~(\ref{eq:E_balance})
control the efficiency of two different feedback processes with two free parameters $\epsilon_{\rm DM}$ and $\epsilon_{f}$. 
The former is the energy introduced into the ICM during major halo mergers via dynamical friction heating,
where $E_{\rm DM}$ is the total energy in the dark matter halo
(i.e., the sum of kinetic and potential energy).
The latter is the energy injected into the ICM due to feedback from SNe and AGN, where $M_*$ is the stellar mass in halo.
Throughout this paper, we parametrize the stellar-to-halo mass relation as
\beq
\frac{M_{*}}{M_{500}} = f_{*}\left(\frac{M_{500}}{3\times10^{14}\,M_{\odot}}\right)^{-S_{*}}, \label{eq:fstar}
\eeq
where $f_{*}$ and $S_{*}$ are free parameters in the model.

The second condition to set $P_0$ and $\rho_0$ is the boundary condition of gas pressure. The boundary condition is determined by setting the total gas pressure $P_{\rm tot}$ at the boundary radius $r_f$ to the effective DM ``pressure'' $P_{\rm DM}$ at the virial radius of the NFW halo scaled by the cosmic baryon fraction:
\beq
P_{\rm tot}(r_f) = \frac{\Omega_{\rm b0}}{\Omega_{\rm m0}}\, P_{\rm DM}(r_{\rm vir}), \label{eq:P_bound}
\eeq
where $\Omega_{\rm b0}$ is the cosmic baryon density parameter and 
$P_{\rm DM}$ is given by the velocity dispersion inside the NFW halo times the NFW density profile
\citep[e.g. see][for details]{shaw10}. 
\ms{
Once $r_f$ is set, we then determine $\rho_{0}$ and $P_{0}$ by solving 
\beq
M_{\rm gas} = 4\pi \rho_{0} \int_{0}^{r_f} {\rm d}r\, r^2 \theta^{n}(r; \rho_{0}/P_{0}) = 
\frac{\Omega_{\rm b0}}{\Omega_{\rm m0}} M_{\rm vir} - M_{*},
\eeq
together with the energy constraint equation in Eq.~(\ref{eq:E_balance}). 
}
\ms{We refer the reader to \citet{ostriker05} for details on setting $P_0$ and $\rho_0$.}

\subsection{Non-thermal pressure}

As in \S\ref{sec:obs}, the thermal pressure of ICM is closely related to the tSZ effect.
The ICM model in this paper assumes a radially-dependent non-thermal pressure calibrated in hydrodynamic simulation \citep{nelson14b} and thus the thermal pressure
can be modeled by the residual between the total and non-thermal pressure profiles:
\begin{align}
P_{\rm th}(r) =& P_{\rm tot}(r) - P_{\rm rand}(r), \\
\frac{P_{\rm rand}(r)}{P_{\rm tot}(r)} =&  1 - A_{\rm nt}\left[1+\exp\left\{-\left(\frac{r}{B_{\rm nt}\, r_{\rm 200m}}\right)^{\gamma_{\rm nt}}\right\}\right], \label{eq:P_nonth}
\end{align}
where $r_{\rm 200m}$ is the spherical over-density radius with respect to 200 times the mean matter density of the universe.
In this paper, we assume the outermost radius of non-thermal pressure component 
to be $r_{\rm max}=4r_{\rm 500}$. 
The non-thermal pressure profile is independent of redshift when the radial scale of the profile is scaled in units of $r_{\rm 200m}(z)$ \citep{nelson14b}. The non-thermal pressure in this model consists of only three free parameters, $A_{\rm nt}$, $B_{\rm nt}$, and $\gamma_{\rm nt}$. 
Note that the thermal electron pressure probed by the tSZ effect is given by $P_{\rm e} = (2X+2)/(5X+3)P_{\rm th}$, where $X=0.76$ is the hydrogen mass fraction.

\subsection{Gas density clumping}

The X-ray emissions from the hot ICM ($>10^7$~K) are dominated by thermal bremsstrahlung emission due to Couloumb interactions between electron and proton. The X-ray emissivity is proportional to the product of electron number density $n_{e}$ and proton number density $n_{p}$:
\beq
\epsilon_X \propto \int n_e n_p \Lambda(T,Z) dV
\eeq
where $\Lambda(T,Z)$ is the bolometric volume emissivity of the ICM, i.e. the number of X-ray photons emitted per volume $dV$ for given gas temperature $T$ and metallicity $Z$. 
In a fully ionized plasma with metallicty with 0.3 solar value, the electron density is related to the proton density as $n_{e} = 1.17 n_{p}$. Thus, the X-ray surface brightness is proportional to the line-of-sight integral over X-ray emissivity or the gas density squared: $x \propto \epsilon_X \propto n_{e}^2 \propto \rho_{g}^2$.  

Gas density is expected to be highly inhomogeneous due to freshly accreted gas inside infalling galaxies and accretion from penetrating filaments \citep{nagai11, vazza13, 2015ApJ...806...43B}. Failing to account for density inhomogeneity will lead to overestimates in gas density and mass by $\sqrt{C}$, where $C \equiv \langle \rho_{g}^2 \rangle/\langle\rho_{g}\rangle^2 \geq 1 $ is the clumping factor that quantifies the degree of gas density inhomogeneities. Gas density inhomogeneity is likely the physical reason for the apparent flattening of measured ICM entropy \cite[e.g.][]{2013MNRAS.432.554W},  as well as for the measured gas mass fraction that exceeded the expected cosmic baryon value \citep{2011Sci...331.1576S}.  

We account of the effect of gas density clumping in our model by including the following generalized NFW form of gas density clumping as
\beq
C(r) = 1 + C_{0}\, \left(\frac{r}{r_{\rm 200m}}\right)^{\alpha_C} \left[1+\left(\frac{r}{r_{\rm 200m}}\right)^{\gamma_C}\right]^{(\beta_C-\alpha_C)/\gamma_C},
\label{eq:cfac}
\eeq
where $C_{0}$, $\alpha_C$, $\beta_C$ and $\gamma_{C}$ are free parameters in the model. This model is motivated from the clumping factor profile computed from the {\em Omega500} cosmological hydrodynamical simulations \citep{nelson14a}. Specifically, the parametrization in Eq.~(\ref{eq:cfac}) provides a flexible fit to the clumping factor profile at $r\simlt 3r_{\rm 200m}$, independent of the implemented baryonic physics, such as gas cooling, star formation and feedback from SNe and AGN. Scaling the radius by $r_{\rm 200m}$ removes mass and redshift dependencies in the clumping profile, as $r_{\rm 200m}$ is shown to be closely tracking the mass accretion rate of the halo \citep{lau15}, which governs the level of gas clumping in cluster outskirts. 
This leads to fewer parameters of our gas clumping model, compared to previous work \citep{2015ApJ...806...43B}.
Note that the gas clumping model includes gas inhomogeneities due to infalling substructures, such as merging halos and accreting filaments, as well as smaller scale gas density inhomogeneities due to turbulent motions in the ICM. We refer the reader to the companion paper (Lau et al. in prep.) for more details about our gas clumping model. 

The X-ray surface brightness profile in a single halo with mass $M$ at redshift $z$ is then given by
\beq
{\cal E}_{h}(r, M, z) \equiv \frac{C(r)\epsilon_{{\rm X}, h}\left(r, M, z\right)}{4\pi(1+z)^4}\, 
n_{{\rm e}, h}\left(r, M, z\right)
n_{{\rm H}, h}\left(r, M, z\right),
\label{eq:XSB_cl}
\eeq
where $\epsilon_{{\rm X}, h}$, $n_{{\rm e}, h}$, and $n_{{\rm H}, h}$ are the volume emissivity profile, the electron number density profile, and the proton density profile of an individual halo, respectively. We use this X-ray surface brightness profile to take into account the effect of gas clumping on the correlation statistics in \S\ref{sec:powerspec}.

\subsection{Summary of model parameters}
Table~\ref{tb:params_ICM} summarizes the 15 free parameters of the ICM model used in the paper and their fiducial values. 
For the normalization of halo concentration, we take $A_C = 1$ as a baseline model which is 
consistent with the dark-matter-only N-body prediction by DK15.
For the polytropic exponent in the cluster core
we refer the best-fit value to recent X-ray measurements of gas density
profiles of clusters, and gas masses of groups and clusters as in \citet{flender17}.
We also adopt the value of parameters for feedback processes, stellar mass fraction as constrained in \citet{flender17} and non-thermal pressure as discussed in \citet{nelson14b}.
The parameters for gas clumping are derived from X-ray angular power spectrum measurements from ROSAT All-Sky Survey (Lau et al.\ in prep.). 

We fix a number of parameters in our parameter forecast.
We fix $x_{\rm break} = 0.195$ because this value is set by the cooling radius where the gas cooling time becomes shorter than the Hubble time. Given the ICM profiles, there is little ambiguity in the location of the cool core radius. 
Furthermore, we fix the shape parameters of of the profiles of non-thermal pressure fraction and the gas clumping model: 
$B_{\rm nt}$, $\gamma_{\rm nt}$, $\alpha_{C}$, $\beta_C$, and $\gamma_C$, as they show little dependence in halo mass, redshift, and baryonic physics in cosmological hydrodynamical simulations. 

As a result, we vary 9 parameters for ICM physics in our fiducial analysis, shown in Table~\ref{tb:params_ICM}. 

\section{POWER SPECTRUM FORMALISM}\label{sec:powerspec}
In this section, we describe the formalism to compute the auto- and cross- power spectra between the tSZ compton parameter $y$, the X-ray surface brightness $x$, and lensing covergence $\kappa$, as shown in Eqs.~(\ref{eq:tSZ_y}), (\ref{eq:XSB}) and (\ref{eq:kappa_delta}), respectively. These three fields are expected to be sensitive to the ICM physics and the dark matter potential well, allowing us to probe the connection between the ICM and dark matter content of galaxy clusters, and the growth of large-scale structure traced by galaxy clusters.

The cross-power spectrum between any two fields is given by: 
\beq
\langle {\cal A}({\bd \ell}_1){\cal B}({\bd \ell}_{2})\rangle \equiv (2\pi)^2 \delta^{(2)}_{\rm D}({\bd \ell}_1-{\bd \ell}_2)
C_{{\cal A}{\cal B}}(\ell_{1}), 
\eeq
where $\langle \cdots \rangle$ indicates the operation of ensemble average,
$\delta^{(n)}_{\rm D}({\bd r})$ represents the Dirac delta function in $n$-dimensional space,
$\cal A$ and $\cal B$ are the two-dimensional (i.e. projected) fields of interest.

\subsection{Halo Model Approach}
The auto and cross power spectra for any two fields $C_{{\cal A}{\cal B}}$, under the flat-sky approximation, can be decomposed into two components within the halo-model framework \citep[e.g.][]{2002PhR...372....1C} 
\beq
C_{{\cal A}{\cal B}}(\ell) = C^{\rm 1h}_{{\cal A}{\cal B}}(\ell) + C^{\rm 2h}_{{\cal A}{\cal B}}(\ell), \label{eq:halo_model_power_ab}
\eeq
where the first term on the r.h.s represents the two-point clustering in a single halo (i.e. the 1-halo term),
and the second corresponds to the clustering term between a pair of halos (i.e. the 2-halo term).
Each term on the r.h.s of Eq.~(\ref{eq:halo_model_power_ab}) is expressed as \citep[e.g.][]{1999ApJ...526L...1K, komatsu02, hill13}
\begin{align}
C^{\rm 1h}_{{\cal A}{\cal B}}(\ell)
=& \int_{z_{\rm min}}^{z_{\rm max}}\, {\rm d}z\, \frac{{\rm d}V}{{\rm d}z{\rm d}\Omega}
\int_{M_{\rm min}}^{M_{\rm max}}\, {\rm d}M\, \frac{{\rm d}n}{{\rm d}M}\, |{\cal A}_{\ell}(M,z) {\cal B}_{\ell}(M,z)| \label{eq:A_B_1h} \\
C^{\rm 2h}_{{\cal A}{\cal B}}(\ell)
=& \int_{z_{\rm min}}^{z_{\rm max}}\, {\rm d}z\, \frac{{\rm d}V}{{\rm d}z
{\rm d}\Omega}\, P_{\rm L}(k=\ell/\chi,z) \nonumber \\
& \qquad \times \left[
\int_{M_{\rm min}}^{M_{\rm max}}\, {\rm d}M\, \frac{{\rm d}n}{{\rm d}M}\, {\cal A}_{\ell}(M,z)\, b(M,z)\right] \nonumber \\
& \qquad \times \left[
\int_{M_{\rm min}}^{M_{\rm max}}\, {\rm d}M\, \frac{{\rm d}n}{{\rm d}M}\, {\cal B}_{\ell}(M,z)\, b(M,z)
\right], \label{eq:A_B_2h}
\end{align}
where $z_{\rm min}=10^{-3}$, $z_{\rm max}=3$, $M_{\rm min}=10^{13}\, h^{-1}M_{\odot}$ and $M_{\rm max}=10^{16}\, h^{-1}M_{\odot}$,
$P_{\rm L}(k,z)$ is the linear matter power spectrum, ${\rm d}n/{\rm d}M$ is the halo mass function, and $b$ is the linear halo bias. We define the halo mass $M$ by virial overdensity \citep{1998ApJ...495...80B}. 
We adopt the simulation-calibrated halo mass function presented in \citet{2008ApJ...688..709T} and linear bias in \citet{2010ApJ...724..878T}.
In Eqs.~(\ref{eq:A_B_1h}) and (\ref{eq:A_B_2h}), ${\cal A}_{\ell}(z, M)$ and ${\cal B}_{\ell}(z, M)$ represent the Fourier transforms of profiles of fields ${\cal A}$ and ${\cal B}$ of a single halo 
with mass of $M$ at redshift $z$, respectively. The fields ${\cal A}$ and ${\cal B}$ can be combinations of the Compton-$y$ signal $y$, X-ray emissivity $x$, and the lensing convergence $\kappa$.  In this paper, we consider the combinations: ${\cal A}{\cal B} = xx, yy, xy, x\kappa, y\kappa$. 
Note that we use a flat-sky approximation in Eqs.~(\ref{eq:A_B_1h}) and (\ref{eq:A_B_2h}), 
The exact expression for the curved sky can be found in Appendix~A of \citet{hill13}. 

The Fourier transforms of the thermal electron pressure profile, $y_{\ell}(M,z)$, the X-ray emissivity profile $x_{\ell}(M,z)$, and the lensing convergence profile $\kappa_{\ell}(M,z)$ of the halo with mass $M$ and redshift $z$ are expressed as
\begin{align}
y_{\ell}(M,z) &= 
\frac{4\pi r_{\rm 500}}{\ell^2_{\rm 500}}
\int\, {\rm d}u\, u^2\, \frac{\sin(\ell u/\ell_{500})}{\ell u/\ell_{500}}\frac{\sigma_{\rm T}}{m_{\rm e}c^2} P_{{\rm e},h}(u, M, z), \label{eq:y_ell} \\
x_{\ell}(M,z) &= 
\frac{4\pi r_{\rm 500}}{\ell^2_{\rm 500}}
\int\, {\rm d}u\, u^2\, \frac{\sin(\ell u/\ell_{500})}{\ell u/\ell_{500}} {\cal E}_{h}(u, M, z), \label{eq:x_ell} \\
\kappa_{\ell}(M, z) &= 
\frac{q(\chi(z))}{\chi^2} \int\, {\rm d}r\, 4\pi r^2\, \frac{\sin(\ell r/\chi)}{\ell r/\chi} \frac{\rho_{\rm m}(r, M, z)}{\bar{\rho}_{\rm m}},
\end{align}
where $r_{500}$ is cluster radius enclosing spherical region centered on the cluster center with average mass density 500 times critical density, 
$P_{{\rm e},h}$ is 3D electron pressure profile in single halo, ${\cal E}_{h}$ is the X-ray emissivity profile defined in Eq.~(\ref{eq:XSB_cl}), $\rho_{\rm m}$ is the mass density profile in a single halo and $\bar{\rho}_{\rm m}$ is the mean matter density in the universe, and $q$ is the lensing kernel defined in Eq.~(\ref{eq:lens_kernel}). We define $u = ar/r_{500}$ and $\ell_{500}=a\chi/r_{500}$.

The auto power spectrum of lensing convergence $\kappa$ can be expressed, under the Limber approximation \citep{Limber:1954zz}, as
\beq
C_{\kappa\kappa}(\ell) = \int_{0}^{\chi_H}\, \frac{\rm d\chi}{\chi^2}\, q^2(\chi)\, P_{\rm m}\left(\frac{\ell}{\chi}, z(\chi)\right),
\label{eq:P_kappa}
\eeq
where $P_{\rm m}(k,z)$ is the non-linear matter power spectrum for wave number $k$ at redshift $z$.
We adopt the fitting formula of $P_{\rm m}$ developed in \citet{takahashi12}, which has been calibrated using a set of dark-matter-only $N$-body simulations for a variety of cosmological models. Note that this model does not capture baryonic effects on the cosmic matter density distribution \citep{jing06,rudd08, duffy10,zentner13,schneider15,shirasaki18,chisari18}. Given the importance of this effect, we decide not to include $C_{\kappa\kappa}$ in the our forecast of ICM physics and cosmology. We defer this to a future work where we account for baryonic effects on the matter power spectrum \citep[e.g.][]{zentner13,eifler15,schneider15,mohammed18,chisari18}. Note that we still use this DM-only based $C_{\kappa\kappa}$ for computing the covariance between the cross spectra between X-ray and lensing, $C_{x\kappa}$ and between tSZ and lensing, $C_{y\kappa}$ (see \S\ref{subsec:cov}).

\begin{figure*}
\centering
\includegraphics[width=0.8\columnwidth]
{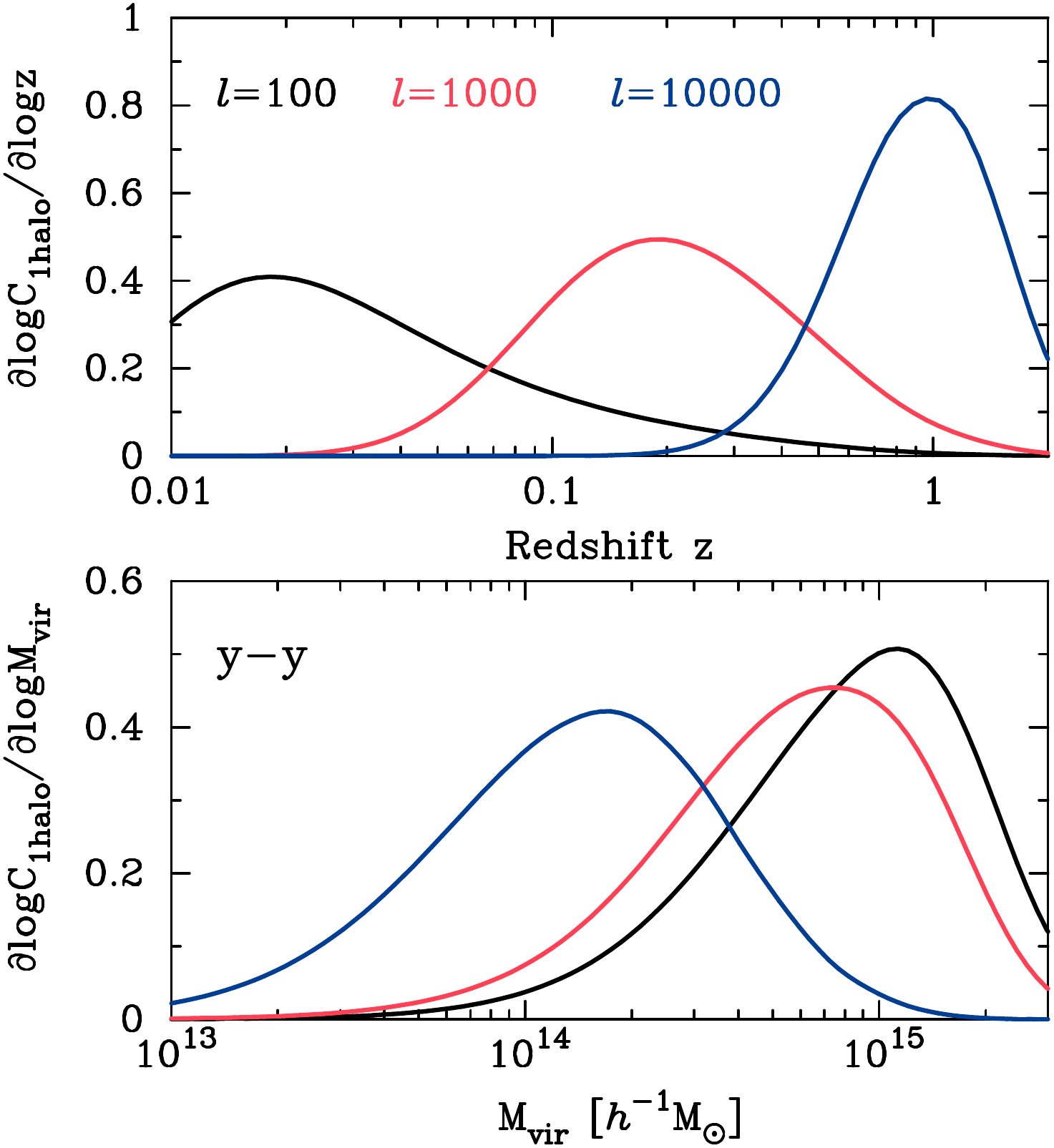}
\includegraphics[width=0.8\columnwidth]
{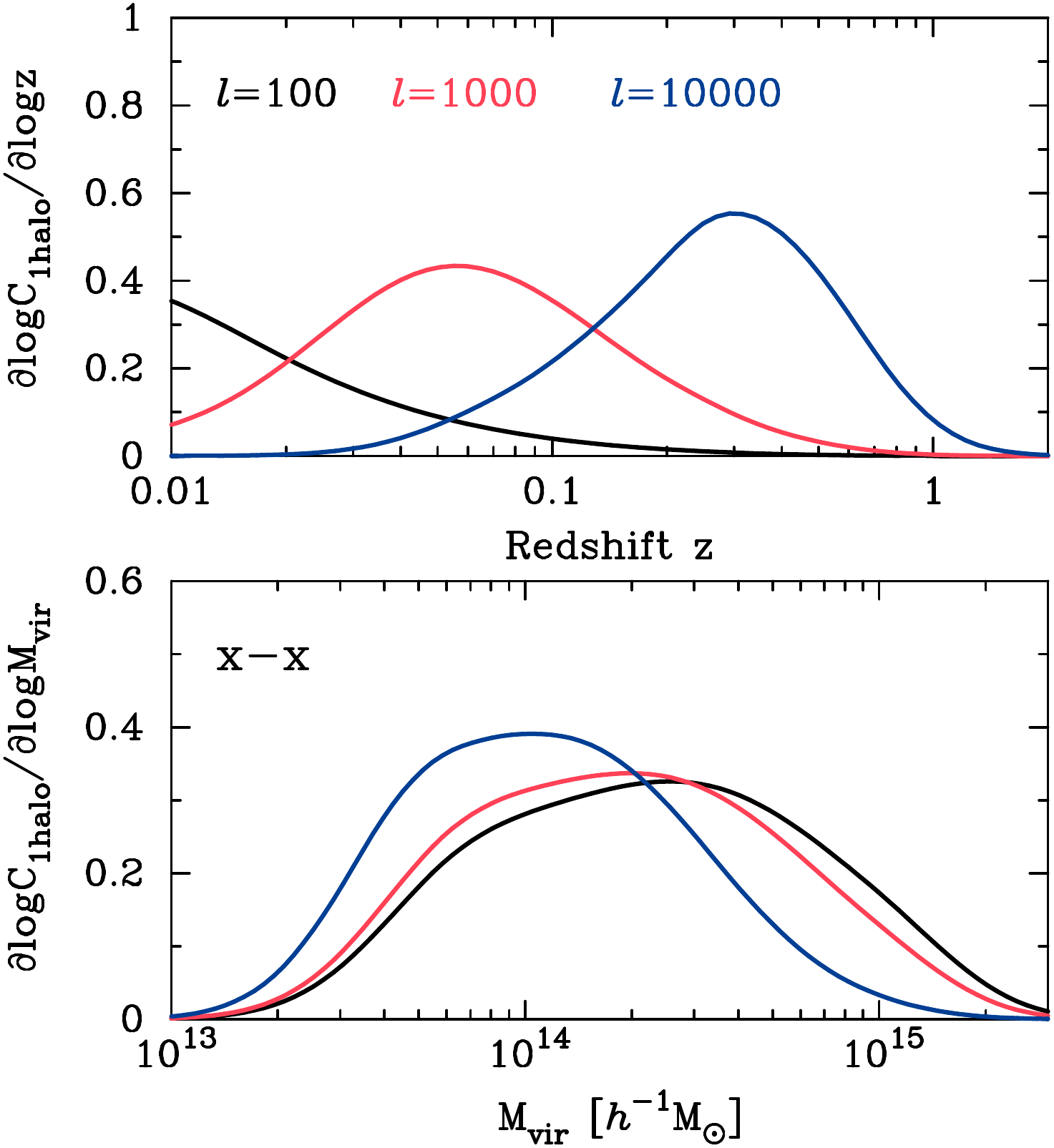}
\caption{
The differential contributions to auto power spectra with respect to redshift (top panels) and halo mass (bottom panels).
The left panels shows the result for one-halo term of $C_{yy}$, while the right is for one-halo term of $C_{xx}$.
In left and right panels, the upper panel represents the derivative of one-halo power spectra in redshift
and the lower is the derivative in virial halo mass.
In each panel, different colored lines show the difference in multipole of $\ell$ (black, red and blue correspond to $\ell=100, 1000$, and 10000, respectively).
}
\label{fig:cl_auto_each_z_M}
\end{figure*} 

\subsection{Effective halo redshift and mass probed by power spectra}

To show the halo redshift and mass dependence in the auto and cross-power spectra, we compute the derivatives of the 1-halo term of the power spectrum $C^{\rm 1h}$ with respect to redshift and mass. For any given fields ${\cal A}{\cal B}$, they are
\begin{align}
\frac{\partial C^{\rm 1h}_{{\cal A}{\cal B}}(\ell)}{\partial z} =& \frac{{\rm d}V}{{\rm d}z{\rm d}\Omega}\,\int_{M_{\rm min}}^{M_{\rm max}}\, {\rm d}M\, \frac{{\rm d}n}{{\rm d}M}\, |{\cal A}_{\ell}(M,z){\cal B}_{\ell}(M,z)|, \\
\frac{\partial C^{\rm 1h}_{{\cal A}{\cal B}}(\ell)}{\partial M} =& \int_{z_{\rm min}}^{z_{\rm max}}\, {\rm d}z\, \frac{{\rm d}V}{{\rm d}z{\rm d}\Omega}
\, \frac{{\rm d}n}{{\rm d}M}\, |{\cal A}_{\ell}(M,z){\cal B}_{\ell}(M,z)|,
\end{align}
We skip the 2-halo term since its contribution is subdominant at angular scales $\ell>100$ for all the cases presented in this paper.

\begin{figure*}
\centering
\includegraphics[width=0.65\columnwidth]
{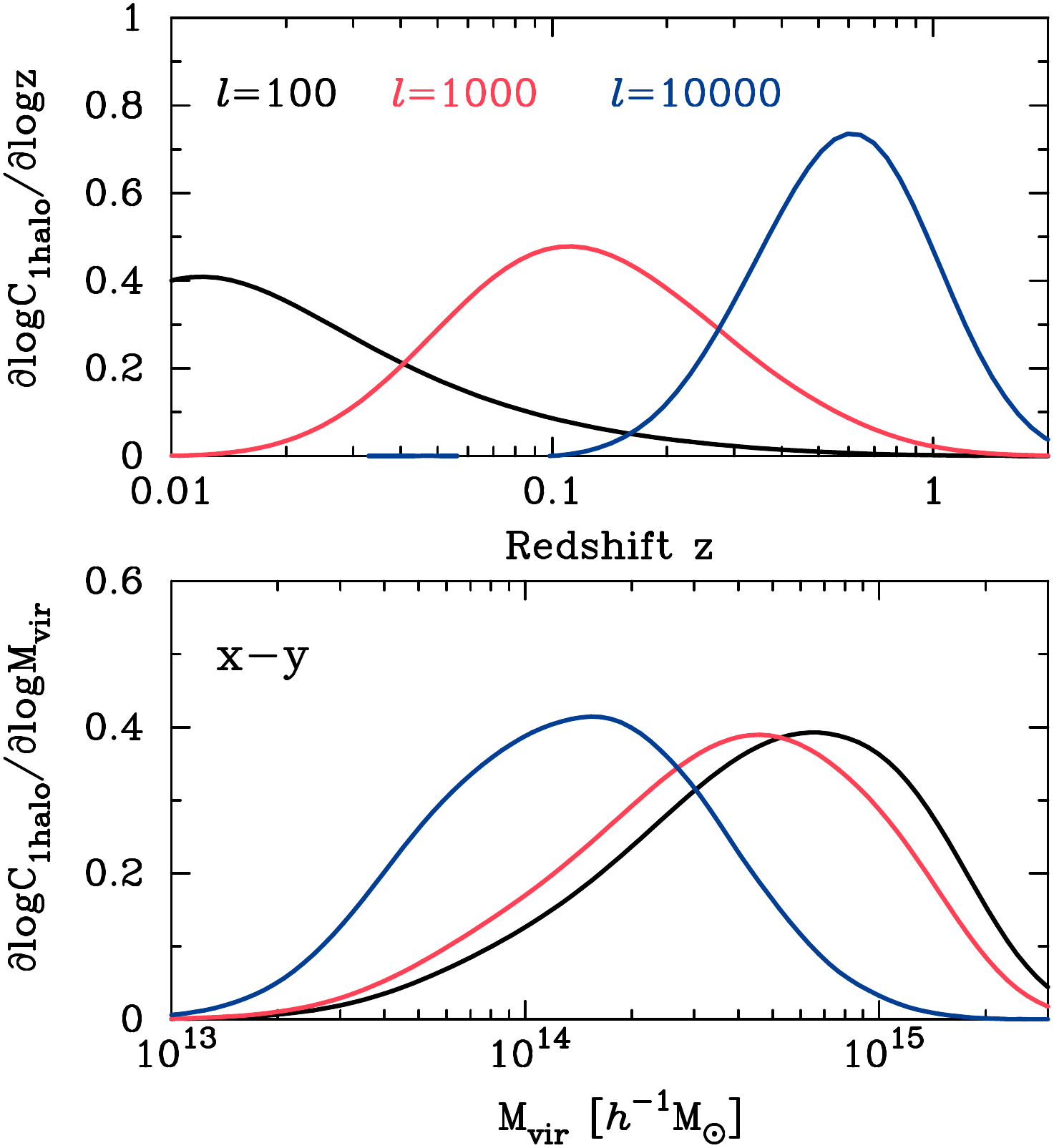}
\includegraphics[width=0.65\columnwidth]
{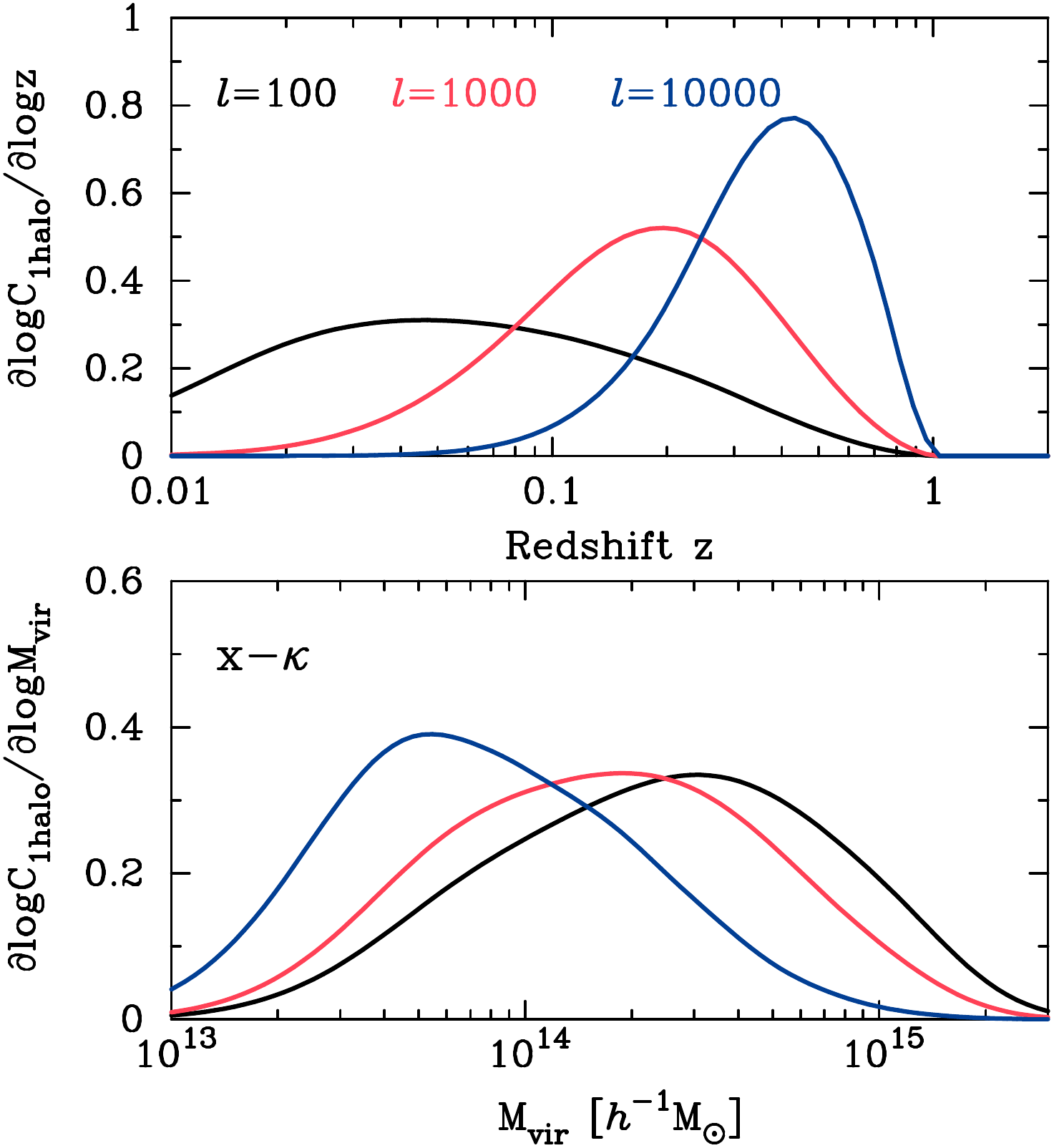}
\includegraphics[width=0.65\columnwidth]
{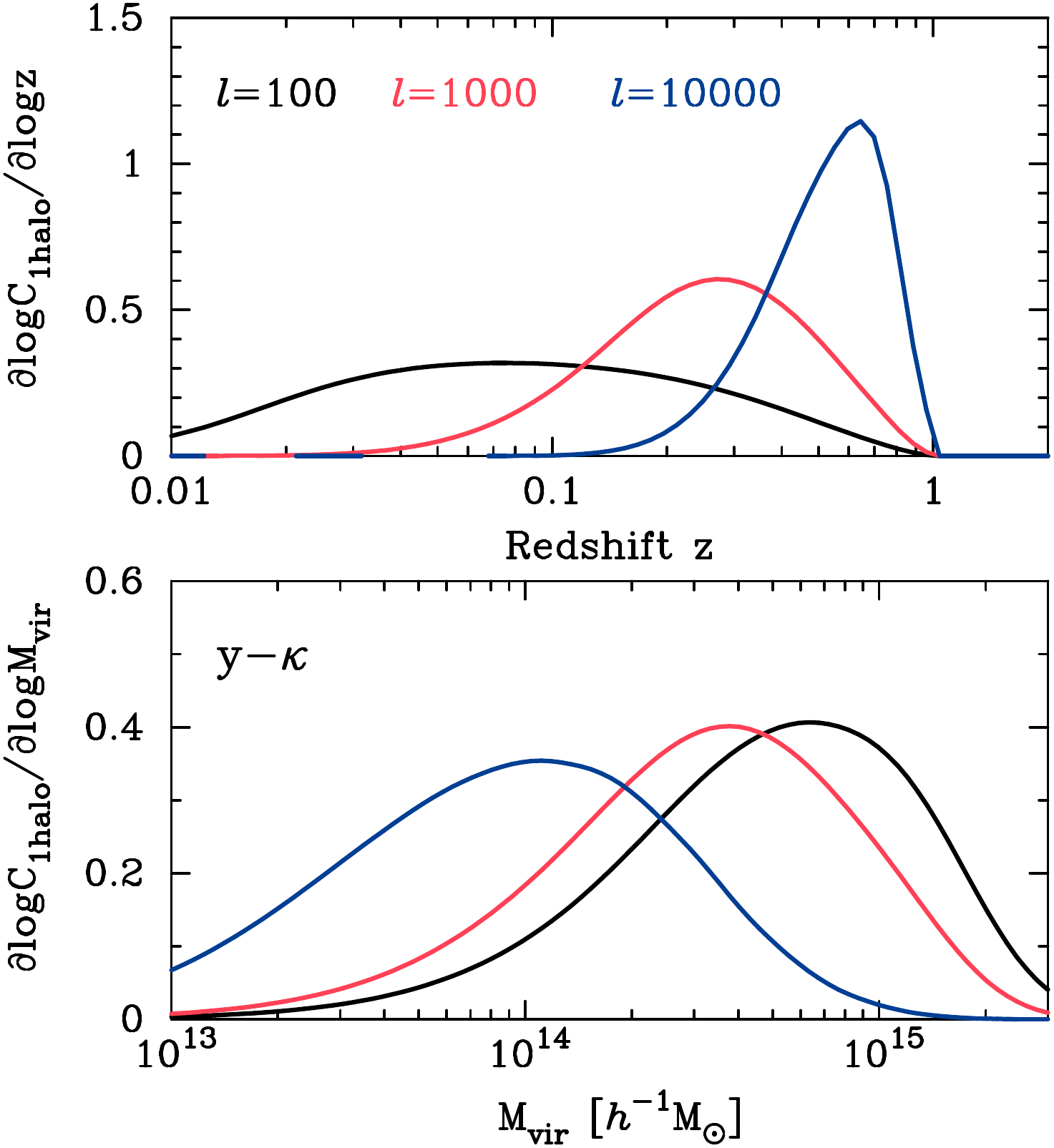}
\caption{
Similar to Figure~\ref{fig:cl_auto_each_z_M}, but we show the differential contribution to cross power spectra with respect to redshift or halo mass. Left, medium, and right panels show the case of $C_{xy}$, $C_{x\kappa}$ and $C_{y\kappa}$, respectively.
Note that we assume every source galaxy locates at $z=1$ for lensing analyses in this figure.
}
\label{fig:cl_cross_each_z_M}
\end{figure*} 

Figure~\ref{fig:cl_auto_each_z_M} shows the redshift- and mass-derivatives of tSZ and X-ray auto power spectra.
In each panel, black, red and blue lines represent the derivative at $\ell=100$, $1000$, and $10000$, respectively.
At larger $\ell$ (i.e. corresponding to the fluctuations at smaller angular separation), the power is dominated by lower mass halos at higher redshifts. Within the halo-model framework, the fluctuations in tSZ and X-ray maps for a given $\ell$ are determined by the halos whose angular size is of $\ell^{-1}$. The typical angular size of a single halo is given by $r_{\rm vir}(z, M)/f_{\rm K}(z)$.
Hence, at larger $\ell$ value, halos with smaller mass and at higher redshift (i.e. with smaller virial radii $r_{\rm vir}$ and larger $f_{\rm K}$) dominates the 1-halo power spectrum.  

Figure~\ref{fig:cl_cross_each_z_M} summarizes the halo mass and redshift dependence in the cross power spectra. The general trend behaves similarly to the auto power spectra in  Figure~\ref{fig:cl_auto_each_z_M}. However, the lensing cross-power spectra $C_{x\kappa}$ and $C_{y\kappa}$ are sensitive to a narrower range of halo redshifts since the lensing does not probe local halos with $z\sim0$ and halos with redshift close to the source $z\sim z_{\rm source}$. For simplicity, we assume the source redshift $z_{\rm source}=1$, with the effective redshift in lensing kernel (Eq.~\ref{eq:lens_kernel}) ranging from $0.2-0.6$.  
To probe the redshift evolution of ICM properties with lensing cross spectra, one need to adopt a ``tomographic'' approach, i.e., by using lensing sources with multiple redshifts. 

\section{STATISTICAL METHOD}\label{sec:method}

In this section, we describe our methods for (1) estimating the statistical uncertainties in the model auto- and cross-power spectra presented in the \S\ref{sec:powerspec} and (2) forecasting ICM physics and cosmological parameter constraints using the auto- and cross-power spectra from upcoming multi-wavelength cluster surveys. 

\subsection{Covariance modeling}\label{subsec:cov}

We first estimate the covariance of a given power spectrum measurements.  Assuming that power spectra follow Gaussian statistics, we can write the covariance matrix between two spectra $C_{\cal AB}$ and $C_{\cal XY}$ as:
\begin{align}
{\rm Cov}&[C_{\cal AB}(\ell_{i})C_{\cal XY}(\ell_{j})]
 = \nonumber \\ 
& \frac{\delta_{ij}}{f_{\rm sky}(2\ell_{i}+1)\Delta \ell}
\left[\tilde{C}_{\cal AX}(\ell_{i})\tilde{C}_{\cal BY}(\ell_{i})
+
\tilde{C}_{\cal AY}(\ell_{i})\tilde{C}_{\cal BX}(\ell_{i})
\right], \label{eq:cov}
\end{align}
where $\delta_{ij}$ is the Kronecker delta,
$f_{\rm sky}$ is the fraction of sky used in the analysis,
and we assume the power spectrum is measured with a finite bin
width of $\Delta \ell$.
In Eq.~(\ref{eq:cov}), $\tilde{C}_{\cal AX} = C_{\cal AX} + N_{\cal AX}$ represents
the observed power spectrum including the observational noise, $N_{\cal AX}$. 
In general, the noise term is non-zero for the auto power spectrum. For instance, the auto power spectrum of tSZ has the noise term associated with the reconstruction method of tSZ map, while the cross power spectrum of tSZ and lensing is noiseless\footnote{\ms{This argument is valid as long as we ignore the correlations among ICM, lensing effects, and astrophysical sources in CMB maps.}}. In the following, we estimate the noise terms in
the auto power spectra of $C_{yy}$, $C_{xx}$ and $C_{\kappa\kappa}$.

\subsubsection{Noise power spectrum in tSZ}\label{subsubsec:noise_tsz}

The noise in tSZ map arises from the reconstruction of a Compton-$y$ map from observed temperature brightness maps at multiple frequencies in the microwave sky. Among several reconstruction methods in the literature \citep[e.g.][]{bobin08,remazeilles11,remazeilles13,hurier13,hurier17,khatri15}, we adopt 
the minimum-variance method as developed in \citet{hill13}.
This method finds the weight of multi-frequency maps in microwave so as to minimize the variance of the tSZ auto power spectrum.

In the framework of minimum-variance method, the noise power spectrum in tSZ can be expressed as the weighted sum of cross-correlation at different frequencies of foreground components:
\beq
N_{yy}(\ell) = \sum_{i,j}w_{{\rm mv}, i} \, w_{{\rm mv}, j} \, C_{ij}. \label{eq:tSZ_noise},
\eeq
where $w_{\rm mv}$ is the minimum variance weight and $C_{ij}$ is the cross-correlation at different frequencies.
Note that the weight of $w_{\rm mv}$ can be set from the observed $C_{ij}$ alone. The detail in the modeling of $C_{ij}$ is found in Appendix~\ref{apdx:noise_tsz}.

\subsubsection{X-ray point sources}\label{subsubsec:noise_x}

In X-ray observations, there are various contributions to the cumulative X-ray surface brightness apart from the ICM.
Among them, AGNs are believed to 
be a dominant component of the extragalactic X-ray emission \citep[e.g.][]{fabian92,treister06,ueda14}, while the X-ray flux count of normal galaxies is found to be close to one of the AGN in the faint end \citep{lehmer12}. Therefore, we consider two populations of X-ray point sources, AGNs and normal galaxies,
as a noise in the cumulative X-ray brightness of ICM (Eq.~\ref{eq:XSB}). 

In this paper, we assume that all resolved point sources are masked.  Assuming that unresolved X-ray point sources follow Poisson statistics, we can express the noise power spectrum of X-ray surface brightness $N_{xx}$ as
\beq
N_{xx}(\ell) = \frac{\left(I_{X, {\rm AGN}}+I_{X, {\rm Galaxy}}\right)^2}{n_{\rm photon}\, B^2_{{\rm X},\ell}},
\eeq
where 
$I_{X,\alpha}$ is the X-ray intensity from unresolved point sources $\alpha$,
$n_{\rm photon}$ is the observed angular number density of photons in the energy range of $E_{\rm min}$ to $E_{\rm max}$,
and $B_{{\rm X}, \ell}$ is the Fourier transform of the point spread function in the X-ray survey (which we assume to be a Gaussian).
In Appendix~\ref{apdx:x-ray-ps}, we summarize a phenomenological model of $I_{X, \alpha}$.

\subsubsection{Shape noises in WL measurement}\label{subsubsec:noise_WL}

In modern galaxy imaging surveys, galaxy ellipticities are commonly used as estimators of the gravitational lensing field. Since the estimator is dominated by the intrinsic 
ellipticities of individual galaxies, 
the observed power spectrum of lensing convergence has the shot noise term, given by
\beq
N_{\kappa\kappa}(\ell, a, b) = 
\frac{\sigma^2_{{\rm int},a}}{2\, n_{{\rm gal},a}}\delta_{ab},
\eeq
where we assume two source galaxy populations ``$a$'' and ``$b$'' with different redshift distributions,
$\sigma_{{\rm int},a}$ is the root-mean-square of the intrinsic ellipticity in the source catalogue of ``$a$'',
and $n_{{\rm gal},a}$ represents the angular number density in the catalogue ``$a$''.

\subsection{Survey specifications}\label{subsec:survey}

We here define the specifications of future multi-wavelength cluster surveys in the microwave, X-ray and optical frequencies with a sky coverage of $20,000$ square degrees. In this work, as an illustration, we will focus on the three major upcoming cluster surveys: CMB-S4 in the microwave, eROSITA in X-ray, and LSST in the optical. 

\begin{table}
\caption{
	The experimental configurations for a hypothetical CMB-S4.
    We set our baseline model with the beam FWHM at 150GHz of 2 arcmin (denoted as ``S4:2arcmin").
	\label{tb:CMBS4}
	}
\begin{tabular}{@{}ccccc}
\hline
Frequency & Noise & & Beam & \\ 
\hline
[GHz] & $[\mu{\rm K}\,{\rm arcmin}]$ & S4:1arcmin & S4:2arcmin & S4:3arcmin \\ 
\hline
21 & 7.9 & 7.1 & 14.3 & 21.4 \\
29 & 5.6 & 5.2 & 10.3 & 15.5 \\
40 & 5.4 & 3.8 & 7.5 & 11.2 \\
95 & 1.5 & 1.6 & 3.2 & 4.7 \\
150 & 1.5 & 1.0 & 2.0 & 3.0 \\
220 & 5.2 & 0.7 & 1.4 & 2.0 \\
270 & 9.0 & 0.6 & 1.1 & 1.7 \\
\hline
\end{tabular}
\end{table}

\subsubsection{CMB Stage 4 (CMB-S4)}

\begin{figure}
\centering
\includegraphics[width=1.0\columnwidth]
{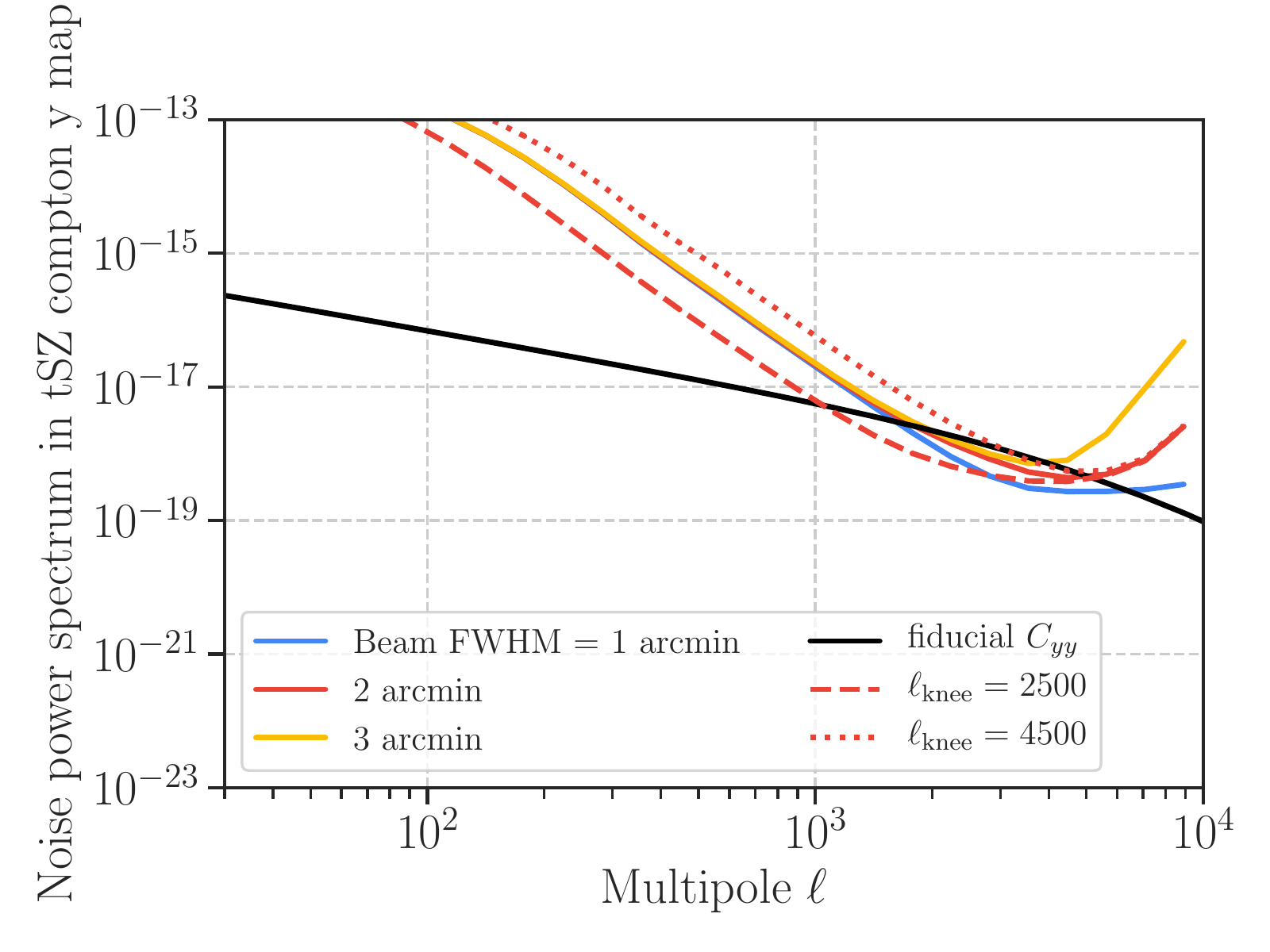}
\caption{
Noise power spectrum of the tSZ analysis in a hypothetical CMB-S4 measurement. 
The black line shows the fiducial model of tSZ auto power spectrum, while other lines are the models of noise power spectrum 
with different parameters in the beam size and pivot angular scale in atmospheric noise.
}
\label{fig:noise_yy_CMBS4}
\end{figure} 

First, we consider the next-generation ground-based CMB experiment, referred to as CMB Stage 4 \citep[CMB-S4;][]{CMBS4-19}.
Following \citet{2017PhRvD..96j3525M}, we adopt the experimental configuration consisting of a single large telescope with seven band-passes.
Table~\ref{tb:CMBS4} summarizes the different experimental configurations of CMB-S4 experiment setups considered in this paper.
The baseline configuration has a white noise level of $s(\nu)=1.5\, \mu {\rm K}\, {\rm arcmin}$ at the 150 GHz and 90 GHz channels. 
We examine different beam full-width half-maximum (FWHM) $\theta_{\rm b}$ in the 150 GHz channel from 1 arcmin to 3 arcmin, 
while we scale the beam FWHM in the other channels at a given $\nu$ as $\theta_{\rm b} \propto 1/\nu$.
For our baseline model, we use the configuration ``S4:2arcmin'' in Table~\ref{tb:CMBS4}.

Using the method in \S\ref{subsubsec:noise_tsz}, we compute the tSZ noise power spectrum for a given experimental configuration.
Figure~\ref{fig:noise_yy_CMBS4} shows the noise tSZ power spectrum for the configurations described in Table~\ref{tb:CMBS4}.
In the figure, the black line is the fiducial model of tSZ power spectrum by the halo-model approach (see Table~\ref{tb:params_fisher} for the fiducial parameters of cosmology and ICM) and different colored lines represent the noise power spectrum for various experimental configurations.
The red solid line is our baseline model of the noise power spectrum, while the red dashed and dotted lines are for the model with different $\ell_{\rm knee}$.
As shown in the figure, the beam size controls the amplitude of noise power spectrum at high multipoles at $\ell\simgt2000$,
while the angular parameter in atmospheric noise $\ell_{\rm knee}$ determines the amplitude of noise at $\ell\simlt1000$.
It is worth noting that the noise power spectrum dominates the observed power spectrum at 
$\ell\simlt1000$ for all the cases considered in this paper.
For simplicity, we assume perfect subtraction of the noise power spectrum from the observed one in the fisher analysis, whereas we discuss the possible biases in parameter estimation due to the imperfect source substraction in  \S\ref{subsubsec:bias_astro_source}.

\subsubsection{eROSITA}

For X-ray, we consider the upcoming eROSITA X-ray all-sky survey \citep{merloni12}.
We adopt a beam size of 30 arcsec FWHM, an energy range of $0.5-2$~keV in the observer's frame, and an effective area of $1365\, {\rm cm}^2$.
The total observation time is assumed to be four years.
The flux limit in the energy range of $0.5-2$~keV is set to be $F_{\rm lim} = 4.4\times10^{-14}\, {\rm erg}\,{\rm cm}^{-2}\,{\rm s}^{-1}$, 
and the mean X-ray intensity from unresolved point sources is estimated to be $14.7\, {\rm cm}^{-2}\,{\rm s}^{-1}\,{\rm str}^{-1}$.
We assume all the point sources with the flux greater then $F_{\rm lim}$ are removed. 

\begin{figure}
\centering
\includegraphics[width=1.0\columnwidth]
{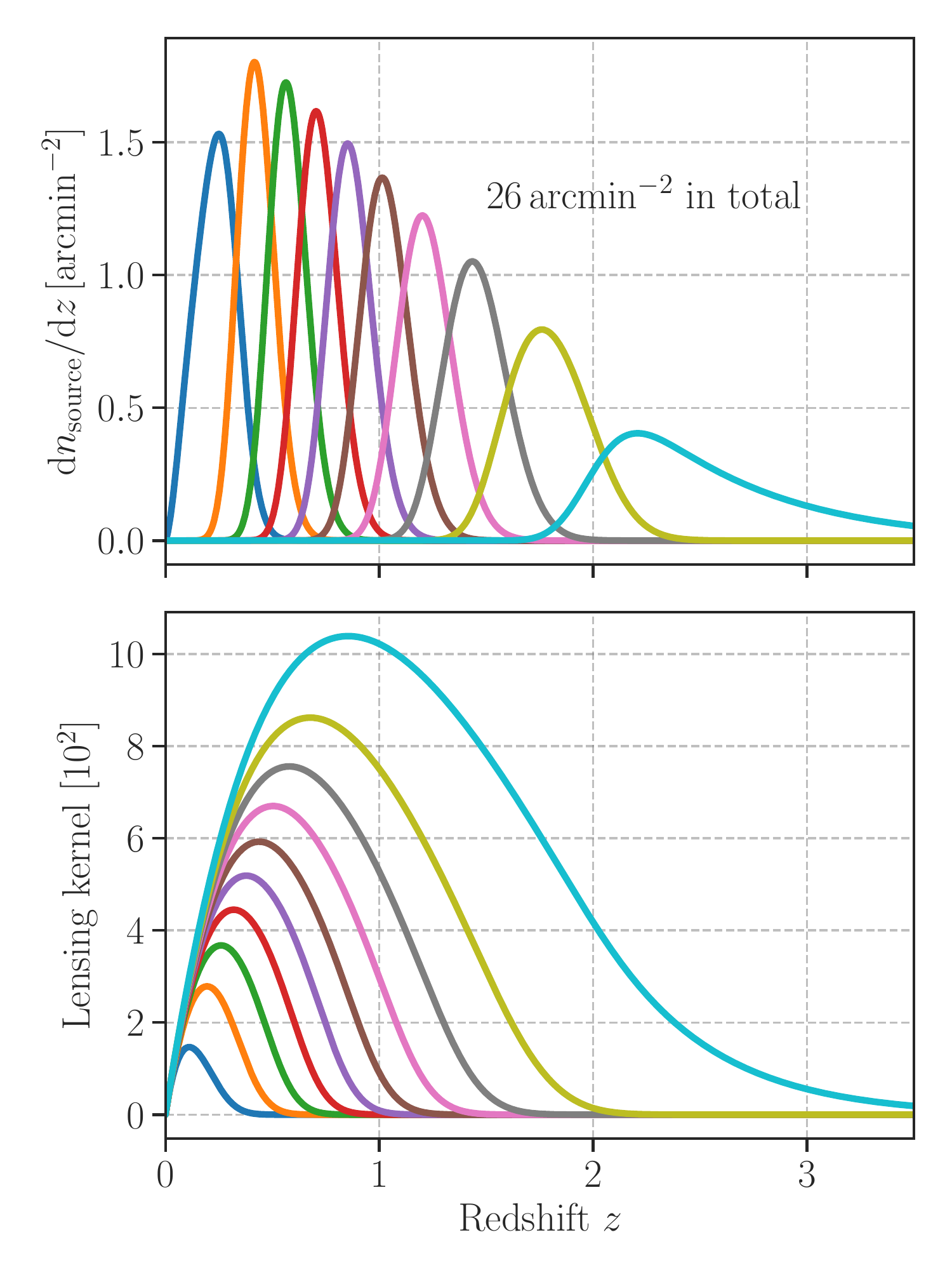}
\caption{
The source redshift distribution in a hypothetical imaging survey with 10 tomographic bins
and its corresponding lensing kernel.
}
\label{fig:dndz_LSST_10bins}
\end{figure} 

\subsubsection{LSST}\label{subsubsection:LSST}

For lensing, we consider a future galaxy imaging survey performed with the Large Synoptic Survey Telescope \citep[LSST;][]{LSST09}.
We assumed the source redshift distribution to follow
\beq
\frac{{\rm d}n_{\rm source}}{{\rm d}z} \propto z^{\alpha}\, \exp\left[-\left(\frac{z}{z_0}\right)^{\beta}\right], \label{eq:dndz_source}
\eeq
where $\alpha=1.27$, $\beta=1.02$, $z_0=0.5$. The normalization is set so that the total number density of the source galaxies is $26\, {\rm arcmin}^{-2}$, which  
represents the largest number density of sources achievable by ground-based imaging surveys. 
For the shape noise, we adopt the RMS of source ellipticity to be $0.367$, the typical value found in the ground-based imaging surveys \citep[e.g., see][]{mandelbaum18}.

To probe the redshift evolution of the ICM and the growth of clusters, we divide the source galaxies into $N$ tomographic subsamples, where we take into account the uncertainties in the photometric redshift $z_{\rm ph}$  with
the following probability distribution for a given true redshift $z$:
\beq
{\rm Prob}(z_{\rm ph}|z) = \frac{1}{\sqrt{2\pi \sigma_z}}\exp\left[-\frac{(z_{\rm ph}-z-z_{\rm b})^2}{2\sigma^2_{z}}\right]. \label{eq:pofz_photoz}
\eeq
We set the scatter of $z_{\rm ph}$ to be $\sigma_{z} = 0.05 (1+z)^{-1}$, a typical estimate in the ongoing imaging survey \citep{takana18}.
The bias parameter $z_{\rm b}$ is assumed to be zero. We discuss possible biases in parameter estimation due to the imperfect knowledge of $z_{\rm b}$ in \S\ref{subsubsec:bias_lens_sys}.
Given Eqs.~(\ref{eq:dndz_source}) and (\ref{eq:pofz_photoz}), we define the redshift distribution of $i$-th source sample as
\beq
\frac{{\rm d}n_{\rm source, i}}{{\rm d}z} = \frac{{\rm d}n_{\rm source}}{{\rm d}z}\,\int_{z_{\rm ph} \in i}\, {\rm d}z_{\rm ph}\, {\rm Prob}(z_{\rm ph}|z),
\eeq
and we set the bin width of $z_{\rm ph}$ so that each subsample of source galaxies has the same number density.
Figure~\ref{fig:dndz_LSST_10bins} shows the source redshift distributions when we adopt 10 tomographic bins ($N=10$)
and its corresponding kernel for lensing convergence (Eq.~\ref{eq:lens_kernel}).

In addition, we also consider the multiplicative uncertainties in the lensing convergence of $i$-th tomographic bin, 
$\kappa_{\rm obs} \rightarrow (1+m_{i}) \, \kappa_{\rm true}$, where $\kappa_{\rm true}$ is given by Eq.~(\ref{eq:kappa_delta})
and $m_{i}$ is the so-called multiplicative bias.
As in our fiducial setup, we set $m_{i}=0$ for each tomographic bin. We then examine the impact of non-zero multiplicative biases in parameter estimation in \S\ref{subsubsec:bias_lens_sys}.

\subsection{Parameter forecast: Fisher analysis framework }\label{subsec:fisher_matrix}

\begin{table}
\caption{
	The list of parameters in our fisher analysis.
    Assuming $N$ tomographic bins in lensing survey, we have $15+2N$ parameters in total, 
    consisting of 6 cosmological parameters, 9 ICM parameters (see Table~\ref{tb:params_ICM} for details), and $2N$ for lensing systematics.
	\label{tb:params_fisher}
	}
\scalebox{1.0}[1.0]{
\begin{tabular}{@{}llc}
\hline
\hline
Parameter & Fiducial & Gaussian prior \\ 
\hline
Cosmology & & \\
\hline
$\Omega_{\rm cdm}h^2$ & 0.1198 & 0.100 \\
$\Omega_{\rm b0}h^2$ & 0.02225 & 0.100 \\
$A_{s}/10^{-9}$ at $k=0.05\, {\rm Mpc}^{-1}$ & 2.2065 & 1.00 \\
$n_{s}$ & 0.9645 & 0.300 \\
$h = H_{0}/\left(100 \, [{\rm km}\, {\rm s}^{-1}\, {\rm Mpc}^{-1}]\right)$ & 0.6727 & 0.500 \\
$w_0$ & -1 & 0.50 \\
\hline
Intracluster medium & & \\
\hline
$A_C$ & 1.0 & 0.50 \\
$\tilde{\Gamma}$ & 0.10 & 0.10 \\
$\gamma$ & 1.72 & 1.72 \\
$x_{\rm break}$ & 0.195 & fixed \\
$\epsilon_{\rm DM}$ & 0.050 & 0.050 \\
$\epsilon_{f}/10^{-6}$ & 1.00 & 1.00 \\
$f_{*}$ & 0.026 & 0.026 \\
$S_{*}$ & 0.37 & 0.37 \\
$A_{\rm nt}$ & 0.452 & 0.452 \\
$B_{\rm nt}$ & 0.841 & fixed \\
$\gamma_{\rm nt}$ & 1.628 & fixed \\
$C_0$ & 0.90 & 0.90 \\
$\alpha_C$ & 1.00 & fixed \\
$\beta_C$ & 6.00 & fixed \\
$\gamma_{C}$ & 3.00 & fixed \\
\hline
Lensing systematics & & \\
\hline
$m_{i}\, (i=1, \cdots, N)$ & 0.00 & 0.10 \\
$z_{bi}\, (i=1, \cdots, N)$ & 0.00 & 0.05 \\
\hline
\end{tabular}
}
\end{table}

We here summarize a Fisher analysis and the fiducial parameters for computing power spectra. 
For a multivariate Gaussian likelihood, the Fisher matrix
$F_{ij}$ can be written as
\beq
F_{ij} = \frac{1}{2}{\rm Tr}\left[A_{i}A_{j}+{\rm Cov}^{-1}H_{ij}\right],\label{eq:fisher_matrix}
\eeq
where
$A_{i}={\rm Cov}^{-1}\partial [{\rm Cov}]/\partial p_i$,
$H_{ij} = 2(\partial \mu/\partial p_i)(\partial \mu/\partial p_j)$,
${\rm Cov}$ is the data covariance matrix,
$\mu$ represents the assumed model, and $p_i$ describes parameters of interest. 
The Fisher matrix provides an estimate of the error covariance
for two parameters as
\beq
\langle \Delta p_{\alpha} \Delta p_{\beta} \rangle = (F^{-1})_{\alpha \beta},
\eeq
where $\Delta p_{\alpha}$ represents the statistical uncertainty of parameter $p_{\alpha}$.
Since the data covariance is expected to scale inversely with the survey area,  the second term in Eq.~(\ref{eq:fisher_matrix}) becomes dominant for a large area survey. We therefore consider only the second term.

We vary 6 cosmological and 9 ICM parameters, and fix the rest as shown in Table~\ref{tb:params_fisher}.
Note that the cosmological parameters consist of 
physical dark matter density $\Omega_{\rm cdm}h^2$,
physical baryon density $\Omega_{\rm b0}h^2$,
the amplitude of initial curvature fluctuations $A_{s}$ (at the pivot scale $k=0.05\, {\rm Mpc}^{-1}$),
the spectrum index parameter of linear power spectrum $n_{s}$,
the parameter of equation of state of dark energy $w_0$,
and the dimensionless Hubble parameter $h$.
In addition, we introduce $2N$ parameters for lensing systematics when using $N$ tomographic bins in lensing field $\kappa$
(see \S\ref{subsubsection:LSST}).

We then construct the data vector $\bd D$ from a set of 
binned spectra $C_{yy}, C_{xx}, C_{xy}$, $C_{x\kappa,i}$
and $C_{y\kappa,i}$, where the index $i$ is for the tomographic redshift bins of the lensing source galaxies.
To compute the power spectra, we perform the logarithmic binning in the multipole range from $\ell_{\rm min}=30$
to $\ell_{\rm max}=3000$. The number of bins in each power spectrum is set to be 10.
Hence, we have $10+10+10+10N+10N=20N+30$ elements in the data vector for $N$ tomographic bins in total.

Owing to the strong degeneracy among cosmological and ICM parameters in the power spectra (especially for $C_{yy}$),
we need to introduce loose priors for realistic forecast of parameter constraints.
In this paper, we compute the Fisher matrix as follows:
\beq
{\bd F} = {\bd F}_{\rm model} + {\bd F}_{\rm prior}, \label{eq:eff_fisher_matrix}
\eeq
where ${\bd F}_{\rm model}$ is computed by Eq.~(\ref{eq:fisher_matrix}) with the power spectra,
and ${\bd F}_{\rm prior}$ represents the prior information. We assume that the loose prior on the $i$-th parameter
is given by Gaussian with the variance, $\sigma_{{\rm prior},i}$. In this case, the prior term is expressed as
${\bd F}_{{\rm prior}, ij} = \sigma^{-2}_{{\rm prior},i}\, \delta_{ij}$. The fiducial value of $\sigma_{{\rm prior},i}$ is found in Table~\ref{tb:params_fisher}. We set the value of $\sigma_{\rm prior}$ to be $\simeq30-450\%$ of fiducial value of parameters.

\section{RESULTS}\label{sec:res}

\begin{figure*}
\centering
\includegraphics[width=2.0\columnwidth]
{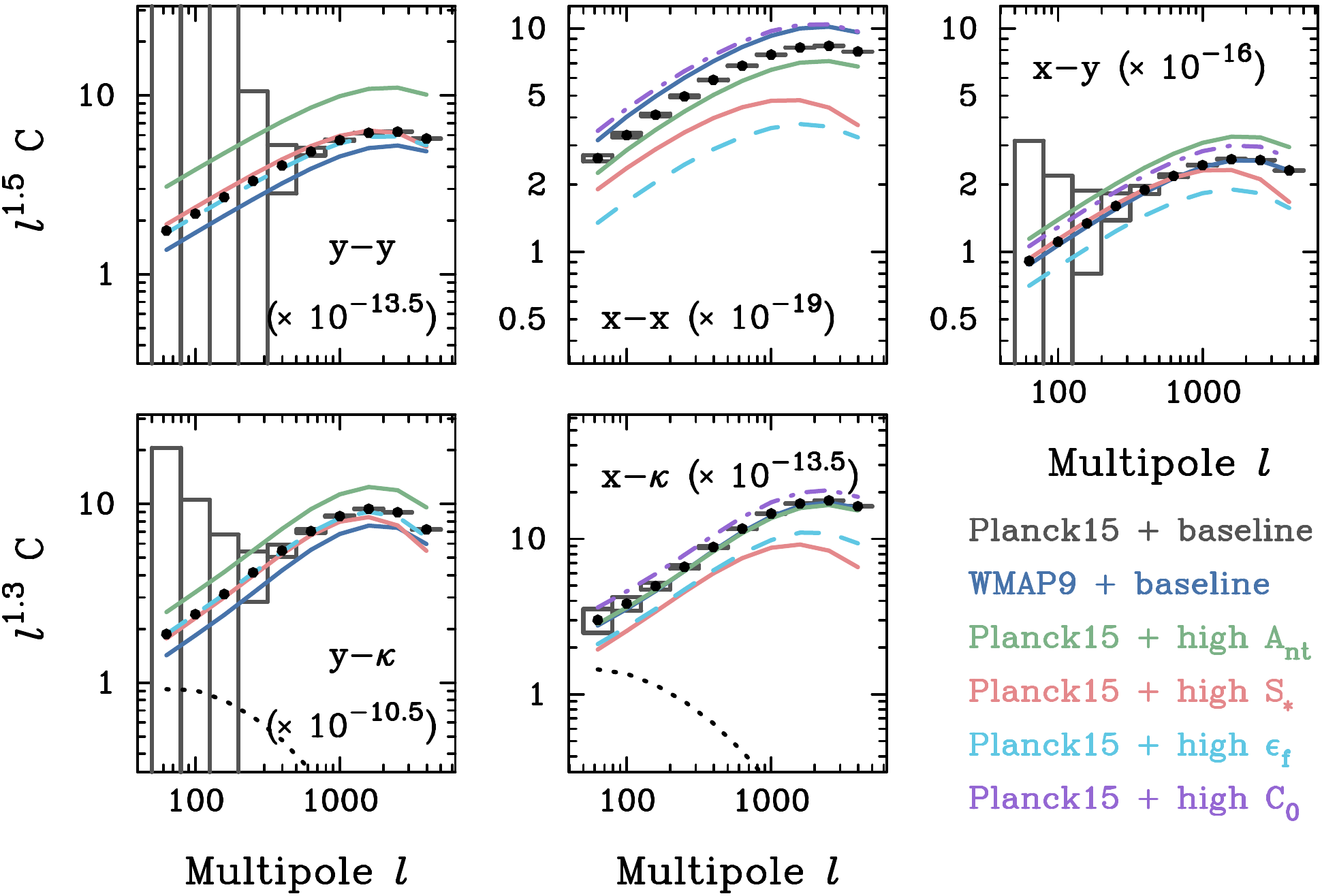}
\caption{
The parameter dependence on power spectra.
Each panel show the power spectrum when one vary the parameters of cosmology and ICM.
{\it Top-left}: tSZ auto power spectrum, {\it Top-middle}: X-ray auto power spectrum, {\it Top-right}: tSZ-X cross power spectrum,
{\it Bottom-left}: tSZ-WL cross power spectrum, and {\it Bottom-right}: X-WL cross power spectrum.
In each panel, the black points show a fiducial model assuming the best-fit cosmology to the Planck CMB measurement \citep{planck16} and ICM parameters (see Table~\ref{tb:params_ICM} for the ICM parameters), 
while the gray squared box represents the expected statistical uncertainty in hypothetical surveys with a sky coverage of 20,000 square degrees (see also \S\ref{subsec:survey}). 
The black dotted line in the bottom panels shows the 2-halo term which is always sub-dominant. 
Different colored lines in each panel show the parameter dependence on the model power spectra.
Blue lines correspond to the model with the best-fit cosmology to the WMAP CMB measurement \citep{hinshaw13} and fiducial ICM,
while green, red, cyan and purple are for the model with the Planck cosmology and 
higher $A_{\rm nt}$, higher $S_{*}$, higher $\epsilon_{f}$, and higher $C_0$, respectively.
Note that $A_{\rm nt}$ represents the parameter about overall amplitude in non-thermal pressure, 
$S_{*}$ is the power-law index in the stellar-to-halo mass relations in cluster-sized halos,
and $\epsilon_{f}$ controls the strength of stellar+AGN feedback inside galaxy clusters.
The parameter of $C_0$ determines amplitude of the  the clumping factor in gas density profile in the cluster outskirts. 
The y-values in the upper and lower panels are multiplied by $\ell^{1.5}$ and $\ell^{1.3}$ respectively to better show the dependence on the  parameters.}
\label{fig:cls_param_depend}
\end{figure*} 

\begin{figure*}
\centering
\includegraphics[width=2.0\columnwidth]
{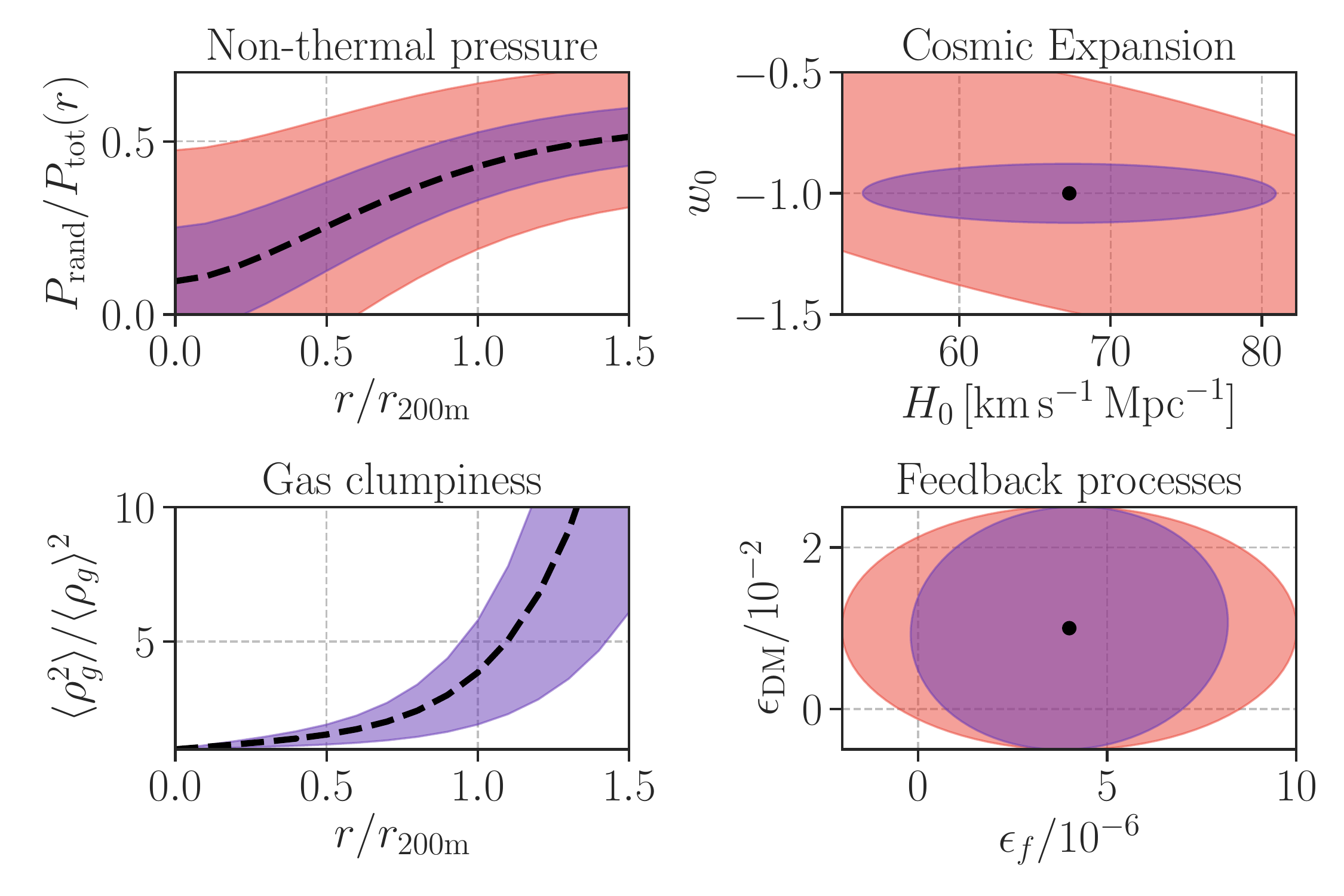}
\caption{
The summary of parameter forecasts with a joint analysis of tSZ, X-ray and lensing observables.
The orange error contours assume the analysis with the power spectra related with tSZ and lensing ($C_{yy}$ and $C_{y\kappa}$),
while the purple one shows the results by the joint analysis of tSZ, X-ray and lensing 
($C_{yy}, C_{xx}, C_{xy}, C_{x\kappa}$ and $C_{y\kappa}$).
In this figure, we assume our fiducial set up of multi-wavelength surveys covering 20,000 square degrees.
All the error contours are for $68\%$ confidence level.
Note that the gas clumpiness can be constrained only if the X-ray information would be available.
}
\label{fig:fid_fourpanels}
\end{figure*} 

\subsection{Parameter dependence}

We begin by studying the dependence of parameters on various power spectra.
Figure~\ref{fig:cls_param_depend} shows some representative examples of predicted power spectra for a range of cosmological models and ICM parameters. In this figure, we consider six cases as follows:
\begin{enumerate}
\item The best-fit cosmological model by \citet{planck16} (denoted as Planck15) and fiducial ICM parameters;
\item The best-fit cosmological model by \citet{hinshaw13} (denoted WMAP9) and fiducial ICM parameters;
\item The Planck15 cosmology and the ICM model with a parameter of non-thermal pressure profile being $A_{\rm nt} = 0.60$ ($A_{\rm nt}=0.451$ for fiducial case);
\item The Planck15 cosmology and the ICM model with a parameter in stellar-to-halo mass relation being $S_{*} = 0.62$ ($S_{*}=0.12$ for fiducial case);
\item The Planck15 cosmology and the ICM model with a parameter of stellar feedbacks being $\epsilon_{f}=10^{-5}$ ($\epsilon_{f}=4\times 10^{-6}$ for fiducial case);
\item The Planck15 cosmology and the ICM model with a parameter of gas clumping being $C_{0}=2.0$ ($C_0= 0.9$ for fiducial case).
\end{enumerate}
Here, we set the weak lensing source redshift to be 1 and source number density to be 10 ${\rm arcmin}^{-2}$
for simplicity.
The black points in the figure show the model (i), while the blue, green, red, cyan and purple lines 
indicate the model (ii), (iii), (iv), (v), and (vi), respectively.
The gray error bars represent the statistical uncertainty evaluated in \S\ref{subsec:cov} for a survey specification in \S\ref{subsec:survey}.

The difference in the black points and blue lines highlights the dependence of cosmological parameters on the power spectra.
Note that the main difference between the Planck15 and WMAP9 cosmological models lies in the cosmic matter density $\Omega_{\rm m0}$
and the present-day normalization of linear matter power spectrum $\sigma_8$ (both being lower in the WMAP9).
Since our power spectra will be mainly determined by the 1-halo term, the dependence on cosmological parameters stems primarily from that of the halo mass function.
In particular, the tSZ power spectrum has a significant dependence on cosmology compared to other spectra, because it contains the information of halo mass function at massive cluster scales (see Figure~\ref{fig:cl_auto_each_z_M}).
Furthermore, the cosmological dependence among power spectra is different, which allows the breaking of the degeneracy among the parameters, thus in turn improving the cosmological constraints.
Interestingly, the X-ray auto power spectrum depends on cosmology in an opposite way as the tSZ power spectrum does. 
We expect that the X-ray auto power spectrum to be sensitive to $\Omega_{\rm m0}$, 
since it depends on the gas density squared in a single halo. Given that the cosmic baryon density $\Omega_{\rm b0}$ is relatively unchanged, as one reduces $\Omega_{\rm m0}$, the gas mass in a single halo increases, thus increasing the overall amplitude of the X-ray auto power spectrum. This explains why the X-ray auto power spectrum amplitude is higher in the WMAP9 cosmology (with lower $\Omega_{\rm m0} = 0.27$) than Planck15 (with $\Omega_{\rm m0} =0.30$). 
While the tSZ power spectrum is more sensitive to $\sigma_8$ (comparing the black points and red lines in Figure~\ref{fig:cls_param_depend}), the X-ray power spectrum is a more efficient probe of $\Omega_{\rm m0}$.

Uncertainties in ICM model will degrade the cosmological constraints from $C_{yy}$ and $C_{xx}$.
The comparison among the black points, green, red, and cyan lines in Figure~\ref{fig:cls_param_depend} demonstrates how ICM physics can hide the cosmological dependence on the power spectra.
Reducing the non-thermal pressure fraction (green lines) can increase the tSZ power spectrum, while the X-ray auto power spectrum is less sensitive to the change in $A_{\rm nt}$. 
On the other hand, increase in the slope in the stellar-to-halo mass relations $S_{*}$ (red lines) can affect the X-ray observables $C_{xx}$ and $C_{x\kappa}$.
The parameter degeneracy in ICM can be seen in the comparison between red and cyan lines in Figure~\ref{fig:cls_param_depend}.
Increasing the amplitude of the stellar+AGN feedback can have a similar effect as increasing the non-thermal pressure in $C_{yy}$ and $C_{y\kappa}$, as both reduce the thermal pressure in less massive group-sized halos. 
Importantly, this degeneracy can be broken if one includes X-ray information, since the X-ray observable is less sensitive to the non-thermal pressure, but still dependent on stellar and AGN feedback.
However, the X-ray observable is dependent on gas clumping, while the tSZ is insensitive to this effect (purple dash-dotted lines).
The cross correlation with lensing will provide additional constraints on cosmology and ICM physics, but the amplitude in lensing observables is subject to the imperfect shape measurement and photo-z accuracy.

\subsection{Fisher forecast for future wide-area surveys}

We present the results of Fisher forecast in our fiducial setup as in \S\ref{subsec:fisher_matrix}.
Figure~\ref{fig:fid_fourpanels} summarizes the forecast for a survey with a sky coverage of 20,000 square degrees when 10 tomographic bins in lening observables are available.
In this figure, each panel shows a set of relevant physical effects in the joint analysis of tSZ, X-ray and lensing observables. In each panel, the orange shaded region is derived by the Fisher matrix analysis when we limit the information of $C_{yy}$ and $C_{y\kappa}$ as examined in \citet{battaglia15} and \citet{osato18}. The purple region represents the constraints if we can combine the information from $C_{yy}$, $C_{y\kappa}$, $C_{xx}$, $C_{xy}$ and $C_{x\kappa}$.

\subsubsection{Cluster Physics Constraints}

In addition to cosmolology, the joint analysis will also provide an important insight to the ICM physics inside virial radius and in the cluster outskirts.
In our ICM model, we vary only the normalization parameter of the non-thermal pressure, $A_{\rm nt}$, while keeping the shape of its profile fixed.
We define the $68\%$ confidence level of the fraction of non-thermal pressure by using the marginalized fisher matrix of $A_{\rm nt}$.
The top left panel of Figure~\ref{fig:fid_fourpanels} shows the non-thermal fraction of ICM in cluster-sized halos as a function of the normalized cluster-centric radius, $r/r_{\rm 200m}$.
The joint analysis allows us to constrain the fraction of non-thermal pressure and gas clumping at the level of $22/50\%$ ($1\sigma$ error), respectively at $r=r_{\rm 200m}$. This represents a factor of two improvements compared to previous studies \citep{battaglia15,osato18}.
Note that there is currently no other method that can observationally constrain these ICM physics for a statistical sample of galaxy clusters.
The bottom right panel shows the constraints on the astrophysical feedback parameters in the ICM model. The ICM model without feedback processes can be ruled out with a $\sim1\sigma$ level. We also find the stellar mass fraction in massive galaxy clusters can be constrained to within $\sim5\%$.

\subsubsection{Cosmological Constraints}

The top right panel of Figure~\ref{fig:fid_fourpanels} shows the constraint on the cosmic expansion model summarized by two parameters of $w_{0}$ and $H_0$.
We find that the dark energy equation-of-state parameter can be constrained at the level of $1\sigma = 0.080$ from 
the joint analysis of tSZ, X-ray and lensing observables. 
That is comparable to the current constraints \citep[e.g.][for review]{2018RPPh...81a6901H},
but we do not use the tightest prior information of cosmology from CMB and SNe measurements.
Therefore, our joint analysis approach is best to be applied at lower redshifts $z\simlt1$,
where the dark energy is expected to dominate the cosmic energy density. 

\subsubsection{Breaking Parameter Degeneracy with X-ray}

\begin{figure*}
\centering
\includegraphics[width=2.2\columnwidth]
{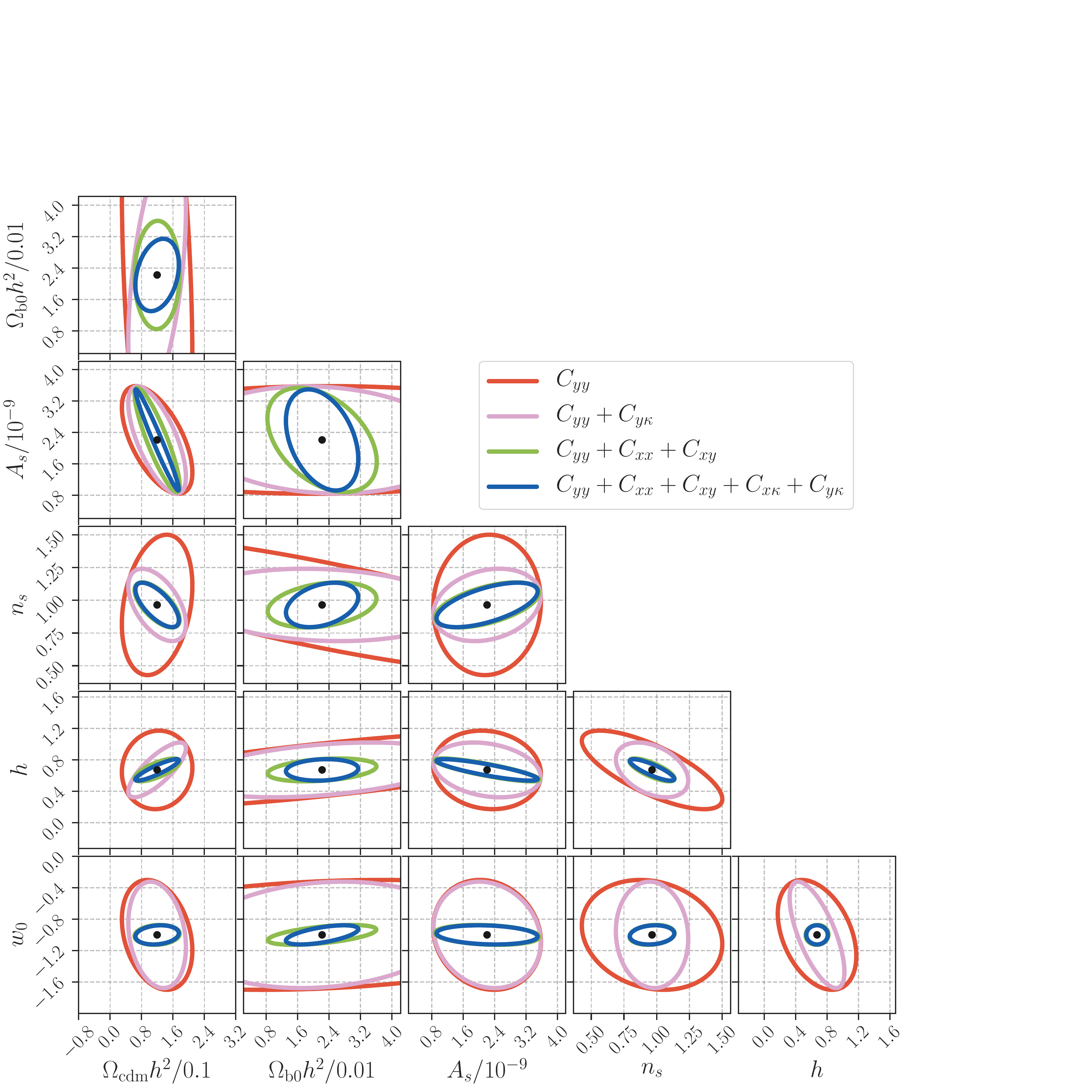}
\caption{
The corner plot of 6 cosmological parameters by different combinations of power spectra.
In this figure, different colored lines show the parameter constraints with a 68\% confidence level
by various combination of tSZ, X-ray and lensing observables (as shown in the legend).
Note that the pink circles are based on the combined analysis with tSZ and lenisng as proposed in the literature \citep{battaglia15,osato18}, while
the blue one shows the significant improvement 
in our joint-analysis method.}
\label{fig:fid_fisher_cosmo}
\end{figure*}

\begin{figure*}
\centering
\includegraphics[width=2.2\columnwidth]
{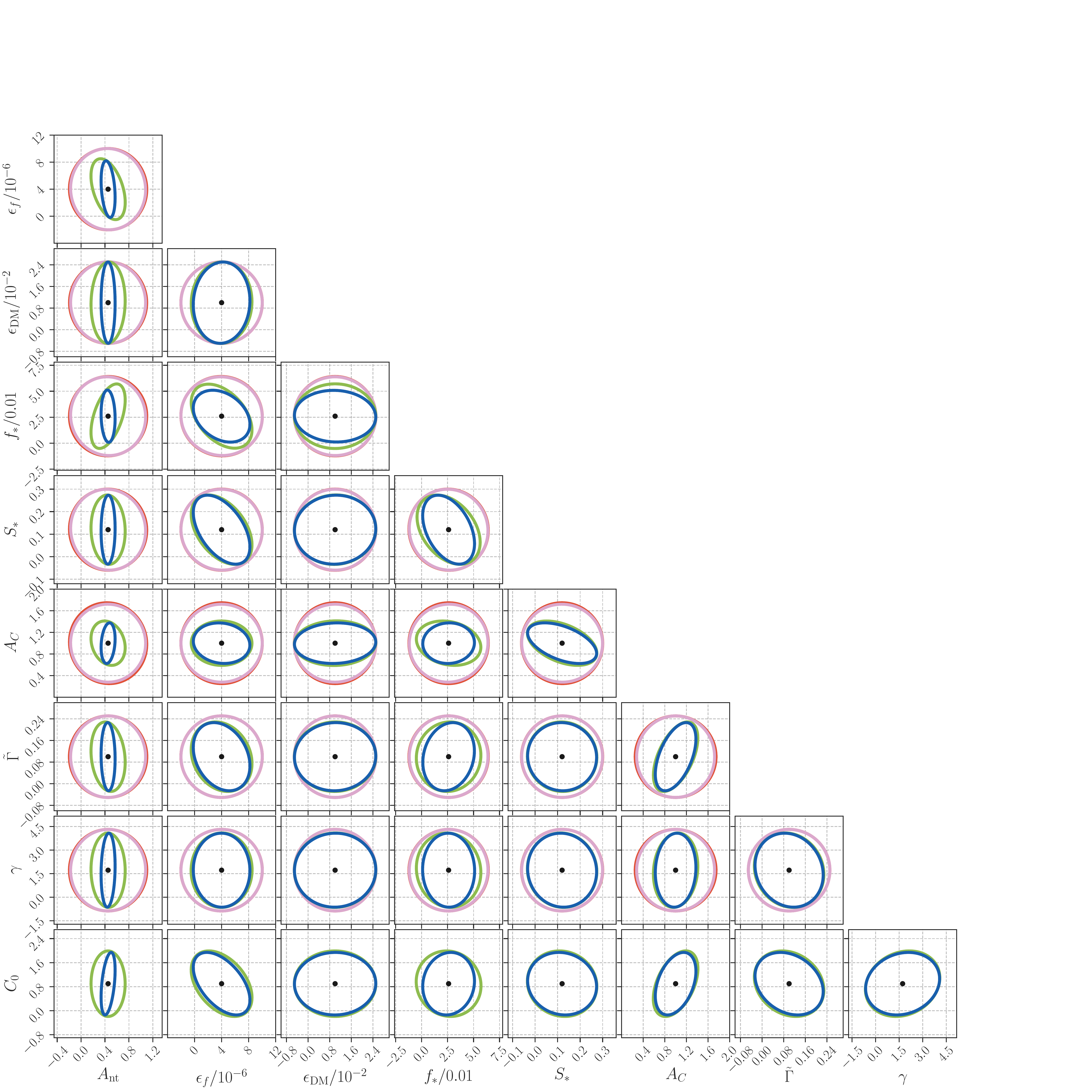}
\caption{
Similar to Figure~\ref{fig:fid_fisher_cosmo}, but focusing on 9 parameters of cluster astrophysics.
The same legend is adopted as in Figure~\ref{fig:fid_fisher_cosmo}.
It would be worth noting that the constraints without X-ray information are mostly determined 
by the priors.}
\label{fig:fid_fisher_gas}
\end{figure*}

As shown in Figure~\ref{fig:fid_fourpanels}, the X-ray information can improve the constraints of ICM model, 
and help break some of the degeneracy among ICM model and cosmological parameters. 
Note that we include the uncertainty of clumping effects in X-ray modeling and marginalize it in the Fisher forecast. 
Although some nuisance parameters can increase the uncertainty in the amplitude of power spectra, the joint analysis
is still able to improve the parameter constraints in ICM and cosmology.
Figures~\ref{fig:fid_fisher_cosmo} and \ref{fig:fid_fisher_gas} 
show the marginalized error circles in different two-parameter planes as we combine various information of the power spectra, highlighting that the X-ray information can play a crucial role in breaking parameter degeneracies.
The marginalized and un-marginalized errors of each parameter in our fiducial Fisher analysis are also provided in Table~\ref{tb:params_fisher_onesigma}. 

\begin{table}
\caption{
	The $1\sigma$ uncertainty evaluated by fisher analysis with a joint analysis
    of tSZ, X-ray, and lensing observables assuming a sky coverage of 20,000 square degrees.
    Note that we assume 10 tomographic bins are available in lensing information. See Table~\ref{tb:params_ICM} for details of the ICM model. \label{tb:params_fisher_onesigma}
	}
\scalebox{0.80}[0.80]{
\begin{tabular}{@{}llcc}
\hline
\hline
Parameter & Fiducial & Marginalized & Un-marginalized \\ 
\hline
Cosmology & & & \\
\hline
$\Omega_{\rm cdm}h^2$ & 0.1198 & $3.66\times10^{-2}$ & $2.48\times10^{-5}$\\
$\Omega_{\rm b0}h^2$ & 0.02225 & $6.10\times10^{-3}$ & $2.23\times10^{-6}$ \\
$A_{s}/10^{-9}$@$k=0.05\, {\rm Mpc}^{-1}$ & 2.2065 & $8.52\times10^{-1}$ & $5.72\times10^{-4}$ \\
$n_{s}$ & 0.9645 & $1.12\times10^{-1}$ & $1.89\times10^{-4}$ \\
$h$ & 0.6727 & $9.04\times10^{-2}$ & $2.34\times10^{-4}$ \\
$w_0$ & -1 & $8.08\times10^{-2}$ & $1.27\times10^{-3}$ \\
\hline
Intracluster medium & & &\\
\hline
$A_C$ & 1.0 & $2.48\times10^{-1}$ & $2.26\times10^{-4}$ \\ 
$\tilde{\Gamma}$ & 0.10 & $8.35\times10^{-2}$ & $5.76\times10^{-4}$ \\ 
$\gamma$ & 1.72 & $1.56$ & $5.82\times10^{-2}$ \\ 
$\epsilon_{\rm DM}$ & 0.010 & $9.96\times10^{-3}$ & $4.01\times10^{-5}$ \\ 
$\epsilon_{f}/10^{-6}$ & 4.00 & $2.77$ & $2.93\times10^{-3}$ \\ 
$f_{*}$ & 0.026 & $1.65\times10^{-2}$ & $5.27\times10^{-6}$ \\ 
$S_{*}$ & 0.12 & $1.01\times10^{-1}$ & $1.78\times10^{-4}$ \\ 
$A_{\rm nt}$ & 0.452 & $7.76\times10^{-2}$ & $1.68\times10^{-4}$ \\ 
$C_0$ & 0.90 & $6.10\times10^{-1}$ & $2.23\times10^{-6}$ \\ 
\hline
\end{tabular}
}
\end{table}

\subsection{Dependence of analysis and survey setups}\label{subsec:depend_analysis_survey}

\begin{figure}
\centering
\includegraphics[width=1.0\columnwidth]
{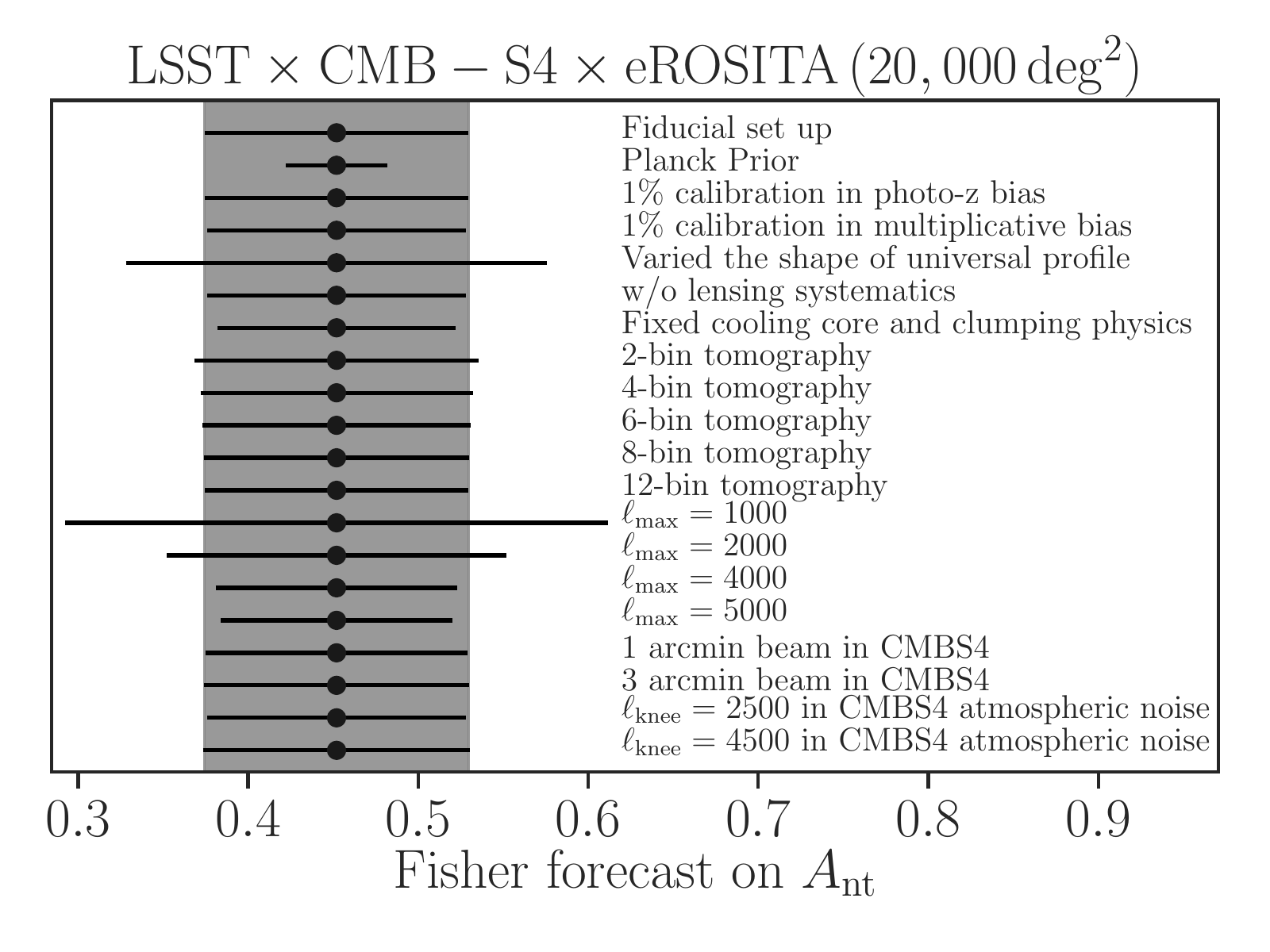}
\caption{
The dependence of survey/analysis setup on the 1$\sigma$ marginalized error in the parameter of non-thermal pressure, $A_{\rm nt}$.
The gray region shows the error with our fiducial setup.
}
\label{fig:test_summary_error_Ant}
\end{figure}

We then discuss how we can tighten the constraints furthermore by changing survey configurations and analysis methods.
In the following, we examine to what extent the constraint on non-thermal presssure $A_{\rm nt}$ will be affected by various choices in prior information, 
number of lensing tomographic bins, smallest scales in the analysis, and the detailed configuration in CMB-S4. 
The results are summarized in Figure~\ref{fig:test_summary_error_Ant}.
Similar tests on constraint for the equation-of-state parameter of dark energy are given in Appendix~\ref{apdx:test_fisher_cosmo}.

\subsubsection{Impact due to priors}

Since our power spectra have a complex parameter dependence and most of them degenerates each other,
it would be important to set appropriate prior information for improving parameter constraints.

We first examine the prior information from the CMB measurements by Planck satellite \citep{planck16}.
To do so, we use the approximate covariance of parameter estimates as shown in \citet{makiya18}, denoted as $C_{\rm P15}$.
After marginalizing the uncertainty of optical depth, we set the prior term in Fisher matrix (Eq.~\ref{eq:eff_fisher_matrix}) to be 
$F_{\rm prior} = \tilde{C}^{-1}_{\rm P15}$, where $\tilde{C}_{\rm P15}$ represents the prior information of covariance over 
five cosmological parameters of $\Omega_{\rm cdm}h^2$, $\Omega_{\rm b0}h^2$, $A_{s}$, $n_{s}$ and $h$.
We find that using the prior information from the Planck CMB measurments can tighten the constraint of 
$A_{\rm nt}$ by a factor of 3.8.

Another possible informative prior information will be included in the systematic effects of lensing measurements.
We change the prior Gaussian uncertainty of multiplicative bias $m$ and photometric redshift bias $z_{\rm b}$ in each tomographic bin to be 0.01,
but fixing other prior scatters same as in Table~\ref{tb:params_fisher}.
When applying a $1\%$ Gaussian scatter in $m$ and $z_{\rm b}$, we find that the expected constraint on $A_{\rm nt}$ is almost unaffected (it differs from the fiducial value by $\simlt0.5\%$ level).
This indicates that the joint analysis of tSZ, X-ray and lenisng power spectra will be insensitive to the choice of prior information of lensing systematics.

Detailed observations of individual galaxy clusters can further improve our understanding of ICM physics in the cluster cores and outskirts.
To see the impact of prior information on the ICM parameters, we simply examine the case with the perfect calibration of ICM parameters for the cooling in core regime ($\tilde{\Gamma}$ and $\gamma$) and gas clumping effects ($C_0$). 
Even if we can determine those parameters with an unprecedented accuracy, 
the joint analysis of tSZ, X-ray and lensing power spectra can constrain $A_{\rm nt}$ 
at a level of 0.078 (i.e. $\sim10\%$ improvement).
In addition, varying the shape of universal profile of non-thermal pressure and gas clumping (i.e. $B_{\rm nt}$, $\gamma_{\rm nt}$, $\alpha_{C}$, $\beta_C$ and $\gamma_C$), 
the expected $1\sigma$ error on $A_{\rm nt}$ will be degraded by $58\%$. 
This highlights the importance of external priors on the ICM model, especially in the X-ray band.

\subsubsection{Efficiency of lensing tomography}

Tomographic lensing analysis is useful for probing redshift evolution in ICM physics. 
Since there is no standard procedure to set the number of tomographic bins, we first examine the different number of tomographic bins from 2 to 12 with a fixed total source number density and check how our fiducial analysis is affected by the number of tomographic bins.
We find that the analysis with smaller tomographic bins slightly degrades the constraint on $A_{\rm nt}$ (by $6\%$ and $3\%$ from the fiducial value when adopting the 2-bin and 4-bin tomography, respectively), 
but the effect is generally small.
Increasing the number of tomographic bins does not always provide better constraints on the redshift evolution of the parameters, since there remains finite cross covariances between different tomographic bins (i.e., the correlation statistics between two different tomographic bins are not independent). 
It is difficult to collect independent information in redshift direction from lening observable alone. As the lensing effect is caused by the intervening gravitational potential between source and observer, the lensing kernels share the redshift information among different tomographic bins. 
From the practical point of view, it is better to set the elements of data vector as small as possible, because 
the inversion of covariance matrix will be inaccurate for large dimensions.

\subsubsection{Importance of small-scale information}

An aggressive analysis of the power spectra with smaller-scale information has the potential to improve the constraints on cluster astrophysics. 
It is, therefore, interesting to examine how much additional information can be gained by increasing the maximum multipole $\ell_{\rm max}$ in the Fisher analysis.
We examine four different values of $\ell_{\rm max}=1000, 2000, 4000$ and $5000$ in order to assess the effect of $\ell_{\rm max}$ on our fiducial analysis.
We find the constraint of $A_{\rm nt}$ is degraded by $104\%$ for $\ell_{\rm max}=1000$, whereas it is improved by $12\%$ for $\ell_{\rm max}=5000$.

\subsubsection{Dependence of noise power spectrum in tSZ measurements}

Finally, we examine the impact of the noise property in tSZ measurements on the Fisher analysis.
We consider four different noise power spectra in the tSZ auto power spectrum $N_{yy}$, indicated using blue solid, yellow solid, red dashed and red dotted lines in Figure~\ref{fig:noise_yy_CMBS4}.
We also vary the covariamce matrix of the power spectra with different $N_{yy}$, while keeping other survey parameters fixed.
We find that the beam size in CMB-S4 is less important for the cosmological constraints, whereas the amplitude of atmospheric noise can change the constraint of $A_{\rm nt}$ at a level of $\sim1\%$.
This indicates that precise measurements of the tSZ auto power spectrum at large scales (smaller $\ell$) 
will be critical for constraining the non-thermal pressure in cluster outskirts.

\subsection{Possible biases in parameter estimation}

Next, we investigate possible sources of systematic effects in parameter estimation.
In particular, we would like to understand how large parameter biases will be induced within our current modeling framework.
The current framework assumes the perfect subtraction of noise in tSZ and X-ray auto power spectra and the systematic effects in lensing measurements are under control.
In practice, astrophysical and cosmological constraints are affected by observational biases.
Here, we consider two main sources of observational biases originating from
(i) imperfect subtraction of noise power spectrum in tSZ auto power spectrum\footnote{The imperfect subtraction of X-ray shot noise is found to have a negligible impact of our Fisher forecast, because the shot noise term is sufficiently small compared to the expected $C_{xx}$ in the range of $\ell\simlt3000$.},
and (ii) mis-estimation of multiplicative and photometric redshift biases in lensing measurements.
To do so, we adopt the method to estimate the bias in parameter estimation \citep{2006MNRAS.366..101H}:
\begin{align}
\delta p_{\alpha}  = \sum_{\beta} F^{-1}_{\alpha \beta} \left( D^{\rm obs}_{i} - D^{\rm fid}_{i}\right) {\rm Cov}^{-1}_{ij} 
\left(\frac{\partial D^{\rm fid}_{j}}{\partial p_{\beta}}\right), \label{eq:param_bias}
\end{align}
where ${\bd p}$ is the parameter of interest, ${\bd D}^{\rm fid}$ is the data vector for fiducial model,
${\bd D}^{\rm obs}$ is the data vector affected by the statistical fluke, and ${\rm Cov}$ represents the data covariance.

\begin{figure}
\centering
\includegraphics[width=1.0\columnwidth]
{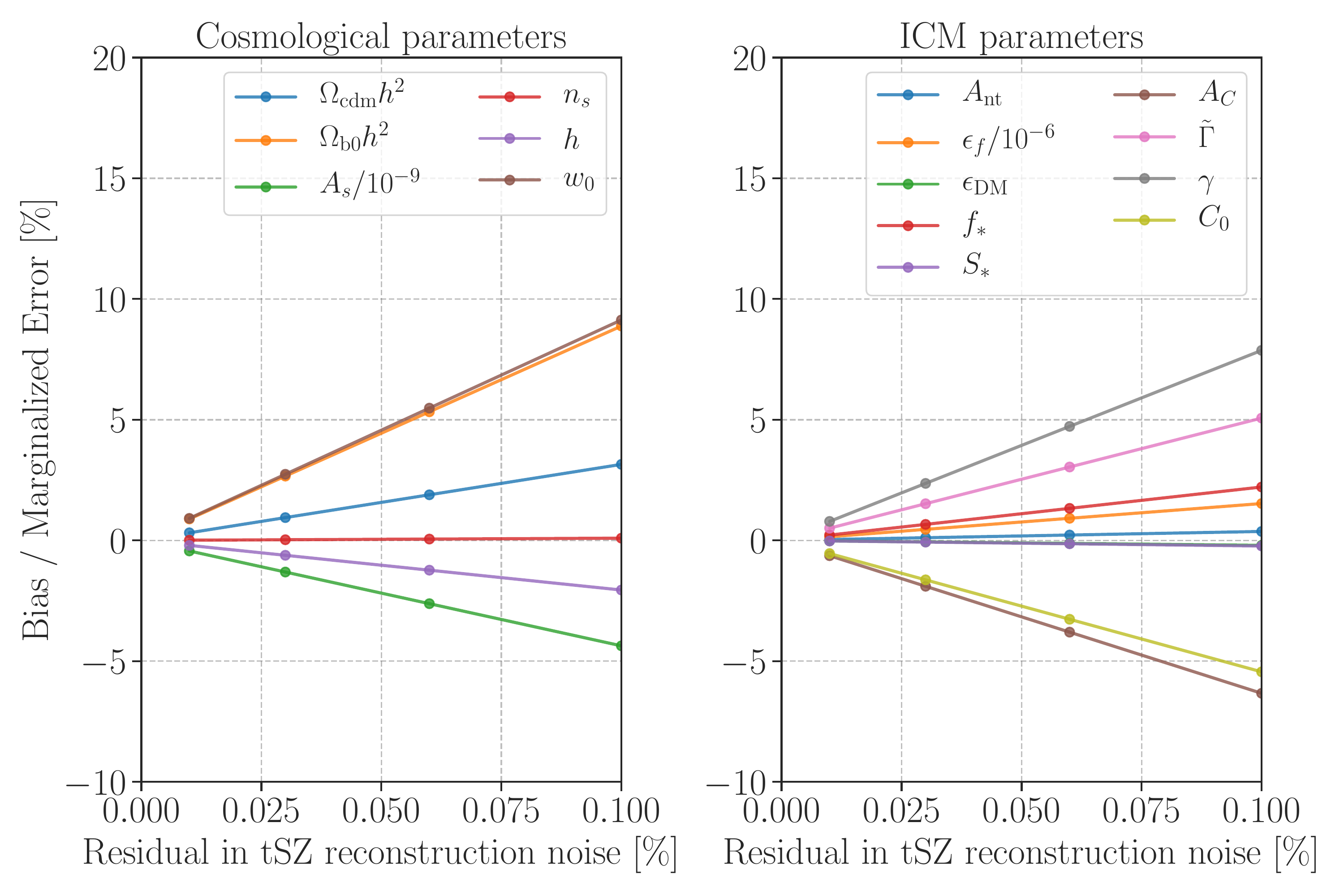}
\caption{
The bias in parameter estimation if the noise power spectrum in tSZ cannot be removed perfectly. The horizontal axis represents the residual of noise power spectrum
and the vertical axis shows the parameter bias with respect to the $1\sigma$ marginalized error.
The left panel shows the bias in cosmological parameters, while the right is for the ICM parameters. }
\label{fig:bias_yy}
\end{figure}

\subsubsection{Impact of astrophysical sources}\label{subsubsec:bias_astro_source}

We first study the impact of imperfect subtraction of noise power spectrum in tSZ measurements.
The expected signal-to-noise ratio of tSZ noise power spectrum is given by
\begin{align}
\left(\frac{\rm S}{\rm N}\right)^2 =& 
\sum_{i} \frac{N^2_{yy}(\ell_i)}{{\rm Var}[N_{yy}(\ell_i)]} = \sum_{i} \frac{f_{\rm sky}(2\ell_{i}+1)\Delta \ell}{2}, \label{eq:sn_noise}
\end{align}
where ${\rm Var}[N_{yy}(\ell_i)]$ represents the variance of $N_{yy}$ at i-th multipole $\ell_i$
and we assume the Gaussian variance in the second line in Eq.~(\ref{eq:sn_noise}).
In our fiducial setup, we find $({\rm S}/\rm N) = 1.47\times 10^{3} \left(\Omega_{\rm survey}/20,000 [{\rm deg}^2]\right)^{1/2}$, 
where $\Omega_{\rm survey}$ is the survey coverage.
Therefore, the amplitude of $N_{\rm yy}$ is estimated to be $\sim1/1470 = 0.068\%$ in the survey covering 20,000 square degrees.
Note that $N_{yy}$ can be estimated with the observational information alone (see Eq.~\ref{eq:tSZ_noise}).

In other words, the imperfect subtraction of $N_{yy}$ will lead to an uncertainty of 0.068\% in the tSZ measurements. This small residual may induce
the biased parameter estimations. We now model the residual of $N_{yy}$ by a simple form of ${\cal A}N_{yy}$, where $\cal A$ is the constant value in multipole $\ell$ and should be an order of 0.068\%.
Assuming there remains a residual, we can express 
the difference of ${\bd D}^{\rm obs}-{\bd D}^{\rm fid}$ in Eq.~(\ref{eq:param_bias}) as ${\cal A}N_{yy}$ for the term of $C_{yy}$ and zero otherwise.
Figure~\ref{fig:bias_yy} shows the possible parameter bias as a function of $\cal A$.
The left panel represents the bias of cosmological parameters with respect to the $1\sigma$ marginalized error, while the right is for some ICM parameters.
We find the residual of $N_{yy}$ with a level of 0.1\% can induce $\simlt0.1\sigma$-level bias in the estimated parameter.

\subsubsection{Photometric redshift estimates and Imperfect shape measurements}\label{subsubsec:bias_lens_sys}

\begin{figure}
\centering
\includegraphics[width=1.1\columnwidth]
{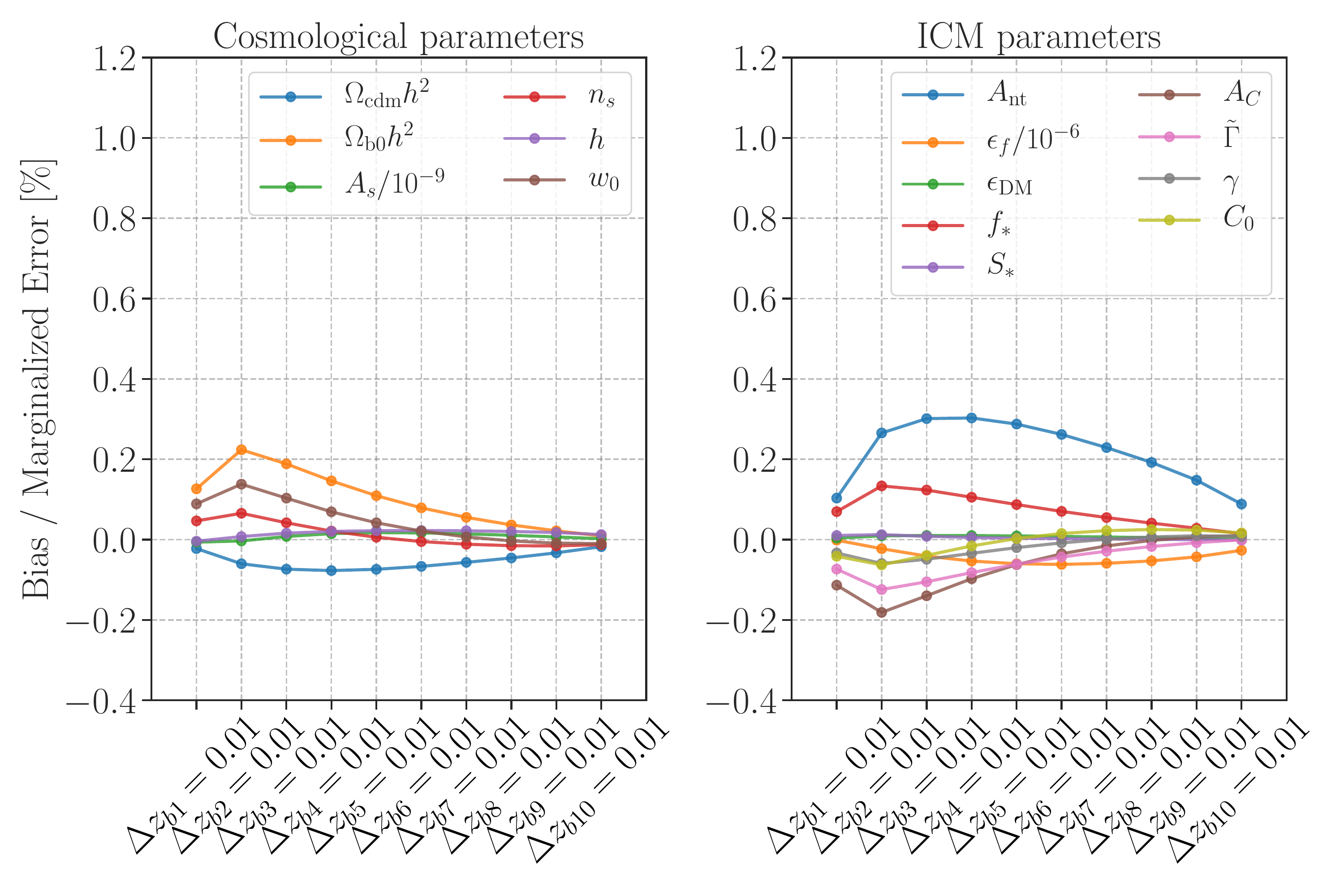}
\caption{
Similar to Figure~\ref{fig:bias_yy}, but we consider the parameter bias if the photometric redshift in single tomographic bins is biased by 0.01. }
\label{fig:bias_pz}
\end{figure}

We next study the impact of systematic effects in lensing observables.
We assume the systematic parameter in lensing ($m$ and $z_{\rm b}$) can be biased in actual observations. Specifically, we consider the bias in $m$ or $z_{\rm b}$ in each tomographic bin with a level of 0.01 and investigate how large bias $\delta {\bd p}$ can be induced by this small bias in $m$ or $z_{\rm b}$. 
Figure~\ref{fig:bias_pz} shows the bias of parameter estimation when the photo-z estimate in a single tomographic bin is biased by $0.01$.
According to this figure, the $1\%$ error
in photo-z bias $z_{b}$ can induce the bias in parameter estimation by $\sim0.005\sigma$.
We also confirm that the $1\%$ mis-estimation of multiplicative bias $m$ causes $\sim0.01\sigma$-level parameter biases.
Hence, we expect the systematic error due to multiplicative and photo-z biases will be subdominant in the joint analysis 
of tSZ, X-ray and lensing observables, as long as we calibrate the systematic bias in lensing at 1\% level.

\section{Limitations}\label{sec:caveat}

We summarize the major limitations in our 
correlation statistics for multi-wavelength cluster surveys.
All of the following issues will be addressed in forthcoming studies.

\subsection{Uncertainties in the ICM model}

Although our semi-analytic ICM model is an important step toward 
an effective utilization of multi-wavelength observables for precision cosmology and understanding ICM physics, it is still insufficient to apply to actual observation data sets.
For instance, our model assumes that there are no contributions 
to the observables from diffuse gas in large-scale filaments.
We also ignore the dependence of mass assembly history on gas profiles \citep[e.g.][]{nelson14b,shi15,lau15} and dark matter density profile \citep[e.g.][]{wechsler02,diemer14} throughout this paper, which might induce the bias in estimation of the parameter of interest in real data analysis.
To study this issue, we require a synthetic observational data generated from cosmological 
hydrodynamic simulations to fully take into account relevant ICM physics and non-linear gravitational growth.

\subsection{Additional correlation from astrophysical sources}

Our model assumes that the observed power spectra are dominated by the ICM. 
In actual observations, correlations can be induced by other faint astrophysical sources, 
including 
the extragalctic radio source and CIB in tSZ, 
and faint unresolved AGNs and galaxies in X-rays. 
Since these sources also correlate with each other and with the lensing observables, it may induce
non-negligible levels of parameter biases. 
The modeling of astrophysical sources in correlation statistics is still developing \citep[e.g.][]{cappelluti13,zentner13,
hurier15,planck16_tSZCIB,li18,shirasaki19} and require appropriate parametrizations to be taken into account in future studies.

\subsection{Intrinsic alignment in cluster-sized halos}

The intrinsic alignment (IA) effect of observed galaxies in galaxy imaging surveys is one of the major contaminants 
in modern lensing analyses \citep[e.g.][for review]{troxel15}.
This effect is likely more prominent in redder galaxies, indicating that member galaxies in cluster regimes will 
be more affected by the IA \citep[e.g. see][for recent observational studies]{singh15,samuroff18,hikage19,johnston19}.
Unfortunately, it is still difficult to include
the IA effect in our halo-model approach 
in a self-consistent way. Although there exists 
a simple halo model of the IA effect \citep[e.g.][]{schneider10}, the halo-model parameters have been poorly constrained so far. Further advances in theoretical modeling and observational studies are needed to implement IA effects in the halo model.

\subsection{Non-Gaussian covariance}

The statistical error in correlation statistics is assumed to be Gaussian throughout this paper, whereas the non-Gaussian covariance may play an important role in actual observations. The non-Gaussian covariance should arise from the non-linear gravitational growth of structures and it is computed as the four-point correlation function among observables of interest.
For instance, the non-Gaussian covariance in tSZ auto power spectrum $C_{yy}$ dominates the Gaussian error at $\ell\simlt100$ in the recent measurement \citep[e.g.][]{horowitz17,bolliet18,osato18}.
Nevertheless, the impact of non-Gaussian covariance on $C_{yy}$ is less critical for our cases, because we work with ground-based experiments and the noise at small $\ell$ is dominated by the atomospheric noises (see Figure~\ref{fig:noise_yy_CMBS4}).
On the other hand, the noise covariance in X-ray auto power spectrum $C_{xx}$ in eROSITA-like survey 
will be dominated by sample variance, 
indicating that non-Gaussian error may become important.
Analogously to the $C_{yy}$ case \citep[e.g.][]{komatsu02,hill13}, 
we expect that our Gaussian covariance in $C_{xx}$ may be underestimated by at most a factor of $\sim10$.
It is still uncertain if the non-Gaussian covariance 
in $C_{xx}$ can induce a large degradation of parameter constraints, since we have $30-40$ model parameters in total.
We will address this issue with a set of synthetic observations to evaluate precise covariance and perform a realistic likelihood analysis.

\section{CONCLUSIONS}\label{sec:conclusions}

Upcoming multi-wavelength galaxy cluster surveys are poised to significantly improve our understanding of astrophysics and cosmology of the large-scale structures. In this paper, we present a theoretical framework to predict the correlation statistics among the different galaxy cluster observables: 
thermal Sunyaev-Zeldovich (tSZ) effect in cosmic microwave background (CMB), 
X-ray emission from ionized intracluster gas, and weak gravitational lensing effect of background galaxies due to the cluster's gravitational field.

Our model framework is based on a physically-motivated semi-analytic model of ICM as presented in \S\ref{sec:SAM}. Using the model, we compute the auto and cross power spectra of the observables and perform a Fisher analysis to forecast the constraints on cosmology and ICM physics in future CMB ground-based experiment, X-ray satellite mission, and galaxy imaging survey. Our findings are summarized as follows:

\begin{itemize}

\item 
With joint tSZ, X-ray, and lensing auto and cross angular power spectra from multiwavelength surveys with 20,000 square degrees sky area coverage, we will be able to constrain the poorly understood cluster astrophysics: non-thermal pressure, gas clumping, and feedback from supernovae (SNe) 
and active galactic nuclei (AGN), at a level of 4.4$\sigma$, 2.0$\sigma$, and $1\sigma$ respectively. Although the improvements on individual cluster astrophysics parameters are modest due to the degeneracy among them, our gas model is sufficiently flexible in marginalising over the poorly understood cluster astrophysics parameters in a physically motivated way. 

\item Marginalising over the aforementioned cluster astrophysics, the joint tSZ, X-ray, and lensing auto- and cross-angular power spectra will allow us to constrain the equation-of-state parameter of Dark Energy, $w_0$, at a precision of $8\%$, which is comparable to the precision achieved by cluster abundance measurements.
\ms{Further improvements of the ICM model are required for its applications to real data sets (\S\ref{sec:caveat}).}

\item Our analyses indicate that tSZ noises will become the major systematic uncertainties, compared to lensing systematics, in affecting the level of bias in the ICM physics constraints. In particular, imperfect subtraction of tSZ noises at a $1\sigma$ level of the signal can induce $\sim 10\%$ bias in the ICM physics parameters with respect to the marginalized uncertainties. Whereas $1\%$-level biases in lensing systematics leads only to $0.5-1\%$ bias. 

\end{itemize}

With data from a plethora of multi-wavelength large-scale astronomical surveys coming online in the next decade, our framework will be useful for extracting valuable information about cosmology and ICM physics (such as feedback from SNe and AGN, non-thermal pressure, gas clumpng in cluster outskirts) simultaneously.
Since these ICM physics are the main source of systematic uncertainties for cluster cosmology, their constraints from the joint analysis can provide meaningful priors for constraining cosmology with measurements of cluster abundance and spatial clustering.

Aside from ICM physics, our physics-based framework can also be extended to smaller-mass haloes, such as groups and galaxies, as well as different cosmological models. In the future works, we plan to extend our model in order to interpret the cross-correlation measurements involving galaxy properties \citep[e.g.][]{planck13,greco15, vikram17,hill18,makiya18} as well as test non-standard cosmologies involving modified gravity, massive neutrinos, and variants of non-cold dark matter models.

\section*{Acknowledgements}
This work was in part supported by Grant-in-Aid for Scientific Research on Innovative Areas
from the MEXT KAKENHI Grant Number (18H04358, 19K14767).
Numerical computations presented in this paper
were in part carried out on the general-purpose PC farm at Center for Computational Astrophysics, CfCA, of National Astronomical Observatory of Japan.




\bibliographystyle{mnras}
\bibliography{ref} 

\begin{thebibliography}{}
\makeatletter
\relax
\def\mn@urlcharsother{\let\do\@makeother \do\$\do\&\do\#\do\^\do\_\do\%\do\~}
\def\mn@doi{\begingroup\mn@urlcharsother \@ifnextchar [ {\mn@doi@}
  {\mn@doi@[]}}
\def\mn@doi@[#1]#2{\def\@tempa{#1}\ifx\@tempa\@empty \href
  {http://dx.doi.org/#2} {doi:#2}\else \href {http://dx.doi.org/#2} {#1}\fi
  \endgroup}
\def\mn@eprint#1#2{\mn@eprint@#1:#2::\@nil}
\def\mn@eprint@arXiv#1{\href {http://arxiv.org/abs/#1} {{\tt arXiv:#1}}}
\def\mn@eprint@dblp#1{\href {http://dblp.uni-trier.de/rec/bibtex/#1.xml}
  {dblp:#1}}
\def\mn@eprint@#1:#2:#3:#4\@nil{\def\@tempa {#1}\def\@tempb {#2}\def\@tempc
  {#3}\ifx \@tempc \@empty \let \@tempc \@tempb \let \@tempb \@tempa \fi \ifx
  \@tempb \@empty \def\@tempb {arXiv}\fi \@ifundefined
  {mn@eprint@\@tempb}{\@tempb:\@tempc}{\expandafter \expandafter \csname
  mn@eprint@\@tempb\endcsname \expandafter{\@tempc}}}

\bibitem[\protect\citeauthoryear{{Abazajian} et~al.,}{{Abazajian}
  et~al.}{2019}]{CMBS4-19}
{Abazajian} K.,  et~al., 2019, arXiv e-prints, \href
  {https://ui.adsabs.harvard.edu/abs/2019arXiv190704473A} {}

\bibitem[\protect\citeauthoryear{{Aird} et~al.,}{{Aird}
  et~al.}{2010}]{2010MNRAS.401.2531A}
{Aird} J.,  et~al., 2010, \mn@doi [\mnras] {10.1111/j.1365-2966.2009.15829.x},
  \href {https://ui.adsabs.harvard.edu/\#abs/2010MNRAS.401.2531A} {401, 2531}

\bibitem[\protect\citeauthoryear{{Allen}, {Evrard}  \& {Mantz}}{{Allen}
  et~al.}{2011}]{allen11}
{Allen} S.~W.,  {Evrard} A.~E.,   {Mantz} A.~B.,  2011, \mn@doi [Annual Review
  of Astronomy and Astrophysics] {10.1146/annurev-astro-081710-102514}, \href
  {https://ui.adsabs.harvard.edu/\#abs/2011ARA&A..49..409A} {49, 409}

\bibitem[\protect\citeauthoryear{{Avestruz}, {Nagai}  \& {Lau}}{{Avestruz}
  et~al.}{2016}]{avestruz16}
{Avestruz} C.,  {Nagai} D.,   {Lau} E.~T.,  2016, \mn@doi [\apj]
  {10.3847/1538-4357/833/2/227}, \href
  {https://ui.adsabs.harvard.edu/abs/2016ApJ...833..227A} {833, 227}

\bibitem[\protect\citeauthoryear{Bartelmann \& Schneider}{Bartelmann \&
  Schneider}{2001}]{Bartelmann2001}
Bartelmann M.,  Schneider P.,  2001, \mn@doi [Physics Reports]
  {10.1016/S0370-1573(00)00082-X}, 340, 291

\bibitem[\protect\citeauthoryear{{Battaglia}, {Bond}, {Pfrommer}, {Sievers}  \&
  {Sijacki}}{{Battaglia} et~al.}{2010}]{battaglia10}
{Battaglia} N.,  {Bond} J.~R.,  {Pfrommer} C.,  {Sievers} J.~L.,   {Sijacki}
  D.,  2010, \mn@doi [\apj] {10.1088/0004-637X/725/1/91}, \href
  {https://ui.adsabs.harvard.edu/abs/2010ApJ...725...91B} {725, 91}

\bibitem[\protect\citeauthoryear{{Battaglia}, {Bond}, {Pfrommer}  \&
  {Sievers}}{{Battaglia} et~al.}{2012a}]{2012ApJ...758...75B}
{Battaglia} N.,  {Bond} J.~R.,  {Pfrommer} C.,   {Sievers} J.~L.,  2012a,
  \mn@doi [\apj] {10.1088/0004-637X/758/2/75}, \href
  {https://ui.adsabs.harvard.edu/\#abs/2012ApJ...758...75B} {758, 75}

\bibitem[\protect\citeauthoryear{{Battaglia}, {Bond}, {Pfrommer}  \&
  {Sievers}}{{Battaglia} et~al.}{2012b}]{battaglia12}
{Battaglia} N.,  {Bond} J.~R.,  {Pfrommer} C.,   {Sievers} J.~L.,  2012b,
  \mn@doi [\apj] {10.1088/0004-637X/758/2/74}, \href
  {https://ui.adsabs.harvard.edu/abs/2012ApJ...758...74B} {758, 74}

\bibitem[\protect\citeauthoryear{{Battaglia}, {Bond}, {Pfrommer}  \&
  {Sievers}}{{Battaglia} et~al.}{2015a}]{2015ApJ...806...43B}
{Battaglia} N.,  {Bond} J.~R.,  {Pfrommer} C.,   {Sievers} J.~L.,  2015a,
  \mn@doi [\apj] {10.1088/0004-637X/806/1/43}, \href
  {https://ui.adsabs.harvard.edu/abs/2015ApJ...806...43B} {806, 43}

\bibitem[\protect\citeauthoryear{{Battaglia}, {Hill}  \& {Murray}}{{Battaglia}
  et~al.}{2015b}]{battaglia15}
{Battaglia} N.,  {Hill} J.~C.,   {Murray} N.,  2015b, \mn@doi [\apj]
  {10.1088/0004-637X/812/2/154}, \href
  {http://adsabs.harvard.edu/abs/2015ApJ...812..154B} {812, 154}

\bibitem[\protect\citeauthoryear{{Biffi} et~al.,}{{Biffi}
  et~al.}{2016}]{biffi16}
{Biffi} V.,  et~al., 2016, \mn@doi [\apj] {10.3847/0004-637X/827/2/112}, \href
  {https://ui.adsabs.harvard.edu/abs/2016ApJ...827..112B} {827, 112}

\bibitem[\protect\citeauthoryear{{Bobin}, {Moudden}, {Starck}, {Fadili}  \&
  {Aghanim}}{{Bobin} et~al.}{2008}]{bobin08}
{Bobin} J.,  {Moudden} Y.,  {Starck} J.~L.,  {Fadili} J.,   {Aghanim} N.,
  2008, \mn@doi [Statistical Methodology] {10.1016/j.stamet.2007.10.003}, \href
  {https://ui.adsabs.harvard.edu/\#abs/2008StMet...5..307B} {5, 307}

\bibitem[\protect\citeauthoryear{{Bocquet} et~al.,}{{Bocquet}
  et~al.}{2019}]{bocquet19}
{Bocquet} S.,  et~al., 2019, \mn@doi [\apj] {10.3847/1538-4357/ab1f10}, \href
  {https://ui.adsabs.harvard.edu/abs/2019ApJ...878...55B} {878, 55}

\bibitem[\protect\citeauthoryear{{Bolliet}, {Comis}, {Komatsu}  \&
  {Mac{\'\i}as-P{\'e}rez}}{{Bolliet} et~al.}{2018}]{bolliet18}
{Bolliet} B.,  {Comis} B.,  {Komatsu} E.,   {Mac{\'\i}as-P{\'e}rez} J.~F.,
  2018, \mn@doi [\mnras] {10.1093/mnras/sty823}, \href
  {https://ui.adsabs.harvard.edu/\#abs/2018MNRAS.477.4957B} {477, 4957}

\bibitem[\protect\citeauthoryear{{Bryan} \& {Norman}}{{Bryan} \&
  {Norman}}{1998}]{1998ApJ...495...80B}
{Bryan} G.~L.,  {Norman} M.~L.,  1998, \mn@doi [\apj] {10.1086/305262}, \href
  {https://ui.adsabs.harvard.edu/\#abs/1998ApJ...495...80B} {495, 80}

\bibitem[\protect\citeauthoryear{{Cappelluti} et~al.,}{{Cappelluti}
  et~al.}{2013}]{cappelluti13}
{Cappelluti} N.,  et~al., 2013, \mn@doi [\apj] {10.1088/0004-637X/769/1/68},
  \href {https://ui.adsabs.harvard.edu/\#abs/2013ApJ...769...68C} {769, 68}

\bibitem[\protect\citeauthoryear{{Chisari} et~al.,}{{Chisari}
  et~al.}{2018}]{chisari18}
{Chisari} N.~E.,  et~al., 2018, \mn@doi [\mnras] {10.1093/mnras/sty2093}, \href
  {https://ui.adsabs.harvard.edu/\#abs/2018MNRAS.480.3962C} {480, 3962}

\bibitem[\protect\citeauthoryear{{Chluba}, {Nagai}, {Sazonov}  \&
  {Nelson}}{{Chluba} et~al.}{2012}]{2012MNRAS.426..510C}
{Chluba} J.,  {Nagai} D.,  {Sazonov} S.,   {Nelson} K.,  2012, \mn@doi [\mnras]
  {10.1111/j.1365-2966.2012.21741.x}, \href
  {http://adsabs.harvard.edu/abs/2012MNRAS.426..510C} {426, 510}

\bibitem[\protect\citeauthoryear{{Chown} et~al.,}{{Chown}
  et~al.}{2018}]{chown18}
{Chown} R.,  et~al., 2018, \mn@doi [The Astrophysical Journal Supplement
  Series] {10.3847/1538-4365/aae694}, \href
  {https://ui.adsabs.harvard.edu/\#abs/2018ApJS..239...10C} {239, 10}

\bibitem[\protect\citeauthoryear{{Cooray} \& {Sheth}}{{Cooray} \&
  {Sheth}}{2002}]{2002PhR...372....1C}
{Cooray} A.,  {Sheth} R.,  2002, \mn@doi [\physrep]
  {10.1016/S0370-1573(02)00276-4}, \href
  {https://ui.adsabs.harvard.edu/\#abs/2002PhR...372....1C} {372, 1}

\bibitem[\protect\citeauthoryear{{Diemer} \& {Kravtsov}}{{Diemer} \&
  {Kravtsov}}{2014}]{diemer14}
{Diemer} B.,  {Kravtsov} A.~V.,  2014, \mn@doi [\apj]
  {10.1088/0004-637X/789/1/1}, \href
  {https://ui.adsabs.harvard.edu/abs/2014ApJ...789....1D} {789, 1}

\bibitem[\protect\citeauthoryear{{Diemer} \& {Kravtsov}}{{Diemer} \&
  {Kravtsov}}{2015}]{2015ApJ...799..108D}
{Diemer} B.,  {Kravtsov} A.~V.,  2015, \mn@doi [\apj]
  {10.1088/0004-637X/799/1/108}, \href
  {http://adsabs.harvard.edu/abs/2015ApJ...799..108D} {799, 108}

\bibitem[\protect\citeauthoryear{{Duffy}, {Schaye}, {Kay}, {Dalla Vecchia},
  {Battye}  \& {Booth}}{{Duffy} et~al.}{2010}]{duffy10}
{Duffy} A.~R.,  {Schaye} J.,  {Kay} S.~T.,  {Dalla Vecchia} C.,  {Battye}
  R.~A.,   {Booth} C.~M.,  2010, \mn@doi [\mnras]
  {10.1111/j.1365-2966.2010.16613.x}, \href
  {https://ui.adsabs.harvard.edu/\#abs/2010MNRAS.405.2161D} {405, 2161}

\bibitem[\protect\citeauthoryear{{Dunkley} et~al.,}{{Dunkley}
  et~al.}{2013}]{2013JCAP...07..025D}
{Dunkley} J.,  et~al., 2013, \mn@doi [Journal of Cosmology and Astro-Particle
  Physics] {10.1088/1475-7516/2013/07/025}, \href
  {https://ui.adsabs.harvard.edu/\#abs/2013JCAP...07..025D} {2013, 025}

\bibitem[\protect\citeauthoryear{{Eifler}, {Krause}, {Dodelson}, {Zentner},
  {Hearin}  \& {Gnedin}}{{Eifler} et~al.}{2015}]{eifler15}
{Eifler} T.,  {Krause} E.,  {Dodelson} S.,  {Zentner} A.~R.,  {Hearin} A.~P.,
  {Gnedin} N.~Y.,  2015, \mn@doi [\mnras] {10.1093/mnras/stv2000}, \href
  {https://ui.adsabs.harvard.edu/\#abs/2015MNRAS.454.2451E} {454, 2451}

\bibitem[\protect\citeauthoryear{{Fabian} \& {Barcons}}{{Fabian} \&
  {Barcons}}{1992}]{fabian92}
{Fabian} A.~C.,  {Barcons} X.,  1992, \mn@doi [\araa]
  {10.1146/annurev.aa.30.090192.002241}, \href
  {http://adsabs.harvard.edu/abs/1992ARA%26A..30..429F} {30, 429}

\bibitem[\protect\citeauthoryear{{Fedeli}}{{Fedeli}}{2012}]{2012MNRAS.424.1244F}
{Fedeli} C.,  2012, \mn@doi [\mnras] {10.1111/j.1365-2966.2012.21302.x}, \href
  {https://ui.adsabs.harvard.edu/\#abs/2012MNRAS.424.1244F} {424, 1244}

\bibitem[\protect\citeauthoryear{{Fixsen}}{{Fixsen}}{2009}]{2009ApJ...707..916F}
{Fixsen} D.~J.,  2009, \mn@doi [\apj] {10.1088/0004-637X/707/2/916}, \href
  {http://adsabs.harvard.edu/abs/2009ApJ...707..916F} {707, 916}

\bibitem[\protect\citeauthoryear{{Flender}, {Nagai}  \& {McDonald}}{{Flender}
  et~al.}{2017}]{flender17}
{Flender} S.,  {Nagai} D.,   {McDonald} M.,  2017, \mn@doi [\apj]
  {10.3847/1538-4357/aa60bf}, \href
  {https://ui.adsabs.harvard.edu/abs/2017ApJ...837..124F} {837, 124}

\bibitem[\protect\citeauthoryear{{Gnedin}, {Kravtsov}, {Klypin}  \&
  {Nagai}}{{Gnedin} et~al.}{2004}]{2004ApJ...616...16G}
{Gnedin} O.~Y.,  {Kravtsov} A.~V.,  {Klypin} A.~A.,   {Nagai} D.,  2004,
  \mn@doi [\apj] {10.1086/424914}, \href
  {https://ui.adsabs.harvard.edu/\#abs/2004ApJ...616...16G} {616, 16}

\bibitem[\protect\citeauthoryear{{Greco}, {Hill}, {Spergel}  \&
  {Battaglia}}{{Greco} et~al.}{2015}]{greco15}
{Greco} J.~P.,  {Hill} J.~C.,  {Spergel} D.~N.,   {Battaglia} N.,  2015,
  \mn@doi [\apj] {10.1088/0004-637X/808/2/151}, \href
  {https://ui.adsabs.harvard.edu/abs/2015ApJ...808..151G} {808, 151}

\bibitem[\protect\citeauthoryear{{Hikage} et~al.,}{{Hikage}
  et~al.}{2019}]{hikage19}
{Hikage} C.,  et~al., 2019, \mn@doi [\pasj] {10.1093/pasj/psz010}, \href
  {https://ui.adsabs.harvard.edu/abs/2019PASJ...71...43H} {71, 43}

\bibitem[\protect\citeauthoryear{{Hildebrandt} et~al.,}{{Hildebrandt}
  et~al.}{2017}]{hildebrandt17}
{Hildebrandt} H.,  et~al., 2017, \mn@doi [\mnras] {10.1093/mnras/stw2805},
  \href {https://ui.adsabs.harvard.edu/\#abs/2017MNRAS.465.1454H} {465, 1454}

\bibitem[\protect\citeauthoryear{{Hill} \& {Pajer}}{{Hill} \&
  {Pajer}}{2013}]{hill13}
{Hill} J.~C.,  {Pajer} E.,  2013, \mn@doi [\prd] {10.1103/PhysRevD.88.063526},
  \href {https://ui.adsabs.harvard.edu/\#abs/2013PhRvD..88f3526H} {88, 063526}

\bibitem[\protect\citeauthoryear{{Hill} \& {Spergel}}{{Hill} \&
  {Spergel}}{2014}]{hill14}
{Hill} J.~C.,  {Spergel} D.~N.,  2014, \mn@doi [\jcap]
  {10.1088/1475-7516/2014/02/030}, \href
  {http://adsabs.harvard.edu/abs/2014JCAP...02..030H} {2, 030}

\bibitem[\protect\citeauthoryear{{Hill}, {Baxter}, {Lidz}, {Greco}  \&
  {Jain}}{{Hill} et~al.}{2018}]{hill18}
{Hill} J.~C.,  {Baxter} E.~J.,  {Lidz} A.,  {Greco} J.~P.,   {Jain} B.,  2018,
  \mn@doi [\prd] {10.1103/PhysRevD.97.083501}, \href
  {http://adsabs.harvard.edu/abs/2018PhRvD..97h3501H} {97, 083501}

\bibitem[\protect\citeauthoryear{{Hilton} et~al.,}{{Hilton}
  et~al.}{2018}]{hilton18}
{Hilton} M.,  et~al., 2018, \mn@doi [The Astrophysical Journal Supplement
  Series] {10.3847/1538-4365/aaa6cb}, \href
  {https://ui.adsabs.harvard.edu/\#abs/2018ApJS..235...20H} {235, 20}

\bibitem[\protect\citeauthoryear{{Hinshaw} et~al.,}{{Hinshaw}
  et~al.}{2013}]{hinshaw13}
{Hinshaw} G.,  et~al., 2013, \mn@doi [The Astrophysical Journal Supplement
  Series] {10.1088/0067-0049/208/2/19}, \href
  {https://ui.adsabs.harvard.edu/\#abs/2013ApJS..208...19H} {208, 19}

\bibitem[\protect\citeauthoryear{{Hojjati} et~al.,}{{Hojjati}
  et~al.}{2017}]{hojjati17}
{Hojjati} A.,  et~al., 2017, \mn@doi [\mnras] {10.1093/mnras/stx1659}, \href
  {http://adsabs.harvard.edu/abs/2017MNRAS.471.1565H} {471, 1565}

\bibitem[\protect\citeauthoryear{{Horowitz} \& {Seljak}}{{Horowitz} \&
  {Seljak}}{2017}]{horowitz17}
{Horowitz} B.,  {Seljak} U.,  2017, \mn@doi [\mnras] {10.1093/mnras/stx766},
  \href {https://ui.adsabs.harvard.edu/\#abs/2017MNRAS.469..394H} {469, 394}

\bibitem[\protect\citeauthoryear{{Hu} \& {Kravtsov}}{{Hu} \&
  {Kravtsov}}{2003}]{2003ApJ...584..702H}
{Hu} W.,  {Kravtsov} A.~V.,  2003, \mn@doi [\apj] {10.1086/345846}, \href
  {http://adsabs.harvard.edu/abs/2003ApJ...584..702H} {584, 702}

\bibitem[\protect\citeauthoryear{{Hurier}}{{Hurier}}{2015}]{hurier15}
{Hurier} G.,  2015, \mn@doi [\aap] {10.1051/0004-6361/201525714}, \href
  {https://ui.adsabs.harvard.edu/\#abs/} {575, L11}

\bibitem[\protect\citeauthoryear{{Hurier}, {Mac{\'{\i}}as-P{\'e}rez}  \&
  {Hildebrandt}}{{Hurier} et~al.}{2013}]{hurier13}
{Hurier} G.,  {Mac{\'{\i}}as-P{\'e}rez} J.~F.,   {Hildebrandt} S.,  2013,
  \mn@doi [\aap] {10.1051/0004-6361/201321891}, \href
  {http://adsabs.harvard.edu/abs/2013A%26A...558A.118H} {558, A118}

\bibitem[\protect\citeauthoryear{{Hurier}, {Aghanim}  \& {Douspis}}{{Hurier}
  et~al.}{2017}]{hurier17}
{Hurier} G.,  {Aghanim} N.,   {Douspis} M.,  2017, arXiv e-prints, \href
  {https://ui.adsabs.harvard.edu/\#abs/2017arXiv170200075H} {p.
  arXiv:1702.00075}

\bibitem[\protect\citeauthoryear{{Huterer} \& {Shafer}}{{Huterer} \&
  {Shafer}}{2018}]{2018RPPh...81a6901H}
{Huterer} D.,  {Shafer} D.~L.,  2018, \mn@doi [Reports on Progress in Physics]
  {10.1088/1361-6633/aa997e}, \href
  {https://ui.adsabs.harvard.edu/\#abs/2018RPPh...81a6901H} {81, 016901}

\bibitem[\protect\citeauthoryear{{Huterer}, {Takada}, {Bernstein}  \&
  {Jain}}{{Huterer} et~al.}{2006}]{2006MNRAS.366..101H}
{Huterer} D.,  {Takada} M.,  {Bernstein} G.,   {Jain} B.,  2006, \mn@doi
  [\mnras] {10.1111/j.1365-2966.2005.09782.x}, \href
  {http://adsabs.harvard.edu/abs/2006MNRAS.366..101H} {366, 101}

\bibitem[\protect\citeauthoryear{{Itoh}, {Kohyama}  \& {Nozawa}}{{Itoh}
  et~al.}{1998}]{1998ApJ...502....7I}
{Itoh} N.,  {Kohyama} Y.,   {Nozawa} S.,  1998, \mn@doi [\apj]
  {10.1086/305876}, \href {http://adsabs.harvard.edu/abs/1998ApJ...502....7I}
  {502, 7}

\bibitem[\protect\citeauthoryear{{Jing}, {Zhang}, {Lin}, {Gao}  \&
  {Springel}}{{Jing} et~al.}{2006}]{jing06}
{Jing} Y.~P.,  {Zhang} P.,  {Lin} W.~P.,  {Gao} L.,   {Springel} V.,  2006,
  \mn@doi [\apj] {10.1086/503547}, \href
  {https://ui.adsabs.harvard.edu/\#abs/2006ApJ...640L.119J} {640, L119}

\bibitem[\protect\citeauthoryear{{Johnston} et~al.,}{{Johnston}
  et~al.}{2019}]{johnston19}
{Johnston} H.,  et~al., 2019, \mn@doi [\aap] {10.1051/0004-6361/201834714},
  \href {https://ui.adsabs.harvard.edu/abs/2019A%26A...624A..30J} {624, A30}

\bibitem[\protect\citeauthoryear{{Khatri}}{{Khatri}}{2015}]{khatri15}
{Khatri} R.,  2015, \mn@doi [\mnras] {10.1093/mnras/stv1167}, \href
  {https://ui.adsabs.harvard.edu/\#abs/2015MNRAS.451.3321K} {451, 3321}

\bibitem[\protect\citeauthoryear{{Komatsu} \& {Kitayama}}{{Komatsu} \&
  {Kitayama}}{1999}]{1999ApJ...526L...1K}
{Komatsu} E.,  {Kitayama} T.,  1999, \mn@doi [\apj] {10.1086/312364}, \href
  {https://ui.adsabs.harvard.edu/\#abs/1999ApJ...526L...1K} {526, L1}

\bibitem[\protect\citeauthoryear{{Komatsu} \& {Seljak}}{{Komatsu} \&
  {Seljak}}{2002}]{komatsu02}
{Komatsu} E.,  {Seljak} U.,  2002, \mn@doi [\mnras]
  {10.1046/j.1365-8711.2002.05889.x}, \href
  {https://ui.adsabs.harvard.edu/abs/2002MNRAS.336.1256K} {336, 1256}

\bibitem[\protect\citeauthoryear{{LSST Science Collaboration} et~al.,}{{LSST
  Science Collaboration} et~al.}{2009}]{LSST09}
{LSST Science Collaboration} et~al., 2009, arXiv e-prints, \href
  {https://ui.adsabs.harvard.edu/\#abs/2009arXiv0912.0201L} {p.
  arXiv:0912.0201}

\bibitem[\protect\citeauthoryear{{Lakey} \& {Huffenberger}}{{Lakey} \&
  {Huffenberger}}{2019}]{lakey19}
{Lakey} V.,  {Huffenberger} K.,  2019, arXiv e-prints, \href
  {https://ui.adsabs.harvard.edu/abs/2019arXiv190208268L} {p. arXiv:1902.08268}

\bibitem[\protect\citeauthoryear{{Lau}, {Kravtsov}  \& {Nagai}}{{Lau}
  et~al.}{2009}]{lau09}
{Lau} E.~T.,  {Kravtsov} A.~V.,   {Nagai} D.,  2009, \mn@doi [\apj]
  {10.1088/0004-637X/705/2/1129}, \href
  {https://ui.adsabs.harvard.edu/abs/2009ApJ...705.1129L} {705, 1129}

\bibitem[\protect\citeauthoryear{{Lau}, {Nagai}  \& {Nelson}}{{Lau}
  et~al.}{2013}]{lau13}
{Lau} E.~T.,  {Nagai} D.,   {Nelson} K.,  2013, \mn@doi [\apj]
  {10.1088/0004-637X/777/2/151}, \href
  {https://ui.adsabs.harvard.edu/abs/2013ApJ...777..151L} {777, 151}

\bibitem[\protect\citeauthoryear{{Lau}, {Nagai}, {Avestruz}, {Nelson}  \&
  {Vikhlinin}}{{Lau} et~al.}{2015}]{lau15}
{Lau} E.~T.,  {Nagai} D.,  {Avestruz} C.,  {Nelson} K.,   {Vikhlinin} A.,
  2015, \mn@doi [\apj] {10.1088/0004-637X/806/1/68}, \href
  {https://ui.adsabs.harvard.edu/abs/2015ApJ...806...68L} {806, 68}

\bibitem[\protect\citeauthoryear{{Le Brun}, {McCarthy}, {Schaye}  \&
  {Ponman}}{{Le Brun} et~al.}{2017}]{lebrun_etal17}
{Le Brun} A.~M.~C.,  {McCarthy} I.~G.,  {Schaye} J.,   {Ponman} T.~J.,  2017,
  \mn@doi [\mnras] {10.1093/mnras/stw3361}, \href
  {http://adsabs.harvard.edu/abs/2017MNRAS.466.4442L} {466, 4442}

\bibitem[\protect\citeauthoryear{{Lehmer} et~al.,}{{Lehmer}
  et~al.}{2012}]{lehmer12}
{Lehmer} B.~D.,  et~al., 2012, \mn@doi [\apj] {10.1088/0004-637X/752/1/46},
  \href {https://ui.adsabs.harvard.edu/\#abs/2012ApJ...752...46L} {752, 46}

\bibitem[\protect\citeauthoryear{{Li}, {Cappelluti}, {Arendt}, {Hasinger},
  {Kashlinsky}  \& {Helgason}}{{Li} et~al.}{2018}]{li18}
{Li} Y.,  {Cappelluti} N.,  {Arendt} R.~G.,  {Hasinger} G.,  {Kashlinsky} A.,
  {Helgason} K.,  2018, \mn@doi [\apj] {10.3847/1538-4357/aad55a}, \href
  {https://ui.adsabs.harvard.edu/\#abs/2018ApJ...864..141L} {864, 141}

\bibitem[\protect\citeauthoryear{Limber}{Limber}{1954}]{Limber:1954zz}
Limber D.~N.,  1954, ApJ, 119, 655

\bibitem[\protect\citeauthoryear{{Ma}, {Van Waerbeke}, {Hinshaw}, {Hojjati},
  {Scott}  \& {Zuntz}}{{Ma} et~al.}{2015}]{ma15}
{Ma} Y.-Z.,  {Van Waerbeke} L.,  {Hinshaw} G.,  {Hojjati} A.,  {Scott} D.,
  {Zuntz} J.,  2015, \mn@doi [\jcap] {10.1088/1475-7516/2015/09/046}, \href
  {http://adsabs.harvard.edu/abs/2015JCAP...09..046M} {9, 046}

\bibitem[\protect\citeauthoryear{{Madhavacheril}, {Battaglia}  \&
  {Miyatake}}{{Madhavacheril} et~al.}{2017}]{2017PhRvD..96j3525M}
{Madhavacheril} M.~S.,  {Battaglia} N.,   {Miyatake} H.,  2017, \mn@doi [\prd]
  {10.1103/PhysRevD.96.103525}, \href
  {https://ui.adsabs.harvard.edu/\#abs/2017PhRvD..96j3525M} {96, 103525}

\bibitem[\protect\citeauthoryear{{Makiya}, {Ando}  \& {Komatsu}}{{Makiya}
  et~al.}{2018}]{makiya18}
{Makiya} R.,  {Ando} S.,   {Komatsu} E.,  2018, \mn@doi [\mnras]
  {10.1093/mnras/sty2031}, \href
  {https://ui.adsabs.harvard.edu/\#abs/2018MNRAS.480.3928M} {480, 3928}

\bibitem[\protect\citeauthoryear{{Makiya}, {Hikage}  \& {Komatsu}}{{Makiya}
  et~al.}{2019}]{makiya19}
{Makiya} R.,  {Hikage} C.,   {Komatsu} E.,  2019, arXiv e-prints, \href
  {https://ui.adsabs.harvard.edu/abs/2019arXiv190707870M} {p. arXiv:1907.07870}

\bibitem[\protect\citeauthoryear{{Mandelbaum} et~al.,}{{Mandelbaum}
  et~al.}{2018}]{mandelbaum18}
{Mandelbaum} R.,  et~al., 2018, \mn@doi [\pasj] {10.1093/pasj/psx130}, \href
  {https://ui.adsabs.harvard.edu/abs/2018PASJ...70S..25M} {70, S25}

\bibitem[\protect\citeauthoryear{{McDonald} et~al.,}{{McDonald}
  et~al.}{2013}]{2013ApJ...774...23M}
{McDonald} M.,  et~al., 2013, \mn@doi [\apj] {10.1088/0004-637X/774/1/23},
  \href {https://ui.adsabs.harvard.edu/\#abs/2013ApJ...774...23M} {774, 23}

\bibitem[\protect\citeauthoryear{{Merloni} et~al.,}{{Merloni}
  et~al.}{2012}]{merloni12}
{Merloni} A.,  et~al., 2012, arXiv e-prints, \href
  {https://ui.adsabs.harvard.edu/\#abs/2012arXiv1209.3114M} {p.
  arXiv:1209.3114}

\bibitem[\protect\citeauthoryear{{Mohammed} \& {Gnedin}}{{Mohammed} \&
  {Gnedin}}{2018}]{mohammed18}
{Mohammed} I.,  {Gnedin} N.~Y.,  2018, \mn@doi [\apj]
  {10.3847/1538-4357/aad3b1}, \href
  {https://ui.adsabs.harvard.edu/\#abs/2018ApJ...863..173M} {863, 173}

\bibitem[\protect\citeauthoryear{{Morrison} \& {McCammon}}{{Morrison} \&
  {McCammon}}{1983}]{mm83}
{Morrison} R.,  {McCammon} D.,  1983, \mn@doi [\apj] {10.1086/161102}, \href
  {https://ui.adsabs.harvard.edu/abs/1983ApJ...270..119M} {270, 119}

\bibitem[\protect\citeauthoryear{{Nagai} \& {Lau}}{{Nagai} \&
  {Lau}}{2011}]{nagai11}
{Nagai} D.,  {Lau} E.~T.,  2011, \mn@doi [\apj] {10.1088/2041-8205/731/1/L10},
  \href {https://ui.adsabs.harvard.edu/\#abs/2011ApJ...731L..10N} {731, L10}

\bibitem[\protect\citeauthoryear{{Nagai}, {Vikhlinin}  \& {Kravtsov}}{{Nagai}
  et~al.}{2007}]{nagai07a}
{Nagai} D.,  {Vikhlinin} A.,   {Kravtsov} A.~V.,  2007, \mn@doi [\apj]
  {10.1086/509868}, \href
  {https://ui.adsabs.harvard.edu/abs/2007ApJ...655...98N} {655, 98}

\bibitem[\protect\citeauthoryear{{Navarro}, {Frenk}  \& {White}}{{Navarro}
  et~al.}{1996}]{NFW}
{Navarro} J.~F.,  {Frenk} C.~S.,   {White} S.~D.~M.,  1996, \mn@doi [\apj]
  {10.1086/177173}, \href {http://adsabs.harvard.edu/abs/1996ApJ...462..563N}
  {462, 563}

\bibitem[\protect\citeauthoryear{{Nelson}, {Lau}, {Nagai}, {Rudd}  \&
  {Yu}}{{Nelson} et~al.}{2014a}]{nelson14a}
{Nelson} K.,  {Lau} E.~T.,  {Nagai} D.,  {Rudd} D.~H.,   {Yu} L.,  2014a,
  \mn@doi [\apj] {10.1088/0004-637X/782/2/107}, \href
  {http://adsabs.harvard.edu/abs/2014ApJ...782..107N} {782, 107}

\bibitem[\protect\citeauthoryear{{Nelson}, {Lau}  \& {Nagai}}{{Nelson}
  et~al.}{2014b}]{nelson14b}
{Nelson} K.,  {Lau} E.~T.,   {Nagai} D.,  2014b, \mn@doi [\apj]
  {10.1088/0004-637X/792/1/25}, \href
  {https://ui.adsabs.harvard.edu/abs/2014ApJ...792...25N} {792, 25}

\bibitem[\protect\citeauthoryear{{Osato}, {Flender}, {Nagai}, {Shirasaki}  \&
  {Yoshida}}{{Osato} et~al.}{2018}]{osato18}
{Osato} K.,  {Flender} S.,  {Nagai} D.,  {Shirasaki} M.,   {Yoshida} N.,  2018,
  \mn@doi [\mnras] {10.1093/mnras/stx3215}, \href
  {https://ui.adsabs.harvard.edu/abs/2018MNRAS.475..532O} {475, 532}

\bibitem[\protect\citeauthoryear{{Ostriker}, {Bode}  \& {Babul}}{{Ostriker}
  et~al.}{2005}]{ostriker05}
{Ostriker} J.~P.,  {Bode} P.,   {Babul} A.,  2005, \mn@doi [\apj]
  {10.1086/497122}, \href
  {https://ui.adsabs.harvard.edu/\#abs/2005ApJ...634..964O} {634, 964}

\bibitem[\protect\citeauthoryear{{Planck Collaboration} et~al.,}{{Planck
  Collaboration} et~al.}{2013}]{planck13}
{Planck Collaboration} et~al., 2013, \mn@doi [\aap]
  {10.1051/0004-6361/201220941}, \href
  {https://ui.adsabs.harvard.edu/abs/2013A&A...557A..52P} {557, A52}

\bibitem[\protect\citeauthoryear{{Planck Collaboration} et~al.,}{{Planck
  Collaboration} et~al.}{2016a}]{planck16}
{Planck Collaboration} et~al., 2016a, \mn@doi [\aap]
  {10.1051/0004-6361/201525830}, \href
  {http://adsabs.harvard.edu/abs/2016A%26A...594A..13P} {594, A13}

\bibitem[\protect\citeauthoryear{{Planck Collaboration} et~al.,}{{Planck
  Collaboration} et~al.}{2016b}]{planck16_tSZCIB}
{Planck Collaboration} et~al., 2016b, \mn@doi [\aap]
  {10.1051/0004-6361/201527418}, \href {https://ui.adsabs.harvard.edu/\#abs/}
  {594, A23}

\bibitem[\protect\citeauthoryear{{Planelles}, {Borgani}, {Fabjan}, {Killedar},
  {Murante}, {Granato}, {Ragone-Figueroa}  \& {Dolag}}{{Planelles}
  et~al.}{2014}]{planelles_etal14}
{Planelles} S.,  {Borgani} S.,  {Fabjan} D.,  {Killedar} M.,  {Murante} G.,
  {Granato} G.~L.,  {Ragone-Figueroa} C.,   {Dolag} K.,  2014, \mn@doi [\mnras]
  {10.1093/mnras/stt2141}, \href
  {http://adsabs.harvard.edu/abs/2014MNRAS.438..195P} {438, 195}

\bibitem[\protect\citeauthoryear{{Pratt}, {Arnaud}, {Biviano}, {Eckert},
  {Ettori}, {Nagai}, {Okabe}  \& {Reiprich}}{{Pratt} et~al.}{2019}]{pratt19}
{Pratt} G.~W.,  {Arnaud} M.,  {Biviano} A.,  {Eckert} D.,  {Ettori} S.,
  {Nagai} D.,  {Okabe} N.,   {Reiprich} T.~H.,  2019, \mn@doi [\ssr]
  {10.1007/s11214-019-0591-0}, \href
  {https://ui.adsabs.harvard.edu/abs/2019SSRv..215...25P} {215, 25}

\bibitem[\protect\citeauthoryear{{Ptak}, {Mobasher}, {Hornschemeier}, {Bauer}
  \& {Norman}}{{Ptak} et~al.}{2007}]{2007ApJ...667..826P}
{Ptak} A.,  {Mobasher} B.,  {Hornschemeier} A.,  {Bauer} F.,   {Norman} C.,
  2007, \mn@doi [\apj] {10.1086/520824}, \href
  {https://ui.adsabs.harvard.edu/\#abs/2007ApJ...667..826P} {667, 826}

\bibitem[\protect\citeauthoryear{{Puchwein}, {Sijacki}  \&
  {Springel}}{{Puchwein} et~al.}{2008}]{puchwein_etal08}
{Puchwein} E.,  {Sijacki} D.,   {Springel} V.,  2008, \mn@doi [\apjl]
  {10.1086/593352}, \href {http://adsabs.harvard.edu/abs/2008ApJ...687L..53P}
  {687, L53}

\bibitem[\protect\citeauthoryear{{Rasia} et~al.,}{{Rasia}
  et~al.}{2014}]{rasia14}
{Rasia} E.,  et~al., 2014, \mn@doi [\apj] {10.1088/0004-637X/791/2/96}, \href
  {https://ui.adsabs.harvard.edu/abs/2014ApJ...791...96R} {791, 96}

\bibitem[\protect\citeauthoryear{{Remazeilles}, {Delabrouille}  \&
  {Cardoso}}{{Remazeilles} et~al.}{2011}]{remazeilles11}
{Remazeilles} M.,  {Delabrouille} J.,   {Cardoso} J.-F.,  2011, \mn@doi
  [\mnras] {10.1111/j.1365-2966.2010.17624.x}, \href
  {https://ui.adsabs.harvard.edu/\#abs/2011MNRAS.410.2481R} {410, 2481}

\bibitem[\protect\citeauthoryear{{Remazeilles}, {Aghanim}  \&
  {Douspis}}{{Remazeilles} et~al.}{2013}]{remazeilles13}
{Remazeilles} M.,  {Aghanim} N.,   {Douspis} M.,  2013, \mn@doi [\mnras]
  {10.1093/mnras/sts636}, \href
  {https://ui.adsabs.harvard.edu/abs/2013MNRAS.430..370R} {430, 370}

\bibitem[\protect\citeauthoryear{{Rudd}, {Zentner}  \& {Kravtsov}}{{Rudd}
  et~al.}{2008}]{rudd08}
{Rudd} D.~H.,  {Zentner} A.~R.,   {Kravtsov} A.~V.,  2008, \mn@doi [\apj]
  {10.1086/523836}, \href
  {https://ui.adsabs.harvard.edu/\#abs/2008ApJ...672...19R} {672, 19}

\bibitem[\protect\citeauthoryear{{Samuroff} et~al.,}{{Samuroff}
  et~al.}{2018}]{samuroff18}
{Samuroff} S.,  et~al., 2018, arXiv e-prints, \href
  {https://ui.adsabs.harvard.edu/abs/2018arXiv181106989S} {}

\bibitem[\protect\citeauthoryear{{Schneider} \& {Bridle}}{{Schneider} \&
  {Bridle}}{2010}]{schneider10}
{Schneider} M.~D.,  {Bridle} S.,  2010, \mn@doi [\mnras]
  {10.1111/j.1365-2966.2009.15956.x}, \href
  {https://ui.adsabs.harvard.edu/\#abs/2010MNRAS.402.2127S} {402, 2127}

\bibitem[\protect\citeauthoryear{{Schneider} \& {Teyssier}}{{Schneider} \&
  {Teyssier}}{2015}]{schneider15}
{Schneider} A.,  {Teyssier} R.,  2015, \mn@doi [Journal of Cosmology and
  Astro-Particle Physics] {10.1088/1475-7516/2015/12/049}, \href
  {https://ui.adsabs.harvard.edu/\#abs/2015JCAP...12..049S} {2015, 049}

\bibitem[\protect\citeauthoryear{{Shaw}, {Nagai}, {Bhattacharya}  \&
  {Lau}}{{Shaw} et~al.}{2010}]{shaw10}
{Shaw} L.~D.,  {Nagai} D.,  {Bhattacharya} S.,   {Lau} E.~T.,  2010, \mn@doi
  [\apj] {10.1088/0004-637X/725/2/1452}, \href
  {https://ui.adsabs.harvard.edu/abs/2010ApJ...725.1452S} {725, 1452}

\bibitem[\protect\citeauthoryear{{Shi} \& {Komatsu}}{{Shi} \&
  {Komatsu}}{2014}]{shi_komatsu14}
{Shi} X.,  {Komatsu} E.,  2014, \mn@doi [\mnras] {10.1093/mnras/stu858}, \href
  {https://ui.adsabs.harvard.edu/abs/2014MNRAS.442..521S} {442, 521}

\bibitem[\protect\citeauthoryear{{Shi}, {Komatsu}, {Nelson}  \& {Nagai}}{{Shi}
  et~al.}{2015}]{shi15}
{Shi} X.,  {Komatsu} E.,  {Nelson} K.,   {Nagai} D.,  2015, \mn@doi [\mnras]
  {10.1093/mnras/stv036}, \href
  {https://ui.adsabs.harvard.edu/abs/2015MNRAS.448.1020S} {448, 1020}

\bibitem[\protect\citeauthoryear{{Shi}, {Komatsu}, {Nagai}  \& {Lau}}{{Shi}
  et~al.}{2016}]{shi16}
{Shi} X.,  {Komatsu} E.,  {Nagai} D.,   {Lau} E.~T.,  2016, \mn@doi [\mnras]
  {10.1093/mnras/stv2504}, \href
  {https://ui.adsabs.harvard.edu/abs/2016MNRAS.455.2936S} {455, 2936}

\bibitem[\protect\citeauthoryear{{Shirasaki}}{{Shirasaki}}{2019}]{shirasaki19}
{Shirasaki} M.,  2019, \mn@doi [\mnras] {10.1093/mnras/sty3162}, \href
  {https://ui.adsabs.harvard.edu/\#abs/2019MNRAS.483..342S} {483, 342}

\bibitem[\protect\citeauthoryear{{Shirasaki}, {Hamana}  \&
  {Yoshida}}{{Shirasaki} et~al.}{2016}]{shirasaki16}
{Shirasaki} M.,  {Hamana} T.,   {Yoshida} N.,  2016, \mn@doi [\pasj]
  {10.1093/pasj/psv105}, \href
  {https://ui.adsabs.harvard.edu/abs/2016PASJ...68....4S} {68, 4}

\bibitem[\protect\citeauthoryear{{Shirasaki}, {Lau}  \& {Nagai}}{{Shirasaki}
  et~al.}{2018}]{shirasaki18}
{Shirasaki} M.,  {Lau} E.~T.,   {Nagai} D.,  2018, \mn@doi [\mnras]
  {10.1093/mnras/sty763}, \href
  {https://ui.adsabs.harvard.edu/\#abs/2018MNRAS.477.2804S} {477, 2804}

\bibitem[\protect\citeauthoryear{{Simionescu} et~al.,}{{Simionescu}
  et~al.}{2011}]{2011Sci...331.1576S}
{Simionescu} A.,  et~al., 2011, \mn@doi [Science] {10.1126/science.1200331},
  \href {https://ui.adsabs.harvard.edu/abs/2011Sci...331.1576S} {331, 1576}

\bibitem[\protect\citeauthoryear{{Singh}, {Mandelbaum}  \& {More}}{{Singh}
  et~al.}{2015}]{singh15}
{Singh} S.,  {Mandelbaum} R.,   {More} S.,  2015, \mn@doi [\mnras]
  {10.1093/mnras/stv778}, \href
  {https://ui.adsabs.harvard.edu/\#abs/2015MNRAS.450.2195S} {450, 2195}

\bibitem[\protect\citeauthoryear{{Takahashi}, {Sato}, {Nishimichi}, {Taruya}
  \& {Oguri}}{{Takahashi} et~al.}{2012}]{takahashi12}
{Takahashi} R.,  {Sato} M.,  {Nishimichi} T.,  {Taruya} A.,   {Oguri} M.,
  2012, \mn@doi [\apj] {10.1088/0004-637X/761/2/152}, \href
  {https://ui.adsabs.harvard.edu/\#abs/2012ApJ...761..152T} {761, 152}

\bibitem[\protect\citeauthoryear{{Tanaka} et~al.,}{{Tanaka}
  et~al.}{2018}]{takana18}
{Tanaka} M.,  et~al., 2018, \mn@doi [\pasj] {10.1093/pasj/psx077}, \href
  {https://ui.adsabs.harvard.edu/abs/2018PASJ...70S...9T} {70, S9}

\bibitem[\protect\citeauthoryear{{Tinker}, {Kravtsov}, {Klypin}, {Abazajian},
  {Warren}, {Yepes}, {Gottl{\"o}ber}  \& {Holz}}{{Tinker}
  et~al.}{2008}]{2008ApJ...688..709T}
{Tinker} J.,  {Kravtsov} A.~V.,  {Klypin} A.,  {Abazajian} K.,  {Warren} M.,
  {Yepes} G.,  {Gottl{\"o}ber} S.,   {Holz} D.~E.,  2008, \mn@doi [\apj]
  {10.1086/591439}, \href {http://adsabs.harvard.edu/abs/2008ApJ...688..709T}
  {688, 709}

\bibitem[\protect\citeauthoryear{{Tinker}, {Robertson}, {Kravtsov}, {Klypin},
  {Warren}, {Yepes}  \& {Gottl{\"o}ber}}{{Tinker}
  et~al.}{2010}]{2010ApJ...724..878T}
{Tinker} J.~L.,  {Robertson} B.~E.,  {Kravtsov} A.~V.,  {Klypin} A.,  {Warren}
  M.~S.,  {Yepes} G.,   {Gottl{\"o}ber} S.,  2010, \mn@doi [\apj]
  {10.1088/0004-637X/724/2/878}, \href
  {http://adsabs.harvard.edu/abs/2010ApJ...724..878T} {724, 878}

\bibitem[\protect\citeauthoryear{{Tozzi} et~al.,}{{Tozzi}
  et~al.}{2006}]{2006A&A...451..457T}
{Tozzi} P.,  et~al., 2006, \mn@doi [\aap] {10.1051/0004-6361:20042592}, \href
  {https://ui.adsabs.harvard.edu/\#abs/2006A&A...451..457T} {451, 457}

\bibitem[\protect\citeauthoryear{{Treister} \& {Urry}}{{Treister} \&
  {Urry}}{2006}]{treister06}
{Treister} E.,  {Urry} C.~M.,  2006, \mn@doi [\apj] {10.1086/510237}, \href
  {https://ui.adsabs.harvard.edu/\#abs/2006ApJ...652L..79T} {652, L79}

\bibitem[\protect\citeauthoryear{{Troxel} \& {Ishak}}{{Troxel} \&
  {Ishak}}{2015}]{troxel15}
{Troxel} M.~A.,  {Ishak} M.,  2015, \mn@doi [\physrep]
  {10.1016/j.physrep.2014.11.001}, \href
  {http://adsabs.harvard.edu/abs/2015PhR...558....1T} {558, 1}

\bibitem[\protect\citeauthoryear{{Troxel} et~al.,}{{Troxel}
  et~al.}{2018}]{troxel18}
{Troxel} M.~A.,  et~al., 2018, \mn@doi [\prd] {10.1103/PhysRevD.98.043528},
  \href {https://ui.adsabs.harvard.edu/\#abs/2018PhRvD..98d3528T} {98, 043528}

\bibitem[\protect\citeauthoryear{{Ueda}, {Akiyama}, {Hasinger}, {Miyaji}  \&
  {Watson}}{{Ueda} et~al.}{2014}]{ueda14}
{Ueda} Y.,  {Akiyama} M.,  {Hasinger} G.,  {Miyaji} T.,   {Watson} M.~G.,
  2014, \mn@doi [\apj] {10.1088/0004-637X/786/2/104}, \href
  {https://ui.adsabs.harvard.edu/\#abs/2014ApJ...786..104U} {786, 104}

\bibitem[\protect\citeauthoryear{{Van Waerbeke}, {Hinshaw}  \& {Murray}}{{Van
  Waerbeke} et~al.}{2014}]{vanwaerbeke14}
{Van Waerbeke} L.,  {Hinshaw} G.,   {Murray} N.,  2014, \mn@doi [\prd]
  {10.1103/PhysRevD.89.023508}, \href
  {http://adsabs.harvard.edu/abs/2014PhRvD..89b3508V} {89, 023508}

\bibitem[\protect\citeauthoryear{{Vazza}, {Eckert}, {Simionescu}, {Br{\"u}ggen}
   \& {Ettori}}{{Vazza} et~al.}{2013}]{vazza13}
{Vazza} F.,  {Eckert} D.,  {Simionescu} A.,  {Br{\"u}ggen} M.,   {Ettori} S.,
  2013, \mn@doi [\mnras] {10.1093/mnras/sts375}, \href
  {https://ui.adsabs.harvard.edu/abs/2013MNRAS.429..799V} {429, 799}

\bibitem[\protect\citeauthoryear{{Vikram}, {Lidz}  \& {Jain}}{{Vikram}
  et~al.}{2017}]{vikram17}
{Vikram} V.,  {Lidz} A.,   {Jain} B.,  2017, \mn@doi [\mnras]
  {10.1093/mnras/stw3311}, \href
  {https://ui.adsabs.harvard.edu/abs/2017MNRAS.467.2315V} {467, 2315}

\bibitem[\protect\citeauthoryear{{Walker}, {Fabian}, {Sanders}, {Simionescu}
  \& {Tawara}}{{Walker} et~al.}{2013}]{2013MNRAS.432.554W}
{Walker} S.~A.,  {Fabian} A.~C.,  {Sanders} J.~S.,  {Simionescu} A.,   {Tawara}
  Y.,  2013, \mn@doi [\mnras] {10.1093/mnras/stt497}, \href
  {http://ads.nao.ac.jp/abs/2013MNRAS.432..554W} {432, 554}

\bibitem[\protect\citeauthoryear{{Walker} et~al.,}{{Walker}
  et~al.}{2019}]{walker19}
{Walker} S.,  et~al., 2019, \mn@doi [\ssr] {10.1007/s11214-018-0572-8}, \href
  {https://ui.adsabs.harvard.edu/abs/2019SSRv..215....7W} {215, 7}

\bibitem[\protect\citeauthoryear{{Wechsler}, {Bullock}, {Primack}, {Kravtsov}
  \& {Dekel}}{{Wechsler} et~al.}{2002}]{wechsler02}
{Wechsler} R.~H.,  {Bullock} J.~S.,  {Primack} J.~R.,  {Kravtsov} A.~V.,
  {Dekel} A.,  2002, \mn@doi [\apj] {10.1086/338765}, \href
  {http://adsabs.harvard.edu/abs/2002ApJ...568...52W} {568, 52}

\bibitem[\protect\citeauthoryear{{Young} et~al.,}{{Young}
  et~al.}{2012}]{2012ApJ...748..124Y}
{Young} M.,  et~al., 2012, \mn@doi [\apj] {10.1088/0004-637X/748/2/124}, \href
  {https://ui.adsabs.harvard.edu/\#abs/2012ApJ...748..124Y} {748, 124}

\bibitem[\protect\citeauthoryear{{Zandanel}, {Weniger}  \& {Ando}}{{Zandanel}
  et~al.}{2015}]{2015JCAP...09..060Z}
{Zandanel} F.,  {Weniger} C.,   {Ando} S.,  2015, \mn@doi [Journal of Cosmology
  and Astro-Particle Physics] {10.1088/1475-7516/2015/09/060}, \href
  {https://ui.adsabs.harvard.edu/\#abs/2015JCAP...09..060Z} {2015, 060}

\bibitem[\protect\citeauthoryear{{Zentner}, {Semboloni}, {Dodelson}, {Eifler},
  {Krause}  \& {Hearin}}{{Zentner} et~al.}{2013}]{zentner13}
{Zentner} A.~R.,  {Semboloni} E.,  {Dodelson} S.,  {Eifler} T.,  {Krause} E.,
  {Hearin} A.~P.,  2013, \mn@doi [\prd] {10.1103/PhysRevD.87.043509}, \href
  {https://ui.adsabs.harvard.edu/\#abs/2013PhRvD..87d3509Z} {87, 043509}

\makeatother
\end{thebibliography}



\appendix

\section{Minimum variance method for thermal Sunyaev-Zel'dovich (tSZ) analysis}\label{apdx:noise_tsz}

We here summarize the minimum-variance method for estimation of tSZ signal from observed surface-brightness temperature maps at multiple wavelengths \citep[e.g., see][]{hill13}.
The method finds the weight of multi-frequency maps in microwave so as to minimize the variance of the tSZ auto power spectrum.

We first define an estimator of Compton-$y$ (Eq.~\ref{eq:T-y}) in Fourier space as
\beq
\hat{y}(\bd{\ell}) = \sum_{i} \frac{w_{i}}{g_{i}}
\frac{T_{i}(\bd{\ell})}{T_0}, \label{eq:y_est_fourier}
\eeq
where $T_{i}$ represents the observed temperature map at $i$-th
frequency channel in the Fourier space, $g_{i}$ is the expected coefficient of tSZ effect at the $i$-th frequency, and the weight function $w_{i}$ must satisfy $\sum_{i} w_{i} = 1$ so that the estimator $\hat{y}$ provides an unbiased estimate of the Compton-$y$. The minimum variance weight $w_{\rm mv}$ can be written as 
\beq
w_{{\rm mv}, i} = \frac{\sum_{j} C^{-1}_{ij} e_{j}}{\sum_{j,k}C^{-1}_{jk}e_{j}e_{k}},
\eeq
where $e_{i}$ is a vector with all ones
and the matrix $C_{ij}$ is defined by
the cross-correlation at different frequencies of 
foreground components in $g^{-1}_{i} \Delta T_{i}/T_0$ \citep{hill13}.
When setting $w_{i} = w_{{\rm mv}, i}$,
one can find the noise term of the auto power spectrum in Eq.~(\ref{eq:y_est_fourier}) as
\beq
N_{yy}(\ell) = \sum_{i,j}w_{{\rm mv}, i} \, w_{{\rm mv}, j} \, C_{ij}.
\eeq

Therefore, we need a model of $C_{ij}$ to compute 
the minimum-variance weight $w_{\rm mv}$ and $N_{yy}(\ell)$.
The term $C_{ij}$ is commonly decomposed into three parts:
\beq
g_{i} \, g_{j} \, C_{ij}(\ell) = 
C_{{\rm CMB}, ij}(\ell) + C_{{\rm sec}, ij}(\ell) + \frac{C_{\rm N}(\ell)}{B^2_{{\rm CMB}, \ell}}\delta_{ij},
\eeq
where $C_{\rm CMB}$ is the primary CMB power spectrum, 
$C_{\rm sec}$ is the secondary power spectrum,
$C_{\rm N}$ is the noise power spectrum in the microwave measurement,
and $B_{{\rm CMB}, \ell}$ is the Fourier transform of the beam
in the CMB measurement (which we assume to be Gaussian).
For the $C_{\rm sec}$ term, we follow the model as developed in \citet{2013JCAP...07..025D}, which is a nine-parameter model including contributions from (i) the thermal and kinematic SZ effects, (ii) the clustered and Poisson-like power from Cosmic Infrared Background (CIB) sources, and their frequency scaling, (iii) the tSZ-CIB correlation, (iv) the extragalactic radio source, 
(v) and thermal dust emission from Galactic cirrus.
We use the best-fit parameters of the model fitted to the data from Atacama Cosmology Telescope \citep[ACT, see][for details]{2013JCAP...07..025D}.
For the $C_{\rm N}$ term, we adopt the form presented in \citet{2017PhRvD..96j3525M} and given by 
\beq
C_{\rm N}(\ell) = s^2(\nu)\left[1+\left(\frac{\ell}{\ell_{\rm knee}}\right)^{\alpha}\right],
\eeq
where $s(\nu)$ is the instrumental white noise at frequency $\nu$
and the remaining term in the right hand side expresses
the effect of atmospheric noise for a ground-based 
experiment. 
Throughout this paper, we fix $\alpha=-4.5$, but
examine three different values of $\ell_{\rm knee}=2500, 3500$ and $4500$.
We adopt $\ell_{\rm knee}=3500$ unless otherwise stated.

\section{A model of X-ray point sources}\label{apdx:x-ray-ps}

In this appendix, we summarize a phenomenological model of X-ray point sources. In the following, we consider two major populations at X-ray bands, AGN and normal galaxies. 

Suppose that all resolved point sources can be masked, 
the remaining X-ray intensity from an unresolved point source $\alpha$ 
is given by \citep[also see][]{2015JCAP...09..060Z}
\begin{align}
I_{X, \alpha} =& \int_{E_{\rm min}}^{E_{\rm max}} {\rm d}E \int_{0}^{\chi_H}\, {\rm d}\chi\, W_{X, \alpha}([1+z]E, z(\chi)),\label{eq:Xray_point_sources} \\
W_{X, \alpha}(E,z) =& \frac{1}{4\pi}
\int_{L_{X, {\rm min}}}^{L_{X, {\rm max}}(z)}\, {\rm d}L_{X}\, 
\Phi_{X, \alpha}(L_{X}, z)\, {\cal L}_{X, \alpha}(E, z),
\label{eq:Xray_point_sources_window}
\end{align}
where $\alpha = \{{\rm AGN}, {\rm Galaxy}\}$, 
$\Phi_{X, \alpha}$ is the X-ray luminosity function of the source $\alpha$, and ${\cal L}_{X, \alpha}$ represents
the differential luminosity at an energy $E$ and redshift $z$.
Note that ${\cal L}_{X, \alpha}$ is defined as
a number of photon emitted per unit time and per unit energy range. 
Hence, it depends on X-ray energy spectrum of the source $\alpha$ and the definition of $L_{X}$.
Also, the upper bound of luminosity $L_{X, {\rm max}}(z)$ is set so as to compute all contributions with X-ray flux 
less than the flux limit of the survey.

To compute Eq.~(\ref{eq:Xray_point_sources}),
we adopt the model of $\Phi_{X, \rm AGN}$ developed by
\citet{2010MNRAS.401.2531A} in the rest-frame energy range of $2-10$ keV, while we use the model in \citet{2007ApJ...667..826P} for $\Phi_{X, \rm Galaxy}$ in the rest-frame energy range of $0.5-2$ keV.
The differential luminosity is then computed as
\beq
{\cal L}_{X,\alpha} = L_{X}\, \frac{E^{-\Gamma_{\alpha}}}{\int_{E_{\rm min,ref}}^{E_{\rm min,ref}}\, {\rm d}E
\, E^{1-\Gamma_{\alpha}}},
\eeq
where 
$E_{\rm min, ref}$ and $E_{\rm max, ref}$ are the minimum and maximum rest-frame energy in the definition of $L_X$, respectively.
We assume the AGN spectrum index to be $\Gamma_{\alpha}=1.7$
\citep[e.g.][]{2006A&A...451..457T} and $\Gamma_{\alpha}=2$ 
for normal galaxies \citep[e.g.][]{2012ApJ...748..124Y} in this paper.

For the lower and upper limit of luminosity integration in Eq.~(\ref{eq:Xray_point_sources_window}), we set $L_{X, {\rm min}}=10^{41}\, {\rm erg}\, {\rm s}^{-1}$ and
$L_{X, {\rm max}}(z) = 4\pi d^2_{L}(z)\, F_{\rm lim}\, K(z, E_{\rm min}, E_{\rm max})$,
where $d_{L}(z)$ is the luminosity function to redshift $z$, $F_{\rm lim}$ is the X-ray flux limit in the survey, and $K$ is the $K$-correction given by
\beq
K(z, E_{\rm min}, E_{\rm max}) = 
\frac{1}{(1+z)^{2-\Gamma}}\, \frac{E_{\rm max}^{2-\Gamma}-E_{\rm min}^{2-\Gamma}}{E_{\rm max, ref}^{2-\Gamma}-E_{\rm min, ref}^{2-\Gamma}},
\eeq
for AGN with $\Gamma=1.7$ and
\beq
K(z, E_{\rm min}, E_{\rm max}) = \frac{\ln(E_{\rm max}/E_{\rm min})}{\ln(E_{\rm max, ref}/E_{\rm min, ref})}.
\eeq
for normal galaxies.

\begin{figure}
\centering
\includegraphics[width=1.0\columnwidth]
{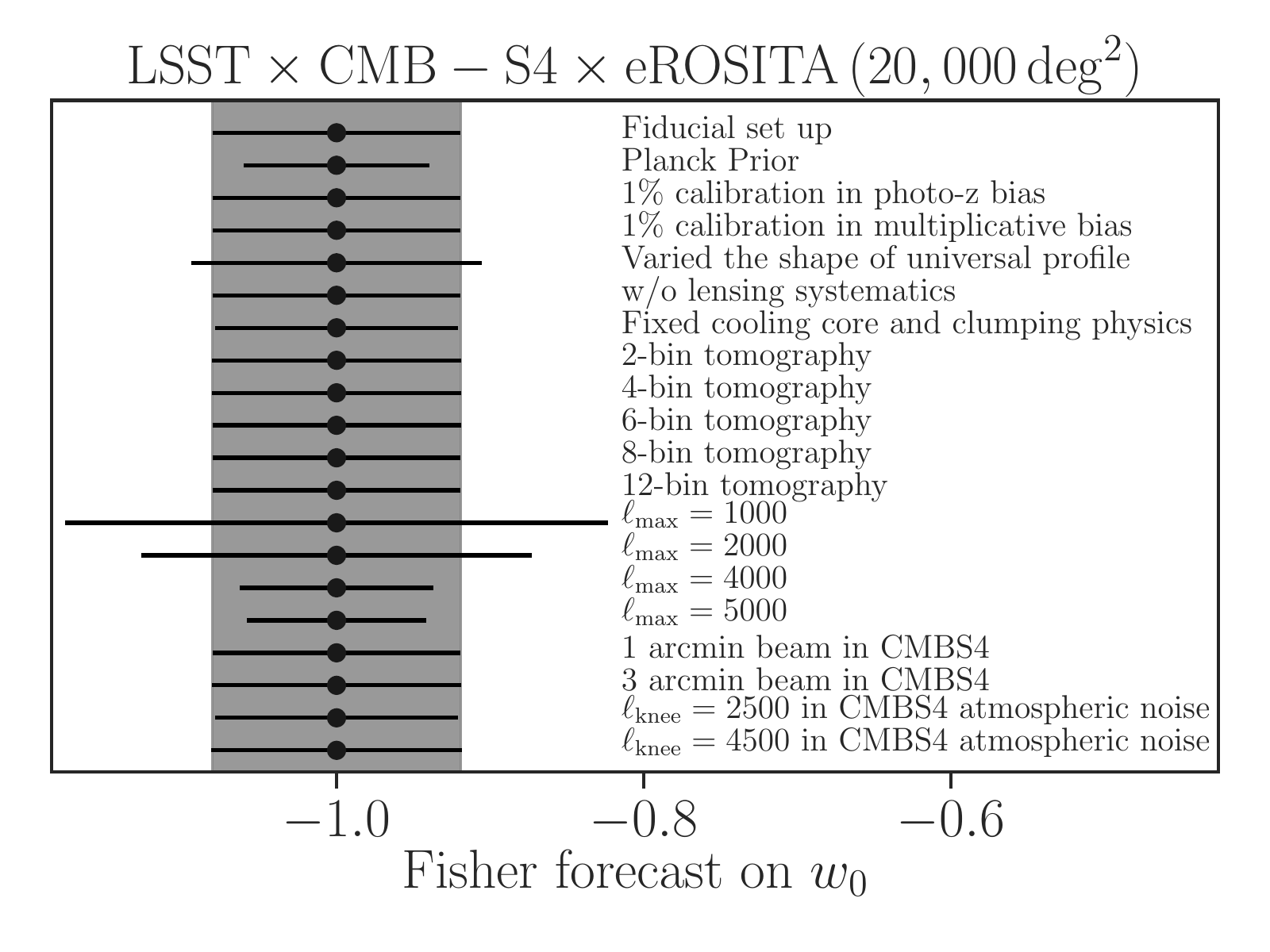}
\caption{
Similar to Figure~\ref{fig:test_summary_error_Ant}, 
but we consider a representative example of cosmological parameters.
The panel shows the dependence of analysis and survey setups on the  parameter of dark energy $w_0$.
}
\label{fig:test_summary_error_w0}
\end{figure} 

\section{Dependence of analysis and survey setups on cosmological constraints} \label{apdx:test_fisher_cosmo}

In this appendix, we provide the results of the expected constraints of the  parameter of dark energy
when varying the analysis and survey setups
as examined in \S\ref{subsec:depend_analysis_survey}. Figure~\ref{fig:test_summary_error_w0} summarizes 
the results. To improve the constraint of $w_0$, we need a prior cosmological information  
from primary CMB fluctuations to reduce the statistical uncertainty in cosmic baryon and mass density, and the spectral index of primordial curvature fluctuations. Also, a small-scale information would be helpful to break the parameter degeneracy among ICM physics.


\bsp	
\label{lastpage}
\end{document}